\documentclass{JFM-FLM_Au}

\usepackage{amsmath}
\usepackage{adjustbox}
\usepackage{multirow}
\usepackage{subcaption}
\usepackage{tabularx}
\usepackage{array}
\usepackage{tikz}
\usetikzlibrary{arrows.meta,positioning}
\graphicspath{{./}{../output/orbit_zoology_phi4_regularity/}}

\newcolumntype{Y}{>{\raggedright\arraybackslash}X}
\definecolor{HydroNew}{RGB}{224,231,255}
\definecolor{HydroNonGeom}{RGB}{209,250,229}
\definecolor{HydroNonGeomBorder}{RGB}{5,150,105}
\newcommand{\glabel}[2]{\(\begin{array}{c}#1\\[-1pt]\scriptstyle #2\end{array}\)}

\definecolor{collapsegray}{gray}{0.9}
\definecolor{nongeom}{rgb}{0.85,0.85,1}
\newcommand{\infcell}[1]{\cellcolor{collapsegray}#1}
\newcommand{\infblank}{\cellcolor{collapsegray}}
\newcommand{\infnongeom}{\cellcolor{nongeom}}

\newcommand{\infmultirowup}[2]{\cellcolor{collapsegray}\multirow{-#1}{4.7em}{\centering #2}}

\newcommand{\OO}{\mathrm{O}}
\newcommand{\SO}{\mathrm{SO}}

\newcommand{\vect}[1]{\boldsymbol{\mathbf{#1}}}

\lefttitle{C. Moreau}
\righttitle{The Stokes resistance of an arbitrary particle}

\title{The Stokes resistance of an arbitrary particle: a classification of hydrodynamic symmetries}

\author{C. Moreau\aff{1}}

\affiliation{\aff{1}Nantes Université, École Centrale Nantes, IMT Atlantique, CNRS, LS2N, UMR 6004, F-44000 Nantes, France}

\corresau{C. Moreau, \email{clement.moreau@cnrs.fr}}

\begin{document}
\maketitle

\begin{abstract}
The linearity of the Stokes equations organises the hydrodynamic response of a rigid particle into a hierarchy of resistance operators, coupling successive truncations of the ambient-flow jet to moments of the surface traction. Since the work of Kelvin and Larmor, it has been known that this response does not resolve particle geometry faithfully: bodies with discrete rotational symmetry may be indistinguishable from bodies of revolution—Brenner’s helicoidal symmetry—and a chiral body may respond isotropically, as in Kelvin’s isotropic helicoid. We regard the resistance operators as elements of finite-dimensional $\OO(3)$-representation spaces and use character formulae to determine, at every level of the hierarchy, which point-group symmetries are hydrodynamically distinguishable and the dimension of each invariant space. This yields an explicit nested sequence of hydrodynamic symmetry-group sets, from the translation–force level to the quadratic-flow level. The framework reveals hydrodynamic classes that no shape can realise geometrically, gives helicoidal symmetry a level-dependent definition, and shows that polyhedral symmetry becomes visible in a strict order: tetrahedral symmetry in shear, octahedral symmetry through the stresslet, and icosahedral symmetry in quadratic flow. Projecting the resistance operators onto force- and torque-free motion provides symmetry-based parameter counts and a constructive route to the corresponding dynamical normal forms. We thereby complete the Jeffery–Bretherton–Ishimoto classification, characterise all hydrodynamic classes producing Jeffery dynamics, and identify chiral tetrahedral and octahedral normal forms that can generate irregular full-attitude dynamics.
\end{abstract}

\begin{keywords}
Stokesian dynamics; suspensions; particle/fluid flow
\end{keywords}


\section{Introduction}

The hydrodynamic resistance of a particle in a zero-Reynolds number flow is one of the oldest problems in fluid mechanics, ever since the derivation of the Stokes equation \citep{Stokes1851}. Finding the force exerted by the fluid on a particle, knowing its motion, is the fundamental building block of many problems related to cell motility, suspensions, and microfluidics.
It has been known since the nineteenth century \citep{Lamb1932} that the linearity of the Stokes equation induces a linear relationship between fluid motion and drag force moments on the particle. 
In the 1960s, Brenner then developed this construction in modern polyadics and tensorial language for
an arbitrary rigid particle in a sequence of papers progressing from uniform
translation and rotation to shear and general ambient Stokes flows
\citep{brenner1964stokes,brenner1964stokesii,brenner1964stokesiii,brenner1964stokesiv,brenner1966stokesv}, and termed the linear operator coupling linear flow to force and torque as the \textit{resistance matrix}. 
Seen as an assembly of coupling tensors of different ranks, it became common to call this object the \textit{grand resistance tensor} in the following decades. The resulting resistance and mobility formulation is now a standard language
of low-Reynolds-number particle mechanics
\citep{happel2012low,kim2013microhydrodynamics}. 

As such, grand resistance matrices and tensors have been studied extensively.
Lorentz reciprocity and viscous dissipation endow these operators with symmetry
and positive definiteness \citep{happel2012low,hinch1972note,masoud2019reciprocal},
whilst their higher blocks include the stresslet governing the leading particle
contribution to suspension stress \citep{batchelor1970stress,hinch1972note} and
the higher force moments used in active-particle hydrodynamics
\citep{elfring2017force,nasouri2018higher}. Closed-form resistance operators are
available only for a handful of shapes: the ellipsoid \citep{oberbeck1876}, and
spheroids and slender bodies through singularity and asymptotic methods
\citep{chwang1975,batchelor1970slender,koens2018}. In general, resistance tensor
entries must be obtained as integrals of the surface traction over the particle
boundary. This has motivated a long line of numerical schemes: boundary-integral
formulations \citep{youngren1975,pozrikidis1992}, induced-force multipole
expansions, which construct the resistance hierarchy to arbitrary order
\citep{felderhof1976,cichocki1994}, Stokesian dynamics for rigid and flexible
bead aggregates \citep{brady1988,gissinger2026resistance}, and bead-shell models
for macromolecules \citep{garciadelatorre1981}. Beyond
suspension rheology, the same operators govern the Brownian diffusion of
arbitrarily shaped particles \citep{brenner1967coupling}, the propulsion matrices
of microswimmers \citep{purcell1997,lauga2009}, chirality-induced drift in
sedimentation \citep{makino2005,witten2020review}, and the orientational
statistics of anisotropic particles in turbulence \citep{voth2017}.

From the outset, a central theme has been that the geometric symmetry of a particle constrains and simplifies the algebraic structure of its resistance tensor -- and, through it, the particle's dynamics. This idea was initially termed \textit{hydrokinetic symmetry}. In \citet{larmor1885hydrokinetic}, we read \begin{quote}
Again, the form which applies to a sphere also applies to a solid having two such axes of [axi]symmetry at right angles to one another. [...] These solids therefore move through the fluid in the same manner as a sphere would move. [...] We can, however, extend these conclusions to solids whose cross-sections are \textit{any} regular figures, and to \textit{any} regular solids, respectively, so that, for example, a right prism or pyramid on an equilateral triangular base [...] has the character of a solid of revolution, and a regular tetrahedron has the character of a sphere.\end{quote}
Later on, the second chapter of Brenner's paper series \citep{brenner1964stokesii} devotes a section to the effect of particle symmetry on the resistance tensor, listing the subsequent cancellations and parameter dependence induced by simple symmetry properties: one, two or three planes of symmetry, one axis of rotational symmetry, \textit{etc}. Then, in agreement with Larmor's observations, it is established that invariance under $n$-fold finite rotation about
an axis, with $n\geqslant 3$, imposes the same structure for the resistance matrix as a body-of-revolution. This is termed \textit{helicoidal symmetry} by Brenner. 
Both Larmor's observations and Brenner's symmetry calculations highlight a rather intriguing phenomenon: from the hydrodynamic point of view of the resistance matrix, some discrete cyclic symmetries are ``invisible'', or indistinguishable from continuous axisymmetry. More generally, resistance operators inherit from particle symmetries by the Curie principle, but the example of helicoidal symmetry shows that there is not a one-to-one correspondence between symmetry classes of physical objects and those of grand resistance tensors. 

Furthermore, it appears that the definition of helicoidal symmetry, or more generally of hydrodynamic symmetry, depends on the type of resistance operator considered. Indeed, consider Jeffery's equations \citep{jeffery1922motion}, which describe the rotational dynamics of spheroids in shear flow. Recently, \citet{fries2017angular} and \citet{ishimoto2020helicoidal} studied an extension of these equations to helicoidal particles, defined as invariance of the particle by some $n$-fold rotation. 
Yet, both studies distinguish the dynamics of helicoidal objects with $n \geqslant 4$, which behave like axisymmetric particles as predicted by Brenner, from objects with $3$-fold rotation symmetry. Additional \textit{triangularity} terms in the rotational equations are required for this class of $3$-fold objects, with a detailed study of their dynamics being carried out in \citet{ishimoto2020jeffery}. 
The discrepancy between that $n \geqslant 4$ helicoidal symmetry and the one in the sense of Brenner requiring only $n\geqslant 3$ is explained by the fact that Brenner considers only the grand resistance matrix without strain component, while Jeffery's equations require additional blocks coupling force and torque to the fluid strain from the grand resistance tensor. 
Hence, objects may or may not be ``hydrodynamically equivalent'' depending on their symmetries and on the retained components of the ambient-flow. 

\begin{figure}
    \centering
    \includegraphics[width=\linewidth]{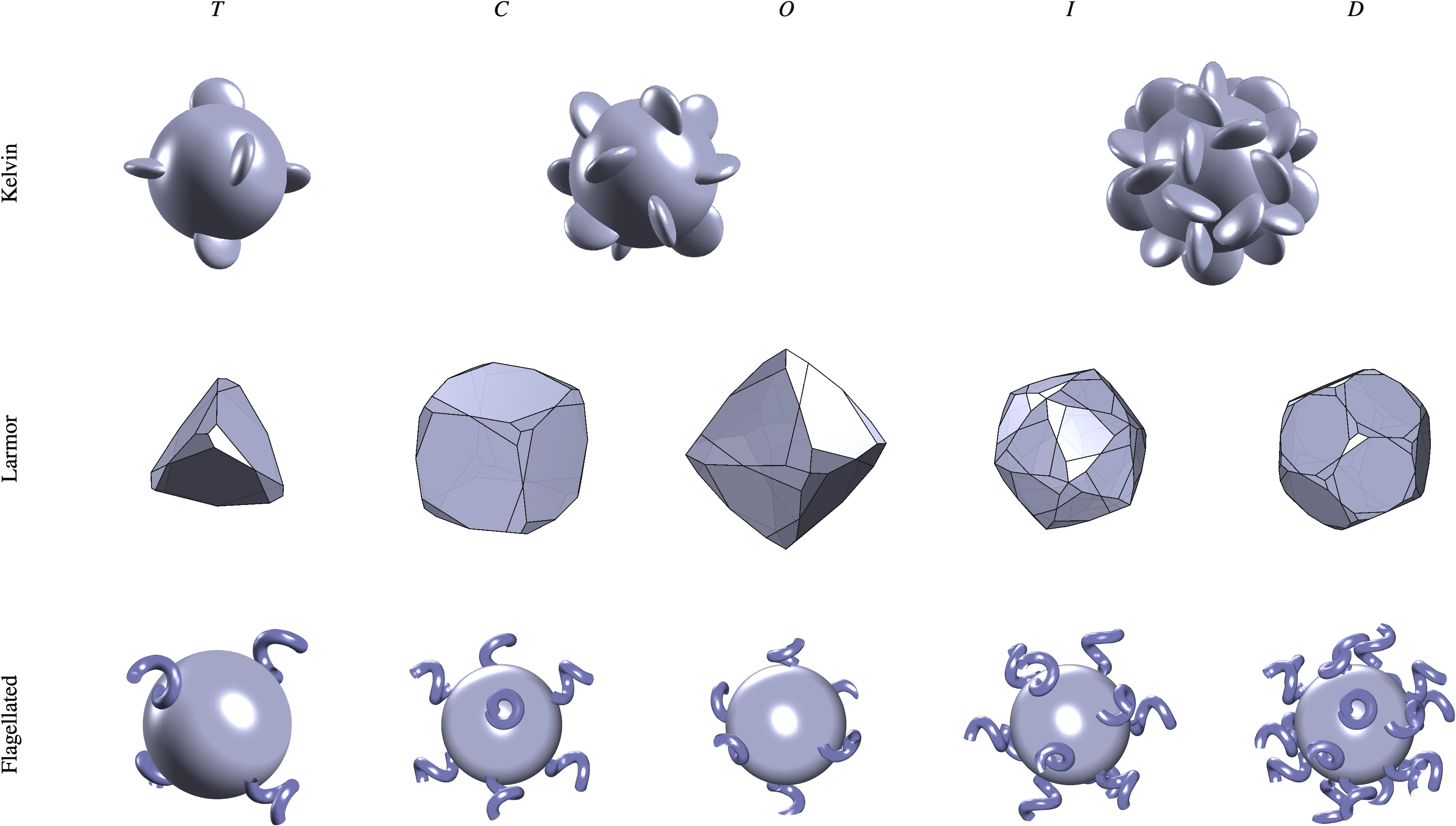}
    \caption{Examples of various attempts at building isotropic helicoids from polyhedral symmetry of the five Platonic solids: tetrahedron (T), cube (C), octahedron (O), icosahedron (I), dodecahedron (D). First row: Kelvin's sphere with vanes placed at the midpoints of polyhedral edges. In this construction, the dual polyhedra (cube/octahedron, icosahedron/dodecahedron) produce identical solids. Original Kelvin helicoid is the C/O one. Second row: Larmor's bevelled polyhedra. Third row: Larmor's spheres with chiral ``decorations'' at polyhedral vertices, called \textit{flagellated}.}
    \label{fig:helicoids}
\end{figure}

The other peculiar emblem of the distinctive character of hydrodynamic symmetry is the \emph{isotropic helicoid}, imagined by Lord Kelvin 
\citep{thomson1871xlvi}:
a chiral body whose resistance is fully isotropic, coupling translation
to rotation identically about every axis, so that it ought to spin as it sinks. A practical realisation of such a particle is described as follows in \citet{thomson1871xlvi}:
\begin{quote}
An isotropic helicoid may be made by attaching projecting
vanes to the surface of a globe in proper positions ; for instance,
cutting at 45° each, at the middles of the twelve quadrants of
any three great circles dividing the globe into eight quadrantal
triangles.
\end{quote}
This isotropic helicoid has octahedral symmetry, while \citet{larmor1885hydrokinetic} proposes an extension to any regular polyhedron as well as other construction schemes:
\begin{quote}
If, however, we take a regular tetrahedron (or other regular solid), and replace the edges by skew bevel faces placed in such wise that when looked at from any corner, they all slope the same way, we have an example of an isotropic helicoid. [...] This would also be the result if three plagiedral faces sloping the same way were imposed on each vertex of the tetrahedron. [...] A form equivalent to [that] is obtained by fixing four equal symmetrical screw-propellers on the surface of a sphere at the corners of an inscribed regular tetrahedron.
\end{quote}
Figure \ref{fig:helicoids} shows the appearance of Kelvin's and Larmor's suggested particles. The original particle of Kelvin was fabricated through 3D printing by \citet{collins2021lord}, showing that its helicoidal behaviour was not observed in experimental conditions due to very weak coupling amplitude. In any case, by analogy with the observations on helicoidal symmetry in generalised Jeffery's equations, it seems that particles with tetrahedral or octahedral symmetry may not be hydrodynamically isotropic under the same conditions (e.g. with or without shear flow). It appears, however, that this distinction has not been systematically characterised beyond the aforementioned case studies.

Overall, the subtle distinction between geometric symmetry and hydrodynamic symmetry is somewhat diffuse in the literature, although it underlies studies on non-spherical particle dynamics, especially in shear flow \citep{gustavsson2016preferential,thorp2019motion}. When studying the influence of rapid particle oscillation on the long-timescale dynamics, \citet{dalwadi2024generalisedI,dalwadi2024generalisedII} explicitly state that the resistance tensor of some particle may induce Jeffery orbits even if the particle is not axisymmetric, and call such particles \textit{Jeffery bodies}. 
However, a clear characterisation of hydrodynamic symmetry and a systematic classification of the resistance operators of particles under various symmetries remains open, to the best of my knowledge.

Hence, the aim of this paper is to provide a framework that separates the symmetry of
the shape from the symmetry of each resistance operator. Representation theory
provides such a framework. Invariant spaces have long been used to classify anisotropic elasticity,
photoelasticity and higher-gradient material tensors
\citep{backus1970geometrical,forte1996symmetry,forte1997symmetry,auffray2013matrix,olive2013symmetry,olive2014symmetry,olive2022characterization,clayton2025symmetries}. In this study, we cast the general hierarchy of hydrodynamic resistance operators as representation-theoretical objects, and derive invariant spaces for all symmetry classes. This allows us to determine the dimension and structure of invariant spaces through character formulae, avoiding tensorial computation. In particular, we are able to systematically identify which types of symmetries are hydrodynamically equivalent at each level of the resistance operator hierarchy. We show that the visible classes of symmetries at each level constitute a finite set that we call the hydrodynamic symmetry group set $\mathfrak{H}_k$. The different notions of helicoidal symmetry and isotropic helicoids naturally fall within this framework. 

Furthermore, we study the rotational and translational dynamics of a particle obtained from the resistance operators by assuming zero net force and torque in shear or quadratic flows. The same representation theory toolbox allows to predict cleanly the number of independent parameters appearing in the equations of motion. For rotational dynamics in shear flow, we recover well-known results: a single parameter for axisymmetric particles \citep{jeffery1922motion,bretherton1962motion}, two for helicoidal particles \citep{ishimoto2020helicoidal}, four for triangular particles \citep{ishimoto2020jeffery}. We refine the characterisation of those particle classes and study new particle types through a systematic classification. Special attention is devoted to the dynamics of tetrahedral and octahedral chiral particles, which exhibit irregular behaviour. As further illustration of the applications of our hydrodynamic symmetry framework, we also investigate dilute suspensions and particle dynamics in quadratic flow. 

The paper is structured as follows. In Section \ref{sec:stokes problem}, we describe the resistance problem in Stokes flow and define the hierarchy of resistance operators. Section \ref{sec:hydrodynamic-symmetry} introduces the necessary concepts of symmetry groups and representation theory, as well as the character formula used to compute the dimension of the invariant subspaces. Section \ref{sec:classification} presents the main results: the structure of the hydrodynamic symmetry group set $\mathfrak{H}_k$ for each resistance level. Applications to particle dynamics are gathered in Section \ref{sec:particle-dynamics}, and Section \ref{sec:discussion} features a few open problems. 

\section{The Stokes resistance problem}
\label{sec:stokes problem}

\subsection{Problem setting and Stokes jet spaces}

We consider a shape space $\mathcal{B}$ consisting of bounded, closed subsets of $\mathbb{R}^3$ with sufficient boundary regularity. Let $B \in \mathcal{B}$ represent a rigid particle immersed in an incompressible Newtonian fluid of viscosity $\mu$ at zero Reynolds number, and let $\mathcal{D}=\mathbb{R}^3\setminus B$ denote the exterior fluid domain. Let $(O,F)$ be a reference origin and a right-handed orthonormal frame, with position vector $\vect{x}$ measured from $O$. The fluid velocity approaches the undisturbed flow $\vect{U}^{\infty}$ in the far field.
The prescribed rigid velocity on the particle boundary $\partial B$ is $\vect{u}=\vect{U}=\vect{V}+\vect{\Omega}\times\vect{x}$, where $\vect{V}$ and $\vect{\Omega}$ are the translational and angular velocities of the particle.

In the fluid domain, the pressure $p(\vect{x})$ and velocity $\vect{u}(\vect{x})$
satisfy the Stokes equations
\begin{align}
    \label{eq:stokes}  \mu \Delta \vect{u} -\nabla p & =0,
  \\
  \label{eq:stokes-incomp} \nabla\cdot \vect{u}& =0.
\end{align}
By linearity, we may subtract the prescribed rigid-body velocity and work in the particle frame. Thus, without loss of generality, the particle is stationary and $\vect{U}=0$; the far-field jet below is understood as the relative incident flow in this frame.

The undisturbed flow $\vect{U}^{\infty}$ may be expanded as an infinite jet $\mathcal{U}$ around $O$, 
\begin{equation}
    \mathcal{U} = \{ \vect{U}^{\infty}(O), \nabla \vect{U}^{\infty}(O), \dots, \nabla^k \vect{U}^{\infty}(O), \dots \}.
\end{equation}
In particular, if $\vect{U}^{\infty}$ is a linear flow, the first two elements in $\mathcal{U}$ constitute an exact representation of $\vect{U}^{\infty}$. The zero-order component $\vect{U}^{\infty}(O)$ is a uniform translation, denoted by $\vect{V}$. The first-order component $\nabla \vect{U}^{\infty}$, which must be traceless by \eqref{eq:stokes-incomp}, is decomposed into its antisymmetric part, represented by the local angular velocity $\vect{\Omega}$ (one half of the vorticity), and its symmetric part $\vect{E}$, the rate-of-strain tensor. Beyond the linear part, the term of polynomial degree $k$ in the expansion is denoted by $\vect{J}_k$.
Then, we define the truncations $\mathcal{U}_k$ and jet spaces sequence $\mathcal{J}_k$ according to the following convention:
\begin{align*}
    \mathcal{U}_0 & =\{ \vect{V} \} \in \mathcal{J}_0, \\ 
    \mathcal{U}_1 & =\{ \vect{V}, \vect{\Omega} \} \in \mathcal{J}_1, \\
    \mathcal{U}_2 & = \{ \vect{V}, \vect{\Omega}, \vect{E} \} \in \mathcal{J}_2, \\
    \mathcal{U}_{k+1} & = \{ \vect{V}, \vect{\Omega}, \vect{E}, \vect{J}_2, \dots, \vect{J}_k \} \in \mathcal{J}_{k+1}, \forall k \geqslant 2.
\end{align*}
The traction on the particle surface is $\vect{t}(\vect{u},p)=\boldsymbol{\sigma}(\vect{u},p)\vect{n}$, where $\boldsymbol{\sigma}=\mu(\nabla\vect{u}+(\nabla\vect{u})^T)-p\mathbf{I}$ is the Newtonian Cauchy stress and $\vect{n}$ is the unit normal pointing into the fluid. The virtual power $P$ for the particle and a virtual velocity field $\hat{\vect{u}}$ is expressed as the surface integral
\begin{equation}
    P(\vect{u},\hat{\vect{u}}) =\int_{\partial B} \vect{t}(\vect{u}) \cdot \hat{\vect{u}} \, \mathrm{d} S.
\end{equation}
By linearity of the Stokes equation, $P(\cdot,\hat{\vect{u}})$ defines a linear form $\mathcal{J}_k \rightarrow \mathbb{R}$ in the dual jet space $\mathcal{J}_k^*$, which identifies the traction surface field $\vect{t}$ as a force moment expansion
\begin{align*}
    \mathcal{F}_0 & =\{ \vect{F} \} \in \mathcal{J}_0^*, \\ 
    \mathcal{F}_1 & =\{ \vect{F},\vect{T}\} \in \mathcal{J}_1^*, \\
    \mathcal{F}_2 & = \{ \vect{F}, \vect{T}, \vect{S} \} \in \mathcal{J}_2^*, \\
    \mathcal{F}_{k+1} & = \{ \vect{F}, \vect{T}, \vect{S}, \vect{G}_2, \dots, \vect{G}_k \} \in \mathcal{J}_{k+1}^*, \forall k \geqslant 2,
\end{align*}
where we used the familiar notations $\vect{F}$ (force), $\vect{T}$ (torque) and $\vect{S}$ (stresslet) for the moments up to linear order. 

\subsection{Resistance operators}

Further, this defines, at each level of the jet space, an operator 
\begin{equation}
    \mathcal{R}_k(B;(O,F)) : \mathcal{J}_k \rightarrow \mathcal{J}_k^*
    \label{eq:resistance-operator}
\end{equation} called the \textit{resistance operator} of order $k$. Equivalently, it is represented by a bilinear form $\mathcal{R}_k : \mathcal{J}_k \times \mathcal{J}_k \rightarrow \mathbb{R}$. By Lorentz reciprocity and viscous dissipation, all resistance operators are symmetric and positive definite. They also scale linearly with fluid viscosity $\mu$; this simple dependence is immaterial to this study's conclusions and will hence be kept implicit henceforth. 

When expressed in coordinates, the linear level operators $\mathcal{R}_1$ and $\mathcal{R}_2$ are represented by the well-known grand resistance matrix and grand resistance tensor:
\begin{equation}\mathcal{R}_1:
\begin{pmatrix} \vect{F} \\ \vect{T} \end{pmatrix}  =   
\begin{pmatrix}
\vect{K} & \vect{C} \\
\vect{C}^T & \vect{Q}  
\end{pmatrix}
 \begin{pmatrix} \vect{V} \\ \vect{\Omega} \end{pmatrix}, \quad
 \mathcal{R}_2:
\begin{pmatrix} \vect{F} \\ \vect{T} \\ \vect{S} \end{pmatrix}  =   
\begin{pmatrix}
\vect{K} & \vect{C} & \vect{\Gamma} \\
\vect{C}^T & \vect{Q} & \vect{\Lambda} \\
\vect{\Gamma}^T & \vect{\Lambda}^T & \vect{\Sigma}
\end{pmatrix}
\begin{pmatrix} \vect{V} \\ \vect{\Omega} \\ \vect{E}\end{pmatrix},
\label{eq:resistance-tensors}
\end{equation}
with each block representing couplings between different moments. More trivially, $\mathcal{R}_0$ is the translation-to-force operator represented by the $3\times 3$ symmetric matrix $\mathbf{K}$. 

A large body of literature pertains to the translational and rotational dynamics of rigid particles in shear flow, dating back to Jeffery's equations \citep{jeffery1922motion}. In that situation, one needs to consider motion moments up to the strain $\vect{E}$, but the stresslet $\vect{S}$ at the same level in $\mathcal{J}_2^*$ is not required to determine the particle dynamics. Therefore, it is common to use a partial version of the grand resistance tensor $\mathcal{R}_2$, coupling a full linear flow to force and torque:
\begin{equation}
     \mathcal{R}_{1.5}:
\begin{pmatrix} \vect{F} \\ \vect{T}  \end{pmatrix}  =   
\begin{pmatrix}
\vect{K} & \vect{C} & \vect{\Gamma} \\
\vect{C}^T & \vect{Q} & \vect{\Lambda} \\
\end{pmatrix}
\begin{pmatrix} \vect{V} \\ \vect{\Omega} \\ \vect{E}\end{pmatrix},
\end{equation}
and we adopt the notation $\mathcal{R}_{1.5}$ for this intermediate resistance operator. In the same spirit, let us define $\mathcal{R}_{0.5}$ as the partial resistance matrix operator $\begin{pmatrix} \vect{K} & \vect{C} \end{pmatrix}$. Higher-order half-integer hydrodynamic levels could be defined similarly, but we will not need them in the present study. We consider $\mathcal{R}_{0.5}$ and $\mathcal{R}_{1.5}$ as part of the resistance operator hierarchy; in the following and unless otherwise stated, the notation $\mathcal{R}_{k}$ refers to those two operators as well as the ones associated with integer values of $k$.

Brenner introduced the resistance hierarchy for an arbitrary particle \citep{brenner1964stokes,brenner1964stokesii,brenner1964stokesiii,brenner1964stokesiv,brenner1966stokesv}; its low-order blocks are now standard in particle mechanics \citep{happel2012low,kim2013microhydrodynamics}. Explicit treatments of higher-order force moments are less common; see, for example, \citet{nasouri2018higher}.

\subsection{Reduced coordinates}
\label{sec:reduced-coordinates}

As highlighted in \eqref{eq:resistance-operator}, a resistance operator depends on the particle shape $B$ and on the reference frame $(O,F)$. Let $O'$ be a new origin and let $\vect{d}$ be the displacement from $O$ to $O'$, with associated skew-symmetric matrix $\hat{\vect{d}}$. Changing the origin from $O$ to $O'$ transforms the block $\vect{C}$ in \eqref{eq:resistance-tensors} into $\vect{C}'=\vect{C}+\vect{K}\hat{\vect{d}}$. Choosing $\hat{\vect{d}}=((\vect{K}^{-1}\vect{C})^T-\vect{K}^{-1}\vect{C})/2$, which cancels the antisymmetric part of $\vect{K}^{-1}\vect{C}$, makes $\vect{K}^{-1}\vect{C}'$ symmetric. This choice is unique and defines the \textit{hydrodynamic centre} $O_B$ of the particle \citep{happel2012low,kim2013microhydrodynamics}. It removes the three gauge-dependent coordinates associated with translations of the reference point. In the following, resistance operators are systematically referred to $O_B$, so that the centred coupling block $\vect{X}=\vect{K}^{-1}\vect{C}'$ is symmetric, carrying six independent components like $\vect{K}$ and $\vect{Q}$, rather than the nine of a general second-rank tensor; from now on, the notation $\mathcal{R}_k(B)$ denotes the resistance tensor transported at the hydrodynamic centre $O_B$.

We also choose an adapted frame $F_B$. Unlike the hydrodynamic centre, this frame is not always unique and depends on the symmetries of $B$. We postpone the precise choice of distinguished axes to \S\ref{sec:hydrodynamic-representation}; depending on the symmetry class, this choice reduces the generic dimension of the block $\vect{K}$ by $0$, $1$ or $3$.

This particular choice of origin $O_B$ and frame $F_B$ allows one to define a \textit{reduced resistance operator} $\tilde{\mathcal{R}}_k(B)$, which is now an intrinsically geometric object depending only on the object shape $B$ and a canonically associated set of coordinates $(O_B,F_B)$. 

\subsection{Irreducible decomposition of the resistance operator}
\label{sec:irreducible-decomposition}

The main purpose of this paper is to study how the resistance operators are affected by orthogonal transformations of $\mathbb{R}^3$, \textit{i.e.} rotations and reflections.
To this end, we propose a representation-theoretic construction of the resistance operators. 
For $\ell \in \mathbb{N}$, introduce $V_\ell$ denoting the irreducible representation of order $\ell$ of the special orthogonal group $\SO (3)$; $V_0$, $V_1$ and $V_2$ respectively represent scalars, vectors and traceless second-rank tensors; more generally, $V_{\ell}$ is a space of dimension $2 \ell + 1$. 
To describe the moments in the jet space in terms of irreducible components, one needs to consider the full orthogonal group $\OO (3)$ including reflection symmetry, which induces two separate representations for each irreducible $\SO (3)$ level. They are usually denoted by adding $-$ exponent for odd parity and $+$ exponent for even parity; for example, at the $V_1$ level, $V_1^{-}$ (odd parity under inversion) and $V_1^{+}$ (even parity) are respectively represented by polar (or true) vectors and axial (or pseudo-) vectors.

Now, the moments in the jet spaces $\mathcal{J}_k$ can be broken down into irreducible representations. Translational velocity $\vect{V}$ is a polar vector in $V_1^{-}$, rotational velocity $\vect{\Omega}$ is an axial vector in $V_1^{+}$, and shear $\vect{E}$ is a traceless symmetric tensor in $V_2^{+}$. Hence the jet spaces may be decomposed into direct sum of irreducible representations:
\begin{equation}
    \mathcal{J}_0 \simeq V_1^-, \quad \mathcal{J}_1 \simeq V_1^- \oplus V_1^+, \quad \mathcal{J}_2 \simeq V_1^- \oplus V_1^+ \oplus V_2^+.
    \label{eq:jet-space-decomposition}
\end{equation}
Using symmetry of the quadratic jet term and incompressibility of the fluid, it is possible to show that the quadratic part of the jet $\mathcal{J}_3$ decomposes into $V_1^- \oplus V_2^- \oplus V_3^-$. Appendix~\ref{app:general-dimension} gives the corresponding harmonic decomposition and its extension to general $\mathcal{J}_k$.

By definition, resistance operators $\mathcal{R}_k$ belong to the space $\mathrm{Sym}^2(\mathcal{J}_k^*)$ of symmetric bilinear forms on $\mathcal{J}_k$. Using the block decomposition \eqref{eq:jet-space-decomposition}, one obtains
\begin{align}
\label{eq:resistance-decomposition-R0}
    \mathcal{R}_0 & \simeq \mathrm{Sym}^2(V_1^-),\\
    \label{eq:resistance-decomposition-R1}
    \mathcal{R}_1 & \simeq \mathrm{Sym}^2(V_1^-) \oplus (V_1^- \otimes V_1^+) \oplus \mathrm{Sym}^2(V_1^+), \\
    \label{eq:resistance-decomposition-R2}
    \mathcal{R}_2 & \simeq \mathcal{R}_1 \oplus (V_1^- \otimes V_2^+) \oplus (V_1^+ \otimes V_2^+) \oplus \mathrm{Sym}^2(V_2^+),
\end{align}
and so on; each term of the direct sums in \eqref{eq:resistance-decomposition-R0}--\eqref{eq:resistance-decomposition-R2} is naturally associated with one of the blocks in the matrix representations \eqref{eq:resistance-tensors}. Such a block decomposition allows a straightforward computation of the dimension of the space $\mathcal{R}_k$, since $\dim (V \oplus W) = \dim V + \dim W$, $\dim (V \otimes W) = \dim V \dim W$ and $\dim (\mathrm{Sym}^2 V) = \dim V (\dim V + 1)/2$. Reduction to the hydrodynamic centre and principal axes reduces the generic dimension by $3$ for $k=0$ and by $6$ for $k>0$, hence the formulas
\begin{equation}
    \dim \tilde{\mathcal{R}}_0 = 3, \; \dim \tilde{\mathcal{R}}_1 = 15, \;
    \dim \tilde{\mathcal{R}}_{1.5} = 45, \; \dim \tilde{\mathcal{R}}_2 = 60, \;
    \dim \tilde{\mathcal{R}}_3 = 345,
\end{equation}
and generally, for $k\geqslant 3$,
\begin{equation}
    \dim \tilde{\mathcal{R}}_{k-1} = \frac{9}{2} k^4 - 18 k^3 + \frac{51}{2} k^2 -15k -3,
    \label{eq:reduced-space-dimension}
\end{equation}
with the detailed calculation given in Appendix~\ref{app:general-dimension}.

\subsection{Action of a shape transformation}

An orthogonal transformation $A \in O(3)$ acting on a shape $B \in \mathcal{B}$ modifies $B$ into $A[B]$. 
Because they are finite-dimensional vector-space representations of the shape space $\mathcal{B}$, the jet spaces $\mathcal{J}_k$ and resistance operators $\mathcal{R}_k$ associated with the modified shape $A[B]$ transform under $A$ through a representation, called the \textit{action} of the orthogonal group $\OO (3)$ on $\mathcal{R}_k$ and denoted by $\rho(A)$. 
Denoting by $Q$ the matrix representation of $A$ in the chosen set of coordinates, the action $\rho(A)$ on irreducible components $V_{\ell}$ explicitly reads as 
\begin{align}
    \rho_{V_1^-} & : \mathbf{V} \mapsto Q \mathbf{V}, \\
    \rho_{V_1^+} & : \vect{\Omega} \mapsto \det (Q) Q \vect{\Omega}, \\
    \rho_{V_2^-} & : \vect{J} \mapsto \det (Q) Q \vect{J} Q^T, \\
    \rho_{V_2^+} & : \vect{E} \mapsto Q \vect{E} Q^T, \\
    \rho_{V_3^-} & : \vect{J} \mapsto (Q_{ia}Q_{jb}Q_{kc}J_{abc})_{ijk}.
\end{align}
Using the definition of the resistance operators and the block decompositions \eqref{eq:resistance-decomposition-R0}--\eqref{eq:resistance-decomposition-R2}, the action of $Q$ on $\mathcal{R}_k$ is represented by a block-diagonal ``grand tensor'' containing the elementary actions $\rho_{V}$ in order; for example,
\begin{equation}
    \rho_{2} : \mathcal{R}_2 \mapsto D(Q) \mathcal{R}_2 D(Q)^T, \quad D(Q) = \begin{pmatrix} Q & 0 & 0 \\ 0 & \det(Q) Q & 0 \\ 0 & 0 & \mathcal{Q} \end{pmatrix},
    \label{eq:action-representation-R2}
\end{equation}
where $\mathcal{Q}$ denotes the tensor representation such that $\mathcal{Q} \vect{E} = Q \vect{E} Q^T$. Here and henceforth, we use the compact notation $\rho_k$ for the action $\rho_{\mathcal{R}_k}$. Actions on individual blocks of $\mathcal{R}_2$ can be deduced from \eqref{eq:action-representation-R2} and straightforwardly adapted to the whole hierarchy of resistance operators.

The language of group actions $\rho(A)$ is particularly useful for manipulating transformations of resistance operators without resorting to explicit tensor calculations. In particular, this greatly simplifies the study of symmetries, \textit{i.e.}\ invariance under transformations $A$. In the following, we will rely on this formalism to study symmetries of the resistance operators. 

\subsection{Mobility operators}

To complete the exposition of the resistance problem, we consider the inverse question of determining the fluid state in $\mathcal{J}_k$ from prescribed force moments $\{\vect{F}, \vect{T}, \dots \}$ in $\mathcal{J}_k^*$. Inverting \eqref{eq:resistance-operator} defines the mobility operator hierarchy
\begin{equation}
    \mathcal{M}_k(B;(O,F)) : \mathcal{J}_k^* \rightarrow \mathcal{J}_k,
    \label{eq:mobility-operator}
\end{equation}
satisfying, for $k \in \mathbb{N}$, $\mathcal{M}_k \mathcal{R}_k = \mathrm{Id}_{\mathcal{J}_k}$ and $\mathcal{R}_k \mathcal{M}_k = \mathrm{Id}_{\mathcal{J}_k^*}$. Block-wise expressions for the mobility operators may be obtained from explicit algebraic inversion of the block resistance operators \citep{happel2012low,kim2013microhydrodynamics}. By duality, the mobility operators transform under the $\OO(3)$ action in the same way as the resistance operators, and their stabilisers coincide. Hence, the symmetry properties and invariant subspace dimensions studied in the following sections are, \textit{mutatis mutandis}, equally applicable to the mobility operator hierarchy.

\section{Hydrodynamic symmetry group}
\label{sec:hydrodynamic-symmetry}

The goal of this section is to exploit the representation structure of $\mathcal{R}_k$ to study its stabilising groups of symmetries. More specifically, we want to answer two questions: which subgroup symmetries are distinguishable at each resistance
level, and what are the dimensions of their fixed spaces? Character formulae
answer both questions without an explicit componentwise tensor calculation.

\subsection{Reminders on the $\OO(3)$ subgroups and shape symmetry}

Up to conjugacy, the closed subgroups of the orthogonal group $\OO(3)$ can be enumerated as follows, where we use the familiar Schoenflies notation \citep{altmann1994point,dresselhaus2008group}:
\begin{itemize}
    \item seven infinite families of finite-order groups, indexed by $n\geqslant 1$: cyclic $C_n$, pyramidal $C_{nv}$, reflection cyclic $C_{nh}$, improper cyclic $S_{2n}$, dihedral $D_n$, prismatic $D_{nh}$ and antiprismatic $D_{nd}$;
    \item seven exceptional polyhedral groups of finite order: three tetrahedral groups $T$, $T_h$, $T_d$, two octahedral groups $O$, $O_h$ and two icosahedral groups $I$, $I_h$;
    \item seven groups of infinite order: five groups of infinite cyclic symmetry $C_{\infty}$, $C_{\infty h}$, $C_{\infty v}$, $D_{\infty}$, $D_{\infty h}$, and two isotropic groups $\SO (3)$ and $\OO (3)$.
\end{itemize}
Note a few particular cases at low symmetry: $C_{1h}$ and $C_{1v}$ are equivalent and are denoted by $C_s$ (simple reflection); $S_2$ contains only the identity and inversion (centrosymmetry) and is denoted by $C_i$; and $D_1$ and $C_2$ are equivalent, with only the latter notation being used. 

In nature, objects such as crystals and biological particles such as micro-organisms and viruses may possess high levels of symmetry and certainly populate most elements of the point group \citep{caspar1962,guasto2012,twarock2019,velhorodrigues2021}, as first wonderfully illustrated by \citet{haeckel1866}. Artificial micro-particles and colloids can be fabricated to exhibit complex symmetry classes which induce a variety of anisotropic behaviours \citep{glotzer2007,damasceno2012,googasian2025532}.

We denote by $\mathfrak{G}$ the set containing all the subgroups. Among these groups, the $C_n$ and $C_{\infty}$, the $D_n$ and $D_{\infty}$, $T$, $O$, $I$, and $\SO(3)$ are chiral, meaning that a shape strictly possessing one of these symmetries is not preserved by any planar reflection. 

This classification is useful to characterise which symmetries an object $B \in \mathcal{B}$ possesses, by defining its \textit{geometric stabiliser}, or simply stabiliser, as the set of orthogonal transformations that preserves $B$:
\begin{equation}
    \mathbb{G}(B) = \left \{ A \in \OO(3) : A[B] = B \right \},
\end{equation}
where $A[B]$ denotes the shape obtained from $B$ after applying $A$.
The stabiliser $\mathbb{G}(B)$ belongs to $\mathfrak{G}$ and we commonly say that $B$ has $\mathbb{G}(B)$-symmetry. 

It is important to note that four elements of $\mathfrak{G}$ -- namely, $C_{\infty}$, $C_{\infty h}$, $D_{\infty}$ and $\SO(3)$ -- cannot be obtained as the exact stabiliser of any shape. In other words, there does not exist a particle $B$ such that $\mathbb{G}(B)$ equals one of these four groups. We call these groups \textit{non-geometric}, in the sense that the symmetry class they represent cannot be strictly realised by a geometric shape. The subset of $\mathfrak{G}$ excluding the four non-geometric groups is denoted by $\mathfrak{G}_{\mathrm{geom}} = \mathfrak{G} \backslash \left \{ C_{\infty}, C_{\infty h}, D_{\infty}, \SO(3) \right \}$.

A characterisation of the non-geometric groups may be formalised as follows. Conversely to $\mathbb{G}(B)$, for a prescribed subgroup $\mathbb{J}$, let $\mathcal{B}^{\mathbb{J}}$ denote the set of shapes whose geometric stabiliser contains $\mathbb{J}$:
\begin{equation}
    \mathcal{B}^{\mathbb{J}} = \left \{ B \in \mathcal{B} : \mathbb{J} \subseteq \mathbb{G}(B) \right \}.
\end{equation}
In representation-theoretic language, $\mathcal{B}^{\mathbb{J}}$ is called the $\mathbb{J}$-invariant set. Then, we define $\hat{\mathbb{J}}^{\mathcal{B}}$ as follows:
\begin{equation}
    \hat{\mathbb{J}}^{\mathcal{B}} = \left \{ A \in \OO (3) : \forall B \in \mathcal{B}^{\mathbb{J}}, A[B] = B \right \}.
\end{equation}
The set $\hat{\mathbb{J}}^{\mathcal{B}}$ is a group and constitutes the largest group in $\mathfrak{G}$ stabilising all shapes in $\mathcal{B}^{\mathbb{J}}$. It is well-known that, for all $\mathbb{J} \in \mathfrak{G}_{\mathrm{geom}}$, $\hat{\mathbb{J}}^{\mathcal{B}} = \mathbb{J}$, but
\begin{equation}
    \hat{C}_{\infty}^{{\mathcal{B}}} = \hat{C}_{{\infty h}}^{{\mathcal{B}}} = C_{\infty v}, \quad \hat{D}_{\infty}^{\mathcal{B}} = D_{\infty h}, \quad \hat{\SO}(3)^{\mathcal{B}} = \OO(3).
\end{equation}
For instance, a shape with $C_{\infty}$-symmetry necessarily possesses $C_{\infty v}$-symmetry; the group $C_{\infty}$ disappears into $C_{\infty v}$ when seen from the point of view of the shape space $\mathcal{B}$. In particular, the three infinite chiral groups $C_{\infty}$, $D_{\infty}$, $\SO (3)$ are non-geometric. In other words, there does not exist a chiral geometric shape with infinite cyclic or spherical symmetry: in the geometric space, the isotropic helicoid does not exist. This notably contrasts with the existence of \textit{hydrodynamically} isotropic helicoids. In the next section, we precisely define the hydrodynamic analogues of $\mathbb{G}(B)$, $\mathcal{B}^{\mathbb{J}}$ and $\hat{\mathbb{J}}^\mathcal{B}$, in order to establish under which conditions such objects can exist.

\subsection{Hydrodynamic representation}
\label{sec:hydrodynamic-representation}

We now turn to the resistance operators and seek to characterise how they inherit symmetry properties as finite-dimensional observables over the shape space $\mathcal{B}$. 

Given a shape $B \in \mathcal{B}$ and the centred resistance operator $\mathcal{R}_k$, we first define the \textit{hydrodynamic stabiliser} of $B$ at level $k$:
\begin{equation}
    \mathbb{H}_k (B)= \left \{ A \in \OO(3) : \mathcal{R}_k(A[B]) = \mathcal{R}_k(B) \right \}.
    \label{eq:hydrodynamic-geometric-stabiliser}
\end{equation}
The action of $A \in \OO(3)$ on $\mathcal{R}_k$ is represented by $\rho(A)$; therefore, the condition for belonging to $\mathbb{H}_k(B)$ in \eqref{eq:hydrodynamic-geometric-stabiliser} becomes $\rho(A) \mathcal{R}_k(B) = \mathcal{R}_k(B)$. Now, instead of seeing $\mathcal{R}_k(B)$ as the image by $\mathcal{R}_k$ of the shape $B$, we can see it as a tensor $R$ in the vector space $\mathcal{R}_k$, and redefine the hydrodynamic stabiliser with respect to $R$, forgetting the underlying shape:
\begin{equation}
    \mathbb{H}_k (R)= \left \{ A \in \OO(3) : \rho(A)R = R \right \}.
\end{equation}
Then, we can naturally construct for the vector space $\mathcal{R}_k$ the same fixed-set and stabiliser
objects as for the shape space $\mathcal{B}$: for $\mathbb{J} \in \mathfrak{G}$, the $\mathbb{J}$-invariant set is given by
\begin{equation}
    \mathcal{R}_k^{\mathbb{J}} = \left \{ R \in \mathcal{R}_k : \forall A \in \mathbb{J}, \rho(A) R = R \right \},
\end{equation}
and the largest group stabilising $\mathcal{R}_k^{\mathbb{J}}$ is given by
\begin{equation}
    \hat{\mathbb{J}}^{\mathcal{R}_k} = \left \{ A \in \OO(3) : \forall R \in \mathcal{R}_k^{\mathbb{J}}, \rho(A) R = R \right \}.
\end{equation}
This terminology allows one to cleanly rephrase statements pertaining to hydrodynamic shape in the literature. For example, the fact that the resistance matrix (${\mathcal{R}_1}$-level) of an object invariant under a $\pi/2$ rotation ($C_4$ symmetry) has so-called helicoidal symmetry (\textit{i.e.} invariance under all rotations, see \citep{brenner1964stokesii}) translates as 
\begin{equation}
    \hat{C}_4^{\mathcal{R}_1} = D_{\infty}.
    \label{eq:example-nongeometric-1}
\end{equation}
The fact that Kelvin’s octahedrally symmetric helicoid has a resistance matrix
invariant under all proper rotations ($\SO (3)$ symmetry) translates as
\begin{equation}
    \hat{O}^{\mathcal{R}_1} = \SO (3).
    \label{eq:example-nongeometric-2}
\end{equation}
From these two examples, we can already highlight an interesting difference between the geometric and hydrodynamic symmetries: $D_{\infty}$ and $\SO (3)$, two \textit{non-geometric} groups, are \textit{visible} at the hydrodynamic level $\mathcal{R}_1$. 

More generally, we saw in the previous section that $\hat{\mathbb{J}}^{\mathcal{B}} = \mathbb{J}$, \textit{i.e.} $\mathbb{J}$ is geometrically visible, for all but four groups in $\mathfrak{G}$. On the other hand, for the hydrodynamic resistance representation spaces $\mathcal{R}_k$, it is not well established, at least systematically, which subgroups of $\mathfrak{G}$ satisfy $\hat{\mathbb{J}}^{\mathcal{R}_k} = \mathbb{J}$, \textit{i.e.} are hydrodynamically \textit{visible} at level $k$. As a matter of fact, since the $\mathcal{R}_k$ spaces are finite-dimensional observations of the flow response, as compared to the infinite-dimensional shape space $\mathcal{B}$, it should not come as a surprise that many more than four groups become invisible, especially for small values of $k$. This warrants a systematic examination of $\hat{\mathbb{J}}^{\mathcal{R}_k}$, for all $\mathbb{J} \in \mathfrak{G}$ and $k \in \mathbb{N}$. For this purpose, we define $\mathfrak{H}_k$ as the set of all groups visible at the $k$-th level of hydrodynamic resistance:
\begin{equation}
    \mathfrak{H}_k  = \left \{ \mathbb{J} \in  \mathfrak{G} : \exists \mathbb{K} \in \mathfrak{G} , \mathbb{J} = \hat{\mathbb{K}}^{\mathcal{R}_k} \right \}.
\end{equation}
Each element $\mathbb{J}$ of $\mathfrak{H}_k$ is associated with a $\mathbb{J}$-invariant set $\mathcal{R}_k^{\mathbb{J}}$, which is a linear subspace of $\mathcal{R}_k$. By construction, the invariant sets associated with elements of $\mathfrak{H}_k$ are all distinct and describe the symmetry structure of the resistance operators $\mathcal{R}_k$. Naturally, one has
\begin{equation}
    \mathfrak{H}_0 \subset \mathfrak{H}_{0.5} \subset \mathfrak{H}_1 \subset \dots \subset \mathfrak{H}_k \subset \dots \subset \mathfrak{G}.
\end{equation}
However, rather remarkably, examples \eqref{eq:example-nongeometric-1}--\eqref{eq:example-nongeometric-2} show that
\begin{equation}
    \forall k > 0, \mathfrak{H}_k \not \subset \mathfrak{G}_{\mathrm{geom}}.
\end{equation}

We now seek to characterise the $\mathcal{R}_k^{\mathbb{J}}$, in particular their dimension as linear subspaces, and provide a systematic description of the $\mathfrak{H}_k$.

\subsection{Characterisation of the hydrodynamic invariant sets}
\label{sec:fixed-space-characterisation}

We rely on two classical methods to study the invariant subspaces $\mathcal{R}_k^{\mathbb{J}}$: character computation and nullspace characterisation. 
Alternatively, their dimensions could be obtained using the more recent \textit{clips}-operator formalism \citep{olive2019effective}, which provides a generic computational framework for linear representations of SO(3) and O(3). For clarity of exposition, we retain the more elementary approach.

\subsubsection{Averaging operator}

In order to compute the dimension of the invariant subspace
$\mathcal{R}_k^{\mathbb J}$, we consider the linear action
$\rho_k$ of $\mathbb J$ on 
$\mathcal{R}_k$. If $\mathbb{J}$ is finite, we define the averaging operator $\Pi_{\mathbb J,k}$ as follows:
\begin{equation}
\begin{array}{l c c c}
      \Pi_{\mathbb J,k}  : & \mathcal{R}_k  & \to & \mathcal{R}_k \\
      & R
  &  \mapsto &
  \frac{1}{|\mathbb J|}
  \sum_{A\in\mathbb J}\rho_k(A)R .
  \end{array}
  \label{eq:fixed-space-projector}
\end{equation}
It is well-known that the operator $\Pi_{\mathbb J,k}$ is a projection on
$\mathcal{R}_k^{\mathbb J}$. Hence,
\begin{equation}
  \dim \mathcal{R}_k^{\mathbb J}
  =
  \operatorname{tr}\Pi_{\mathbb J,k}
  =
  \frac{1}{|\mathbb J|}
  \sum_{A\in\mathbb J}\chi_k(A),
  \label{eq:fixed-space-dimension}
\end{equation}
where we introduced $\chi_k(A)=\operatorname{tr}\rho_k(A)$, called the character of the action $\rho_k$. For the compact infinite groups in $\mathfrak G$, the finite average \eqref{eq:fixed-space-projector} can be
replaced by the Haar average such that $
  \Pi_{\mathbb J,k}R
  =
  \int_{\mathbb J}\rho_k(A)R\,\mathrm d\nu_{\mathbb J}(A),
$
where \(\nu_{\mathbb J}\) is called the normalized Haar probability measure on
\(\mathbb J\); it generalises the averaging formulas and allows one to work with continuous groups likewise \citep{folland2016course}.
The characters satisfy the following identities: for all representation spaces $V,W$,
\begin{equation}
    \chi_{V \oplus W} = \chi_V + \chi_W, \quad \chi_{V \otimes W} = \chi_V \chi_W, \quad \textstyle \chi_{\rm{Sym}^2(V)} (A) = \frac{\chi_V(A)^2 +\chi_V(A^2)}{2}.
    \label{eq:symmetric-square-character}
\end{equation}
These formulae allow the dimensions of the invariant subspaces of $\mathcal{R}_k$ to be computed conveniently by decomposing it over the irreducible blocks detailed in \eqref{eq:resistance-decomposition-R0}--\eqref{eq:resistance-decomposition-R2}.

Let us illustrate this computation on an example: we will calculate $\dim \mathcal{R}_1^{C_{3h}}$, \textit{i.e.} the dimension of the space of resistance matrices invariant under $3$-fold reflection cyclic symmetry. The group $C_{3h}$ contains six elements 
\begin{equation}
    C_{3h} = \{\mathrm{id},r_{2\pi/3},r_{-2\pi/3},\sigma_h, \sigma_h r_{2\pi/3}, \sigma_h r_{-2\pi/3} \},
\end{equation} where $r_{\theta}$ and $\sigma_h$ denote, respectively, rotation by $\theta$ about the symmetry axis and reflection in the horizontal plane. By \eqref{eq:resistance-decomposition-R1}, $\mathcal{R}_1  \simeq \mathrm{Sym}^2(V_1^-) \oplus (V_1^- \otimes V_1^+) \oplus \mathrm{Sym}^2(V_1^+)$, so the identities \eqref{eq:symmetric-square-character} reduce \eqref{eq:fixed-space-dimension} to
\begin{equation}
    \dim \mathcal{R}^{C_{3h}}_1 = \frac{1}{6} \sum_{A \in C_{3h}} \left ( \frac{\chi_{V_1^-}(A)^2 +\chi_{V_1^-}(A^2)}{2} + \chi_{V_1^-}(A) \chi_{V_1^+}(A) + \frac{\chi_{V_1^+}(A)^2 +\chi_{V_1^+}(A^2)}{2} \right ).
\end{equation}
Since the representations of $A$ on $V_1^-$ and $V_1^+$ are respectively given by $A$ and $\det(A)A$, we deduce
\begin{align}
    \chi_{V_1^-}(r_{\theta})  = \chi_{V_1^+}(r_{\theta}) &= 2 \cos\theta + 1, \\
    \chi_{V_1^-}(\sigma _h r_{\theta}) & =  2 \cos\theta - 1, \\
    \chi_{V_1^+}(\sigma _h r_{\theta}) & = 1 - 2\cos \theta,
\end{align}
which yields after straightforward computations
\begin{equation}
    \dim \mathcal{R}^{C_{3h}}_1 = \frac{1}{6}  \left ( 18 + 0 + 0 + 2+ 2+ 2\right ) = 4.
\end{equation}
We conclude that resistance matrices invariant under $C_{3h}$ symmetry have four independent (generally) non-zero parameters. This agrees with the resistance tensors in \citet{brenner1964stokesii} for shapes with an axis of ``helicoidal symmetry'' and a reflection plane, for which $\mathbf{C}=0$, $\mathbf{K}=\operatorname{diag}(k_1,k_1,k_2)$ and $\mathbf{Q}=\operatorname{diag}(q_1,q_1,q_2)$ .

The character computation method has the advantage of completely avoiding tensorial computation, which makes it relatively easy to compute by hand and derive analytical expressions for invariant-subspace dimensions, even when generalised to higher levels of hydrodynamic operators. 

In particular, we can obtain generic formulae for the dimension of $\mathcal{R}_k^{\mathbb{J}}$ for any $k$ and $\mathbb{J}$, similar to the dimension of the full $\mathcal{R}_k$ space in \eqref{eq:reduced-space-dimension}. Detailed class-sum and residue formulae, including the necessary quasi-polynomial congruence subcases, are provided in Appendix~\ref{app:family-dimensions}. For finite groups, the dimensions grow with $k^4$; for infinite axial groups, they grow with $k^3$; for the two spherical groups $\SO(3)$ and $\OO(3)$, they grow linearly. 

\subsubsection{Nullspace}

Whilst the character method allows clean computation of invariant-subspace dimensions, it does not provide detailed information on the structure of these subspaces, \textit{i.e.} where the independent parameters lie in the resistance-operator sub-blocks.
To gather such information, we adopt a different viewpoint which exploits the vector space structure of $\mathcal{R}^{\mathbb{J}}_k$. Let us first deal with a finite group $\mathbb{J} \in \mathfrak{G}$. 
Recall that, by definition, any tensor element $R$ in $\mathcal{R}^{\mathbb{J}}_k$ satisfies, for all $A \in \mathbb{J}$, $\rho(A) R = R$, which rewrites as a linear equation $( \rho(A)-\mathrm{Id} ) R = 0$. Denoting the elements of $\mathbb{J}$ by $\{a_1, \dots, a_p \}$, we may assemble all the linear constraints on $R$ into a linear system
\begin{equation}
    \begin{pmatrix} \rho(a_1)-\mathrm{Id}  \\ \rho(a_2)-\mathrm{Id} \\ \vdots \\ \rho(a_p)-\mathrm{Id} \end{pmatrix} R = \mathbf{L}R = 0.
\end{equation}
Then, one simply has 
\begin{equation}
    \dim \mathcal{R}^{\mathbb{J}}_k = \dim \ker \mathbf{L} = \dim \mathcal{R}_k - \operatorname{rank} \mathbf{L}.
\end{equation}
Continuous groups can be dealt with similarly by sampling them in the same linear system using Lie algebra generators (see Appendix \ref{app:numerical-audits}). This provides a robust numerical framework: we perform a singular value decomposition of $\mathbf{L}$ and count its non-zero singular values. Further, any basis of $\ker \mathbf{L}$ obtained from this decomposition provides a set of elementary basis tensors spanning the invariant subspace of $\mathbb{J}$. 

The calculation through both character and nullspace methods can be automated formally and numerically. The complete dimension audit is detailed in Appendix~\ref{app:numerical-audits}, and Appendix~\ref{app:normal-forms} explains the construction of blockwise normal forms.

\section{Classification of hydrodynamic symmetries}
\label{sec:classification}

The full structure of $\mathfrak{H}_k$ for $k$ up to $3$, as well as the dimensions of the invariant subspaces, is presented in table~\ref{tab:master-observability}. In the following sections, we provide comments and interpretations for each hydrodynamic level. 

Before proceeding, we explicitly set, for each symmetry group of $B$, the coordinate choice defining the adapted axes associated to the reduced hydrodynamic operators $\tilde{\mathcal{R}}_k$. We must distinguish three cases:
\begin{itemize}
    \item \textit{$B$ has (strictly) $C_1$ or $C_i$ symmetry}. No particular coordinates are privileged by symmetry. Then, we choose an orthogonal frame $F_B$ in which the block $\mathbf{K}$ is diagonal. This reduces the generic dimension of $\mathbf{K}$ by 3. 
    \item \textit{$B$ has (strictly) $C_s$, $C_n$, $C_{nh}$ or $S_{2n}$ symmetry}. The first axis is fixed by the symmetry axis (or the axis normal to the reflection plane for $C_s$). Then, no particular coordinates are privileged by symmetry in the remaining plane; we choose the remaining two axes to diagonalise the remaining $2 \times 2$ block in $\mathbf{K}$. Within the subspace of shapes with such symmetry class, this operation reduces the generic dimension of $\mathbf{K}$ by 1. 
    \item In any other case, symmetries provide enough distinguished directions to make an unambiguous (although usually non-unique) choice of coordinates, and this frame choice does not impact the dimension of the invariant subspace. 
\end{itemize}
The reduced operator $\tilde{\mathcal{R}}_k$ corresponds to the most relevant coordinate choice for a given particle $B$. For a given class of symmetry, it reflects the physically relevant dimension of the invariant space. However, unlike the hydrodynamic centre, the frame choice $F_B$ is \textit{not} preserved by action of a transformation $A$ in $\OO (3)$, and therefore the $\tilde{\mathcal{R}}_k$ are not embedded with a linear space structure. For that reason, the stabilisers $\mathcal{R}_k^{\mathbb{J}}$ in \S\ref{sec:hydrodynamic-representation} were defined on the centred operators $\mathcal{R}_k$. Then, we subtract $s_k^\mathbb{J}$ from $\dim \mathcal{R}_k^{\mathbb{J}}$, where $s_k^\mathbb{J}$ equals 0, 1 or 3 according to the list above. The final dimensions reported in table \ref{tab:master-observability} correspond to $\dim \mathcal{R}_k^{\mathbb{J}} - s_k^\mathbb{J}$. When referring to the quotient spaces obtained after setting the reduced choice of coordinates, we will occasionally use the notations $\tilde{\mathcal{R}}_k^{\mathbb{J}}$ and $\mathbb{J}^{\tilde{\mathcal{R}}_k}$.

\begin{table}
\centering
\footnotesize
\setlength{\tabcolsep}{2pt}
\renewcommand{\arraystretch}{1.05}
\begin{adjustbox}{max totalsize={\textwidth}{0.8\textheight},center}
\begin{tabular}{@{}c @{}c|*{6}{c}@{}}
\toprule
\multicolumn{2}{c|}{Group \( \mathbb{J} \in \mathfrak{G} \)} &
${\tilde{\mathcal{R}}_0}$ & ${\tilde{\mathcal{R}}_{0.5}}$ & ${\tilde{\mathcal{R}}_1}$ &
${\tilde{\mathcal{R}}_{1.5}}$ & ${\tilde{\mathcal{R}}_2}$ & ${\tilde{\mathcal{R}}_3}$ \\
\hline
\multirow{4}{*}{Elementary} & \(C_1\) & \infblank & \(9\) & \(15\) & \(45\) & \(60\) & \(345\)\\
& \(C_2\) & \infblank & \(7\) & \(11\) & \(25\) & \(34\) & \(181\)\\
& \(C_s\) & \infblank & \(5\) & 9 & 24 & \(33\) & \(180\)\\
& \(C_i\) & \infmultirowup{4}{\(D_{2 h}\)} & \infblank \(D_{2h}\) & 9 & 24 & \(39\) & \(204\)\\
\hline
\multirow{4}{*}{Cyclic} & \(C_3\) & \infblank & \infblank & \infblank & \(15\) & \(20\) & \(115\)\\
& \(C_4\) & \infblank & \infblank & \infblank & \infblank & \(16\) & \(89\)\\
& $C_5, C_6$ & \infblank & \infblank & \infblank & \infblank & \infblank & $71, 67$ \\
& \(C_7, C_8\) & \infmultirowup{4}{\(D_{\infty h}\)} & \infmultirowup{4}{\(D_\infty\)} & \infmultirowup{4}{\(D_\infty\)} & \infmultirowup{3}{\(C_\infty\)} & \infmultirowup{2}{\(C_\infty\)} & \infcell{\(C_\infty\)}\\
\hline
\multirow{5}{*}{Pyramidal} & \(C_{2v}\) & \infblank \(D_{2h}\)  & \(4\) & \(7\) & \(14\) & \(20\) & \(98\)\\
& \(C_{3v}\) & \infblank & \infblank & \infblank & \(9\) & \(13\) & \(65\)\\
& \(C_{4v}\) & \infblank & \infblank & \infblank & \infblank & \(11\) & \(52\)\\
& \(C_{5v},C_{6v}\) & \infblank & \infblank & \infblank & \infblank & \infblank & \(43,41\)\\
& \(C_{7v},C_{8v}\) & \infmultirowup{4}{\(D_{\infty h}\)} & \infmultirowup{4}{\(D_{\infty h}\)} & \infmultirowup{4}{\(D_{\infty h}\)} & \infmultirowup{3}{\(C_{\infty v}\)} & \infmultirowup{2}{\(C_{\infty v}\)} & \infcell{\(C_{\infty v}\)}\\
\hline
\multirow{5}{*}{Reflection cyclic} & \(C_{2h}\) & \infblank \(D_{2h}\) & \infblank \(D_{2h}\) & \(7\) & \(14\) & \(23\) & \(110\)\\
& \(C_{3h}\) & \infblank & \infblank & \infblank & \(8\) & \(11\) & \(60\)\\
& \(C_{4h}\) & \infblank & \infblank & \infblank & \infblank & \(11\) & \(54\)\\
& \(C_{5h},C_{6h}\) & \infblank & \infblank & \infblank & \infblank & \infblank & \(40, 40\)\\
& \(C_{7h},C_{8h}\) & \infmultirowup{4}{\(D_{\infty h}\)} & \infmultirowup{4}{\(D_{\infty h}\)} & \infmultirowup{4}{\(D_{\infty h}\)} & \infmultirowup{3}{\(C_{\infty h}\)} & \infmultirowup{2}{\(C_{\infty h}\)} & \infcell{\(C_{\infty h}\)}\\
\hline
\multirow{3}{*}{Rotoreflection} & \(S_4\) & \infblank & \infcell{$D_{2d}$} & \infcell{$D_{2d}$} & \(12\) & \(17\) & \(90\)\\
& \(S_6\) & \infblank & \infblank & \infblank & \(8\) & \(13\) & \(68\)\\
& \(S_8, S_{10}\) & \infmultirowup{3}{\(D_{\infty h}\)} & \infmultirowup{2}{\(D_{\infty h}\)} & \infmultirowup{2}{\(D_{\infty h}\)} & \infcell{\(C_{\infty h}\)} & \infcell{\(C_{\infty h}\)} & \(46, 42\)\\
\hline
\multirow{5}{*}{Dihedral} & \(D_2\) & \infblank \(D_{2h}\) & \(6\) & \(9\) & \(15\) & \(21\) & \(99\)\\
& \(D_3\) & \infblank & \infblank & \infblank & \(10\) & \(14\) & \(66\)\\
& \(D_4\) & \infblank & \infblank & \infblank & \infblank & \(12\) & \(53\)\\
& \(D_{5},D_{6}\) & \infblank & \infblank & \infblank & \infblank & \infblank & \(44, 42\)\\
& \(D_{7},D_{8}\) & \infmultirowup{4}{\(D_{\infty h}\)} & \infmultirowup{4}{\(D_\infty\)} & \infmultirowup{4}{\(D_\infty\)} & \infmultirowup{3}{\(D_\infty\)} & \infmultirowup{2}{\(D_\infty\)} & \infcell{\(D_\infty\)}\\
\hline
\multirow{5}{*}{Prismatic} & \(D_{2h}\) & \(3\) & \(3\) & \(6\) & \(9\) & \(15\) & \(63\)\\
& \(D_{3h}\) & \infblank & \infblank & \infblank & \(6\) & \(9\) & \(38\)\\
& \(D_{4h}\) & \infblank & \infblank & \infblank & \infblank & \(9\) & \(35\)\\
& \(D_{5h},D_{6h}\) & \infblank & \infblank & \infblank & \infblank & \infblank & \(28, 28\)\\
& \(D_{7h},D_{8h}\) & \infmultirowup{4}{\(D_{\infty h}\)} & \infmultirowup{4}{\(D_{\infty h}\)} & \infmultirowup{4}{\(D_{\infty h}\)} & \infmultirowup{3}{\(D_{\infty h}\)} & \infmultirowup{2}{\(D_{\infty h}\)} & \infcell{\(D_{\infty h}\)}\\
\hline
\multirow{4}{*}{Antiprismatic} & \(D_{2d}\) & \infblank & $3$ & \(5\) & \(8\) & \(12\) & \(53\)\\
& \(D_{3d}\) & \infblank & \infblank & \infblank & \(6\) & \(10\) & \(42\)\\
& \(D_{4d}, D_{5d}\) & \infblank & \infblank & \infblank & \infblank & \infblank & \(31, 29\)\\
& \(D_{6d}, D_{7d}\) & \infmultirowup{4}{\(D_{\infty h}\)} & \infmultirowup{3}{\(D_{\infty h}\)} & \infmultirowup{2}{\(D_{\infty h}\)} & \infmultirowup{2}{\(D_{\infty h}\)} & \infmultirowup{2}{\(D_{\infty h}\)} & \infmultirowup{1}{\(D_{\infty h}\)}\\
\hline
\multirow{3}{*}{Polyhedral chiral} & \(T\) & \infblank & \infblank & \infblank & \(5\) & \(7\) & \(33\)\\
& \(O\) & \infblank & \infblank & \infblank & \infblank & \(5\) & \(20\)\\
& \(I\) & \infmultirowup{3}{\(\OO(3)\)} & \infmultirowup{3}{\(\SO(3)\)} & \infmultirowup{3}{\(\SO(3)\)} & \infmultirowup{2}{\(\SO(3)\)} & \infcell{\(\SO(3)\)} & \(11\)\\
\hline
\multirow{4}{*}{Polyhedral achiral} & \(T_d\) & \infblank & \infblank & \infblank & 3 & 5 & \(20\)\\
& \(T_h\) & \infblank & \infblank & \infblank & 3 & 5 & \(21\)\\
& \(O_h\) & \infblank & \infblank & \infblank & \infblank & \(4\) & \(14\)\\
& \(I_h\) & \infmultirowup{4}{\(\OO(3)\)} & \infmultirowup{4}{\(\OO(3)\)} & \infmultirowup{4}{\(\OO(3)\)} & \infmultirowup{2}{\(\OO(3)\)} & \infcell{\(\OO(3)\)} & \(8\)\\
\hline
\multirow{5}{*}{Infinite cyclic} & \infnongeom \(C_\infty\) & \infblank & \infcell{$D_{\infty}$} & \infcell{$D_{\infty}$} & \(12\) & \(15\) & \(66\)\\
& \infnongeom \(C_{\infty h}\) & \infblank & \infblank & \infblank & 7 & 10 & \(39\)\\
& \(C_{\infty v}\) & \infblank & \infmultirowup{2}{$D_{\infty h}$}  & \infmultirowup{2}{$D_{\infty h}$} & 7 & 10 & \(40\)\\
& \infnongeom \(D_{\infty}\) & \infmultirowup{4}{\(D_{\infty h}\)} & \(4\) & \(6\) & \(8\) & \(11\) & \(41\)\\
& \(D_{\infty h}\) & \(2\) & \(2\) & \(4\) & \(5\) & \(8\) & \(27\)\\
\hline
\multirow{2}{*}{$\quad$Infinite spherical$\quad$} & \infnongeom \(\SO(3)\) & \infcell{\(\OO(3)\)} & \(2\) & \(3\) & \(3\) & \(4\) & \(10\)\\
& \(\OO(3)\) & \(1\) & \(1\) & \(2\) & \(2\) & \(3\) & \(7\)\\
\bottomrule
\end{tabular}
\end{adjustbox}
\caption{Hydrodynamic observability table. The table shows the visibility of $\mathbb{J} \in \mathfrak{G}$ among a selected list on the first column, and the dimension of its invariant subspace for hydrodynamic levels from $k=0$ to $k=3$. If $\hat{\mathbb{J}}^{\tilde{\mathcal{R}}_k} \neq \mathbb{J}$, then $\hat{\mathbb{J}}^{\tilde{\mathcal{R}}_k}$ is shown under $\tilde{\mathcal{R}}_k$ column, on a gray background. 
If $\hat{\mathbb{J}}^{\tilde{\mathcal{R}}_k} = \mathbb{J}$, then the reduced dimension of the invariant subspace $\dim \tilde{\mathcal{R}}_k^\mathbb{J}$ is shown. Non-geometric groups in the first column are shaded in blue.
}
\label{tab:master-observability}
\end{table}

\subsection{Pure translation level}
\label{sec:pure-translation}

\begin{figure}
    \centering
    $\quad$
    \begin{subfigure}{0.09\textwidth}
        \centering
        \begin{tikzpicture}[
          x=1.15cm,y=0.78cm,
          every node/.style={align=center, font=\scriptsize, inner sep=3pt},
          new/.style={text=blue},
          hydro/.style={text=red},
          arr/.style={-{Latex[length=1.6mm]}, thin}
        ]
        \node(0) at (0,0) {$\,$};
        \node (D2h) at (0,0.5) {\glabel{D_{2h}}{3}};
        \node (Dinfh) at (0,3.5) {\glabel{D_{\infty h}}{2}};
        \node (O3) at (0,6.5) {\glabel{\OO(3)}{1}};
        
        \draw[arr] (D2h) -- (Dinfh);
        \draw[arr] (Dinfh) -- (O3);
        \end{tikzpicture}
        \caption{$\mathfrak{H}_0$}
        \label{fig:diagram-R0}
    \end{subfigure}
    $\qquad$
    \begin{subfigure}{0.4\textwidth}
        \begin{tikzpicture}[
          x=0.75cm,y=0.7cm,
          every node/.style={align=center, font=\scriptsize, inner sep=0pt},
          new/.style={text=blue},
          hydro/.style={text=red},
          arr/.style={-{Latex[length=1.6mm]}, thin}
        ]
        \node[new] (C1) at (0,-0.2) {\glabel{C_1}{9}};
        \node[new] (Cs) at (-2,1) {\glabel{C_s}{5}};
        \node[new] (C2) at (2,1) {\glabel{C_2}{7}};
        \node[new] (C2v) at (-2,3) {\glabel{C_{2v}}{4}};
        \node[new] (D2) at (2,3) {\glabel{D_2}{6}};
        \node (D2h) at (-2,5) {\glabel{D_{2h}}{3}};
        \node[new] (D2d) at (0,5) {\glabel{D_{2d}}{3}};
        \node[hydro] (Dinf) at (1,6) {\glabel{D_\infty}{4}};
        \node (Dinfh) at (-1,7) {\glabel{D_{\infty h}}{2}};
        \node[hydro] (SO3) at (3,7) {\glabel{\SO(3)}{2}};
        \node (O3) at (1,8) {\glabel{\OO(3)}{1}};
        
        \draw[arr] (C1) -- (Cs);
        \draw[arr] (C1) -- (C2);
        \draw[arr] (Cs) -- (C2v);
        \draw[arr] (C2) to[out=165,in=-30] (C2v);
        \draw[arr] (C2) -- (D2);
        \draw[arr] (C2v) -- (D2h);
        \draw[arr] (D2) to[out=165,in=-30] (D2h);
        \draw[arr] (D2) -- (D2d);
        \draw[arr] (Dinf) -- (Dinfh);
        \draw[arr] (Dinf) -- (SO3);
        \draw[arr] (D2h) -- (Dinfh);
        \draw[arr] (D2d) -- (Dinfh);
        \draw[arr] (Dinfh) -- (O3);
        \draw[arr] (SO3) -- (O3);
        
        \draw[arr] (C2v) -- (D2d);
        \draw[arr] (D2) -- (Dinf);
        \end{tikzpicture}
        \caption{$\mathfrak{H}_{0.5}$}
        \label{fig:diagram-R0.5}
    \end{subfigure}
    \begin{subfigure}{0.4\textwidth}
        \begin{tikzpicture}[
          x=0.75cm,y=0.7cm,
          every node/.style={align=center, font=\scriptsize, inner sep=0pt},
          new/.style={text=blue},
          hydro/.style={text=red},
          arr/.style={-{Latex[length=1.6mm]}, thin}
        ]
        \node (C1) at (0,-0.2) {\glabel{C_1}{15}};
        \node (Cs) at (-2,1) {\glabel{C_s}{9}};
        \node[new] (Ci) at (0,1) {\glabel{C_i}{9}};
        \node (C2) at (2,1) {\glabel{C_2}{11}};
        \node (C2v) at (-2,3) {\glabel{C_{2v}}{7}};
        \node[new] (C2h) at (0,3) {\glabel{C_{2h}}{7}};
        \node (D2) at (2,3) {\glabel{D_2}{9}};
        \node (D2h) at (-2,5) {\glabel{D_{2h}}{6}};
        \node (D2d) at (0,5) {\glabel{D_{2d}}{5}};
        \node[hydro] (Dinf) at (1,6) {\glabel{D_\infty}{6}};
        \node (Dinfh) at (-1,7) {\glabel{D_{\infty h}}{4}};
        \node[hydro] (SO3) at (3,7) {\glabel{\SO(3)}{3}};
        \node (O3) at (1,8) {\glabel{\OO(3)}{2}};
        
        \draw[arr] (C1) -- (Cs);
        \draw[arr] (C1) -- (Ci);
        \draw[arr] (C1) -- (C2);
        \draw[arr] (Cs) -- (C2v);
        \draw[arr] (Cs) -- (C2h);
        \draw[arr] (Ci) -- (C2h);
        \draw[arr] (C2) to[out=165,in=-30] (C2v);
        \draw[arr] (C2) -- (C2h);
        \draw[arr] (C2) -- (D2);
        \draw[arr] (C2v) -- (D2h);
        \draw[arr] (C2h) -- (D2h);
        \draw[arr] (D2) to[out=165,in=-30] (D2h);
        \draw[arr] (D2) -- (D2d);
        \draw[arr] (Dinf) -- (Dinfh);
        \draw[arr] (Dinf) -- (SO3);
        \draw[arr] (D2h) -- (Dinfh);
        \draw[arr] (D2d) -- (Dinfh);
        \draw[arr] (Dinfh) -- (O3);
        \draw[arr] (SO3) -- (O3);
        
        \draw[arr] (C2v) -- (D2d);
        \draw[arr] (D2) -- (Dinf);
        \end{tikzpicture}
        \caption{$\mathfrak{H}_{1}$}
        \label{fig:diagram-R1}
    \end{subfigure}
    \caption{Lattice diagrams of the hydrodynamic symmetry groups at levels $0$, $0.5$ and $1$. An arrow from $\mathbb{J}_1$ to $\mathbb{J}_2$ indicates that $\tilde{\mathcal R}_k^{\mathbb J_2} \subset \tilde{\mathcal R}_k^{\mathbb J_1}$. The dimension of $\tilde{\mathcal R}_k^{\mathbb J}$ is written under each node. Red nodes indicate non-geometric groups; blue nodes indicate groups that become visible at the current level.}
    \label{fig:diagram-R0-0.5-1}
\end{figure}
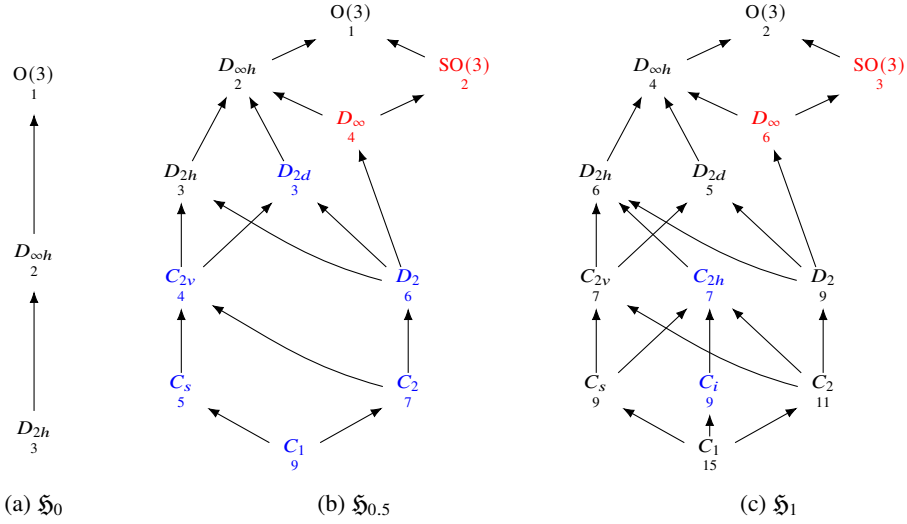

Unsurprisingly, this level of pure translation-force coupling shows a rather trivial hydrodynamic symmetry group containing three elements:
\begin{equation}
    \mathfrak{H}_0 = \{ D_{2h}, D_{\infty h}, \OO (3) \}.
\end{equation}
The corresponding lattice diagram, reduced to a simple chain, is shown in figure~\ref{fig:diagram-R0-0.5-1}(a). The three associated invariant spaces respectively contain unconstrained diagonal matrices $\operatorname{diag}(k_1,k_2,k_3)$, diagonal matrices with a double eigenvalue $\operatorname{diag}(k_1,k_1,k_2)$, and scalar matrices $k \,\mathrm{Id}$. This means that at this level, any object behaves either as an ellipsoid ($D_{2h}$), a spheroid ($D_{\infty h}$), or a sphere ($\OO (3)$).

One should underline again that belonging to $\mathcal{R}^{\mathbb{J}}_k$ merely indicates hydrodynamic symmetry, and does not imply any particular geometric symmetries of the underlying object. Here, for example,
objects in $\hat{\OO}(3)^{\mathcal{R}_0}$, \textit{i.e.} behaving as spheres at the $\mathcal{R}_0$-level, need not have geometrical spherical symmetry. To get a better intuition, imagine an axisymmetric dumbbell-like object with $D_{\infty h}$ symmetry. Its $\mathcal{R}_0$ operator then also has $D_{\infty h}$-symmetry, meaning that its resistance matrix can be written as $\mathbf{K} = \operatorname{diag}(k_1,k_1,k_2)$. But, depending on the size and profile of the dumbbell, it may very well, by accident, satisfy $k_1=k_2$, in which case the dumbbell's resistance operator would fall into the $\OO(3)$-invariant subspace, and the dumbbell would behave strictly like a sphere at $k=0$ hydrodynamic level. This concept of an accidental symmetry of the resistance operator can occur at all levels, and we will discuss its role at higher levels in the following. 

Figure~\ref{fig:diagram-R0-0.5-1} and the subsequent higher-level diagrams illustrate, in particular, which accidental symmetries are allowed. For an object $B$ with $\mathbb{J}$ symmetry and $\mathbb{K} \in \mathfrak{H}_k$, the hydrodynamic operator $\mathcal{R}_k(B)$ may accidentally belong to the invariant subspace $\mathcal{R}^{{\mathbb{K}}}_k$ if and only if $\mathbb{K}$ can be reached by upward arrows from $\hat{\mathbb{J}}^{\tilde{\mathcal{R}_k}}$; precisely, $\mathcal{R}^{{\mathbb{K}}}_k \subset \mathcal{R}^{{\mathbb{J}}}_k$.

Of note, the three groups in $\mathfrak{H}_0$ contain reflection symmetry, so any geometric chirality at this level is completely invisible to particle dynamics. 

\subsection{Rotating fluid}

Adding the force-rotation coupling block $\mathbf{C}$ to obtain the intermediate operator $\mathcal{R}_{0.5}$ reveals chirality effects. The hydrodynamic symmetry group grows by eight elements:
\begin{equation}
    \mathfrak{H}_{0.5} = \mathfrak{H}_0 \cup \{ C_1, C_2, C_s, C_{2v}, D_2, D_{2d}, D_{\infty}, \SO (3) \},
\end{equation} 
represented in figure~\ref{fig:diagram-R0-0.5-1}(b). It now contains two additional infinite groups, both accounting for chirality: dihedral $D_{\infty}$ and chiral isotropic $\SO (3)$.
All chiral cyclic groups $C_n$ and $D_n$ for $n\geqslant 3$ are invisible and satisfy $\hat{C}_n^{\mathcal{R}_{0.5}}=\hat{D}_n^{\mathcal{R}_{0.5}}=D_\infty$, while the achiral cyclic families $C_{nv}$, $C_{nh}$, $S_{2n}$, $D_{nh}$, $D_{nd}$ all vanish behind $D_{\infty h}$.

Geometrically, this means that any shape with an axis of threefold discrete rotational symmetry behaves hydrodynamically like an axisymmetric object ($D_{\infty h}$) or a chiral axisymmetric object ($D_{\infty}$). This is the first occurrence of a non-geometric group and formalises the notion of helicoidal symmetry, or more precisely $\mathcal{R}_{0.5}$-\textit{helicoidal symmetry} -- since it is now clear that this notion is level-dependent.

The infinite $D$ groups represent objects with a form of fore-aft symmetry; the fact that only these groups are visible at this level indicates that
the fore-aft asymmetric character of a particle in the preimage of ${\mathcal{R}_{0.5}^{D_{\infty}}}$ is not hydrodynamically relevant at this level.

An additional (non-parallel) axis of 3-fold discrete rotational symmetry brings the hydrodynamic behaviour either to the spherical class $\OO(3)$ or to the isotropic helicoid class $\SO(3)$. In particular, the chiralised polyhedra of Larmor and Kelvin shown in figure~\ref{fig:helicoids} all belong to this class. Further, six finite groups of low symmetry, including the identity group $C_1$, appear at this level, representing new possible polar-axial coupling effects. 

\subsection{Grand resistance level}

At the full grand resistance level, the hydrodynamic symmetry group grows by two elements:
\begin{equation}
    \mathfrak{H}_{1} = \mathfrak{H}_{0.5} \cup \{ C_i, C_{2h} \}.
\end{equation} 
This means the hydrodynamic resistance at $k=1$ level, and therefore any induced dynamics, of an arbitrary particle in Stokes flow belongs to exactly one of the 13 symmetry types in $\mathfrak{H}_{1}$. 
The two newcomers $C_i, C_{2h}$ are exactly the ones outside of $\mathfrak{H}_{0}$ containing \textit{inversion} (or centrosymmetry, defined as $\pi$-rotation followed by reflection with respect to a plane perpendicular to the rotation axis), excluding the $n$-cyclic groups with $n\geqslant3$ (invisible at this level). Inversion implies that the polar--axial coupling block $\vect{X}$ vanishes, naturally collapsing $C_i$ and $C_{2h}$ onto $D_{2h}$ at the $k=0.5$ level; independent parameters in $\vect{Q}$ then distinguish these three groups.

Besides the low-symmetry cyclic groups ($n\leqslant 2$), cyclic groups at this level still disappear behind dihedral groups: one has 
\begin{equation}
    \forall n \geqslant 3, \quad \hat{C}_n^{\mathcal{R}_1} = \hat{C}_{\infty}^{\mathcal{R}_1} = D_{\infty}, \quad 
    \hat{C}_{nv}^{\mathcal{R}_1} = \hat{C}_{nh}^{\mathcal{R}_1} =\hat{C}_{\infty h}^{\mathcal{R}_1} =\hat{C}_{\infty v}^{\mathcal{R}_1} = D_{\infty h},
\end{equation}
meaning that the fore-aft asymmetry character of a geometric object is still not hydrodynamically visible at this level.

At this level, any chiral polyhedral group $P \in \{ T,O,I \}$ still satisfies $\hat{P}^{\mathcal{R}_1} = \SO(3)$ and any of the four achiral polyhedral groups $P \in \{T_h, T_d, O_h, I_h \}$ satisfies $\hat{P}^{\mathcal{R}_1} = \OO(3)$; in particular, polyhedral-chiral particles are isotropic-chiral, and polyhedral-achiral particles are hydrodynamically spherical.

\subsection{Shear flow level}

We now go up to level $k=1.5$, adding the force-strain and torque-strain couplings $\vect{\Gamma}$ and $\vect{\Lambda}$ into the resistance operator. The hydrodynamic symmetry group grows by 14 elements, so that $|\mathfrak{H}_{1.5}| = 27$, namely:
\begin{equation}
    \mathfrak{H}_{1.5} = \mathfrak{H}_{1} \cup \{ C_3, C_{3h}, C_{3v}, D_3,D_{3h},D_{3d}, S_4, S_6, T, T_h, T_d, C_{\infty},C_{\infty h}, C_{\infty v} \},
\end{equation} 
The associated diagram is shown in figure~\ref{fig:phi1.5}, and several interesting phenomena are noteworthy. The newcomers mostly highlight the fact that $3$-fold symmetries become visible at this level. In the context of particle dynamics in a shear flow, this agrees precisely with the conclusion of \citet{ishimoto2020jeffery}. On the other hand, $4$-fold symmetries are still invisible at this level and vanish behind infinite cyclic groups: this may be unambiguously qualified as $\mathcal{R}_{1.5}$-helicoidal symmetry. 

The three infinite cyclic groups, $C_{\infty}$, $C_{\infty v}$ and $C_{\infty h}$, are visible at this level, meaning that hydrodynamics now sees fore-aft asymmetry and other finer axial effects.
Note that all infinite groups in $\mathfrak{G}$ are now visible. 

Another interesting family of newcomers, also linked to $3$-fold symmetry, is the trio of tetrahedral point groups $T$, $T_h$ and $T_d$. Since they have now become visible, it means that, at this $k=1.5$ level, the resistance operators of objects possessing $T$, $T_h$ or $T_d$ symmetries are, in general, distinct from isotropic-class tensors, \textit{i.e.} $\mathcal{R}_{1.5}^{\SO(3)}$ and $\mathcal{R}_{1.5}^{\OO(3)}$. This is seen in table~\ref{tab:master-observability} and figure~\ref{fig:phi1.5} from the different dimensions of the invariant subspaces: for chiral groups, $\dim \mathcal{R}_{1.5}^{T} = 5$ whilst $\dim \mathcal{R}_{1.5}^{\SO(3)} = 3$; for achiral groups, $\dim \mathcal{R}_{1.5}^{T_h} = \dim \mathcal{R}_{1.5}^{T_d} = 3$ whilst $\dim \mathcal{R}_{1.5}^{\OO(3)} = 2$. Note that $\mathcal{R}_{1.5}^{T_h}$ and $\mathcal{R}_{1.5}^{T_d}$ have the same dimension, but different normal forms: the extra nonzero coefficient of $\mathcal{R}_{1.5}^{T_h}$ is in the strain--torque block $\vect{\Lambda}$, while that of $\mathcal{R}_{1.5}^{T_d}$ is in the strain--force block $\vect{\Gamma}$, which implies different dynamics; see Section~\ref{sec:shear-dynamics}.

In particular, Larmor's proposal of an isotropically helicoidal tetrahedron, and more generally any construction based on tetrahedral symmetry like the $T$ row in figure~\ref{fig:helicoids}, is, in fact, not isotropic at this level. On the other hand, one has $\hat{O}^{\mathcal{R}_{1.5}} = \SO (3)$ and $\hat{O}_h^{\mathcal{R}_{1.5}} = \OO (3)$: octahedral symmetry groups $O$ and $O_h$ still vanish behind $\SO(3)$ and $\OO(3)$ at this level; hence Kelvin's isotropic helicoid, which has $O$ symmetry, is indeed hydrodynamically isotropic.
Moreover, $\dim \mathcal{R}_{1.5}^{\SO(3)} = 3$, which is also equal to $\dim \mathcal{R}_{1}^{\SO(3)}$. 
This means that the particle dynamics of an $\mathcal{R}_{1.5}$-isotropic chiral object are unaffected by the strain part of the background flow $\mathbf{E}$.

\begin{figure}
    \centering
        \begin{tikzpicture}[
          x=0.57cm,y=1cm,
          every node/.style={align=center, font=\scriptsize, inner sep=0pt},
          new/.style={text=blue},
          inf/.style={text=red},
          newinf/.style={text=orange},
          old/.style={text=gray!80},
          arr/.style={-{Latex[length=1.6mm]}, thin},
          oldarr/.style={-{Latex[length=1.6mm]}, thin, dashed, draw=gray!80}
        ]
        \node[new] (C3) at (0,0) {\glabel{C_3}{15}};
        
        \node[new] (C3v) at (-4,1.4) {\glabel{C_{3v}}{9}};
        \node[new] (D3)  at (-1,1.4) {\glabel{D_3}{10}};
        \node[new] (C3h) at (1.8,1.4)  {\glabel{C_{3h}}{8}};
        \node[new] (S6)  at (5.5,1.4) {\glabel{S_6}{8}};
        \node[new] (T)   at (0,3.5) {\glabel{T}{5}};
        
        \node[new] (Td)  at (-4,4.5) {\glabel{T_d}{3}};
        \node[new] (D3h) at (-2.8,3) {\glabel{D_{3h}}{6}};
        \node[new] (D3d) at (3,3)  {\glabel{D_{3d}}{6}};
        \node[new] (Th)  at (5.5,4.5)  {\glabel{T_h}{3}};
        
        \node[old] (Ci) at (4.7,0) {\glabel{C_i}{24}};
        \node[old] (D2) at (1,2.8) {\glabel{D_2}{15}};
        \node[old] (C2) at (-5.5,2) {\glabel{C_2}{15}};
        \node[new] (S4) at (-5.5,3.9) {\glabel{S_4}{12}};
        
        \node[newinf] (Cinf)  at (-7,3) {\glabel{C_\infty}{12}};
        \node[new] (Cinfv) at (-7,5) {\glabel{C_{\infty v}}{7}};
        \node[inf] (Dinf)  at (-1.0,5) {\glabel{D_\infty}{8}};
        \node[newinf] (Cinfh) at (3,5)  {\glabel{C_{\infty h}}{7}};
        
        \node (Dinfh) at (-2.8,6.1) {\glabel{D_{\infty h}}{5}};
        \node[inf] (SO3)   at (3,6.1)  {\glabel{\SO(3)}{3}};
        \node (O3)    at (0.2,7.6)  {\glabel{\OO(3)}{2}};
        
        \draw[oldarr] (Ci) -- (S6);
        \draw[oldarr] (C2) -- (S4);
        \draw[oldarr] (D2) to[out=90,in=-15] (T);
        \draw[arr] (S4) to[out=70,in=190] (Td);
        
        \draw[arr] (C3) -- (C3v);
        \draw[arr] (C3) -- (D3);
        \draw[arr] (C3) -- (C3h);
        \draw[arr] (C3) -- (S6);
        \draw[arr] (C3) -- (T);
        \draw[arr] (C3) to[out=170,in=-90] (Cinf);
        
        \draw[arr] (C3v) -- (D3h);
        \draw[arr] (C3v) -- (Td);
        \draw[arr] (D3) -- (D3h);
        \draw[arr] (D3) to[out=20,in=-120] (D3d);
        \draw[arr] (C3h) -- (D3h);
        \draw[arr] (S6) -- (D3d);
        \draw[arr] (S6) -- (Th);
        
        \draw[arr] (C3v) to[out=110,in=-80] (Cinfv);
        \draw[arr] (D3) -- (Dinf);
        \draw[arr] (C3h) to[out=100,in=-110] (Cinfh);
        \draw[arr] (S6) -- (Cinfh);
        
        \draw[arr] (Cinf) -- (Cinfv);
        \draw[arr] (Cinf) to[out=25,in=-130] (Dinf);
        \draw[arr] (Cinf) to[out=15,in=-150] (Cinfh);
        
        \draw[arr] (Cinfv) -- (Dinfh);
        \draw[arr] (Dinf) -- (Dinfh);
        \draw[arr] (D3h) -- (Dinfh);
        \draw[arr] (D3d) to[out=120,in=-35] (Dinfh);
        
        \draw[arr] (Dinf) -- (SO3);
        \draw[arr] (T) to[out=70,in=-130] (SO3);
        
        \draw[arr] (Dinfh) -- (O3);
        \draw[arr] (SO3) -- (O3);
        \draw[arr] (Td) to[out=120,in=190] (O3);
        \draw[arr] (Th) to[out=100,in=-10] (O3);
        
        \draw[arr] (C3v) to[out=25,in=-150] (D3d);
        \draw[arr] (T) -- (Td);
        \draw[arr] (T) -- (Th);
        \draw[arr] (Cinfh) -- (Dinfh);
        \end{tikzpicture}
    \caption{Lattice diagram of $\mathfrak{H}_{1.5}$. Only infinite groups and the groups appearing at the current level are shown. The three connections going from already existing nodes (visible at $k=1$ level) to new finite nodes (in blue)   are shown as grey dashed arrows (omitting $C_1 \rightarrow C_3$). Orange nodes indicate the two non-geometric infinite groups becoming visible at this level.}
    \label{fig:phi1.5}
\end{figure}

\subsection{Full linear level}

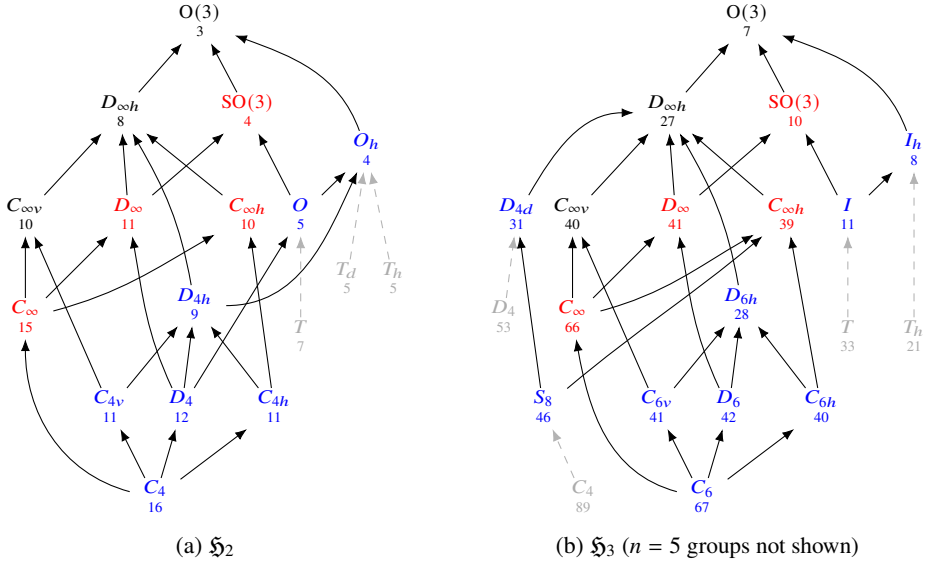
\begin{figure}
\centering
\begin{subfigure}{0.45\textwidth}
    \centering
\begin{tikzpicture}[
  x=0.35cm,y=0.85cm,
  every node/.style={align=center, font=\scriptsize, inner sep=0pt},
  new/.style={text=blue},
  inf/.style={text=red},
  old/.style={text=gray!65},
  arr/.style={-{Latex[length=1.6mm]}, thin},
  oldarr/.style={-{Latex[length=1.6mm]}, thin, dashed, draw=gray!60}
]
\node[new] (C4) at (-1.5,0.6) {\glabel{C_4}{16}};

\node[new] (C4v) at (-3.2,2.0) {\glabel{C_{4v}}{11}};
\node[new] (D4)  at (-0.5,2.0) {\glabel{D_4}{12}};
\node[new] (C4h) at (3.0,2.0)  {\glabel{C_{4h}}{11}};
\node[new] (D4h) at (0,3.6) {\glabel{D_{4h}}{9}};

\node[inf] (Cinf)  at (-6.4,3.4) {\glabel{C_\infty}{15}};
\node (Cinfv)       at (-6.4,5.0) {\glabel{C_{\infty v}}{10}};
\node[inf] (Dinf)  at (-2.5,5.0) {\glabel{D_\infty}{11}};
\node[inf] (Cinfh) at (2.0,5.0)  {\glabel{C_{\infty h}}{10}};

\node (Dinfh) at (-2.8,6.6) {\glabel{D_{\infty h}}{8}};
\node[inf] (SO3) at (2.0,6.6) {\glabel{\SO(3)}{4}};
\node (O3) at (0.2,8.0) {\glabel{\OO(3)}{3}};

\node[old] (T)  at (4.0,3.1) {\glabel{T}{7}};
\node[old] (Td) at (5.75,4) {\glabel{T_d}{5}};
\node[old] (Th) at (7.5,4) {\glabel{T_h}{5}};
\node[new] (O)  at (4,5.0) {\glabel{O}{5}};
\node[new] (Oh) at (6.5,6.0) {\glabel{O_h}{4}};

\draw[oldarr] (T) -- (O);
\draw[oldarr] (Td) -- (Oh);
\draw[oldarr] (Th) -- (Oh);

\draw[arr] (C4) -- (C4v);
\draw[arr] (C4) -- (D4);
\draw[arr] (C4) -- (C4h);
\draw[arr] (C4) to[out=165,in=-90] (Cinf);

\draw[arr] (C4v) -- (D4h);
\draw[arr] (D4) -- (D4h);
\draw[arr] (C4h) -- (D4h);

\draw[arr] (C4v) -- (Cinfv);
\draw[arr] (C4h) -- (Cinfh);
\draw[arr] (D4h) to[out=100,in=-60] (Dinfh);

\draw[arr] (Cinf) -- (Cinfv);
\draw[arr] (Cinf) -- (Dinf);
\draw[arr] (Cinf) to[out=15,in=-150] (Cinfh);

\draw[arr] (Cinfv) -- (Dinfh);
\draw[arr] (Dinf) -- (Dinfh);
\draw[arr] (Dinf) -- (SO3);

\draw[arr] (O) -- (SO3);
\draw[arr] (Oh) to[out=110,in=-20] (O3);
\draw[arr] (Dinfh) -- (O3);
\draw[arr] (SO3) -- (O3);

\draw[arr] (D4) to[out=115,in=-80] (Dinf);
\draw[arr] (D4) -- (O);
\draw[arr] (D4h) to[out=0,in=-115] (Oh);
\draw[arr] (O) -- (Oh);
\draw[arr] (Cinfh) -- (Dinfh);
\end{tikzpicture}
\caption{$\mathfrak{H}_2$}
\end{subfigure}
$\quad$
\begin{subfigure}{0.45\textwidth}
    \centering
\begin{tikzpicture}[
  x=0.35cm,y=0.85cm,
  every node/.style={align=center, font=\scriptsize, inner sep=0pt},
  new/.style={text=blue},
  inf/.style={text=red},
  old/.style={text=gray!65},
  arr/.style={-{Latex[length=1.6mm]}, thin},
  oldarr/.style={-{Latex[length=1.6mm]}, thin, dashed, draw=gray!60}
]
\node[new] (C6) at (-1.5,0.6) {\glabel{C_6}{67}};

\node[new] (C6v) at (-3.2,2.0) {\glabel{C_{6v}}{41}};
\node[new] (D6)  at (-0.5,2.0) {\glabel{D_6}{42}};
\node[new] (C6h) at (3.0,2.0)  {\glabel{C_{6h}}{40}};
\node[new] (D6h) at (0,3.6) {\glabel{D_{6h}}{28}};

\node[inf] (Cinf)  at (-6.4,3.4) {\glabel{C_\infty}{66}};
\node (Cinfv)       at (-6.4,5.0) {\glabel{C_{\infty v}}{40}};
\node[inf] (Dinf)  at (-2.5,5.0) {\glabel{D_\infty}{41}};
\node[inf] (Cinfh) at (1.7,5.0)  {\glabel{C_{\infty h}}{39}};

\node (Dinfh) at (-2.8,6.6) {\glabel{D_{\infty h}}{27}};
\node[inf] (SO3) at (2.0,6.6) {\glabel{\SO(3)}{10}};
\node (O3) at (0.2,8.0) {\glabel{\OO(3)}{7}};

\node[old] (T)  at (4.0,3.1) {\glabel{T}{33}};
\node[old] (Th) at (6.5,3.1) {\glabel{T_h}{21}};
\node[new] (I)  at (4,5.0) {\glabel{I}{11}};
\node[new] (Ih) at (6.5,6.0) {\glabel{I_h}{8}};

\node[old] (C4) at (-6,0.6) {\glabel{C_4}{89}};
\node[old] (D4) at (-9,3.4) {\glabel{D_4}{53}};
\node[new] (S8) at (-7.5,2) {\glabel{S_8}{46}};
\node[new] (D4d) at (-8.5,5.0) {\glabel{D_{4d}}{31}};

\draw[oldarr] (T) -- (I);
\draw[oldarr] (Th) -- (Ih);
\draw[oldarr] (C4) -- (S8);
\draw[oldarr] (D4) -- (D4d);

\draw[arr] (C6) -- (C6v);
\draw[arr] (C6) -- (D6);
\draw[arr] (C6) -- (C6h);
\draw[arr] (C6) to[out=165,in=-80] (Cinf);

\draw[arr] (C6v) -- (D6h);
\draw[arr] (D6) -- (D6h);
\draw[arr] (C6h) -- (D6h);

\draw[arr] (C6v) -- (Cinfv);
\draw[arr] (C6h) -- (Cinfh);
\draw[arr] (D6h) to[out=100,in=-60] (Dinfh);

\draw[arr] (Cinf) -- (Cinfv);
\draw[arr] (Cinf) -- (Dinf);
\draw[arr] (Cinf) to[out=15,in=-150] (Cinfh);

\draw[arr] (Cinfv) -- (Dinfh);
\draw[arr] (Dinf) -- (Dinfh);
\draw[arr] (Dinf) -- (SO3);

\draw[arr] (S8) -- (D4d);
\draw[arr] (D4d) to[out=60,in=-180] (Dinfh);

\draw[arr] (I) -- (SO3);
\draw[arr] (Ih) to[out=110,in=-20] (O3);
\draw[arr] (Dinfh) -- (O3);
\draw[arr] (SO3) -- (O3);

\draw[arr] (S8) to[out=40,in=-140] (Cinfh);
\draw[arr] (D6) to[out=115,in=-80] (Dinf);
\draw[arr] (I) -- (Ih);
\draw[arr] (Cinfh) -- (Dinfh);
\end{tikzpicture}
\caption{$\mathfrak{H}_3$ ($n=5$ groups not shown)}
\end{subfigure}
    \caption{Lattice diagram at full linear ($k=2$) and quadratic ($k=3$) levels. Only infinite groups and the groups appearing at the current level are shown. Selected existing connections with the rest of the diagram (\textit{i.e.} the part already visible at the lower levels) are shown as grey dashed arrows.}
    \label{fig:phi34}
\end{figure}

This level ($k=2$) is the full grand resistance tensor level, including the shear-stresslet coupling block $\vect{\Sigma}$. Seven elements are added to the hydrodynamic symmetry group:
\begin{equation}
    \mathfrak{H}_{2} = \mathfrak{H}_{1.5} \cup \{ C_4, C_{4h}, C_{4v}, D_4,D_{4h}, O, O_h \}.
\end{equation} 
The set $\mathfrak{H}_{2}$ contains 34 elements. The lattice diagram of its new elements is shown in figure~\ref{fig:phi34}a.

The $4$-fold cyclic groups are now visible. Notably, this means that objects with $4$-fold symmetry generally no longer behave like objects with the corresponding infinite symmetry at this level. The notion of helicoidal symmetry must be understood here as $\mathcal{R}_2$-helicoidal symmetry: the vanishing of $n$-fold symmetry for $n\geqslant 5$ behind the infinite cyclic groups, \textit{e.g.} $\hat{C}_5^{\mathcal{R}_{2}} = C_{\infty}$. 

For objects with $4$-fold symmetry in group $P$ among $\{ C_{4}, C_{4h}, C_{4v}, D_{4}, D_{4h} \}$, one has $\dim \mathcal{R}_{2}^{P} = \dim \mathcal{R}_{2}^{P_\infty} + 1$: in every case, the grand resistance tensor of $P$ contains a single extra independent parameter compared to that of the corresponding infinite axial group $P_{\infty}$.

Quite strikingly, the octahedral groups $O$ and $O_h$ also become hydrodynamically visible at this full linear level. Consequently, the $O$-symmetric Kelvin helicoid is no longer an isotropic helicoid at this response level. The hydrodynamic response of an octahedral object is distinct from that of a sphere or chiral-isotropic object. Recall that this distinction is not visible at the $k=1.5$ level: it only appears in the stresslet--strain coupling block, which does not affect the translational and rotational dynamics of a passive particle in shear flow (see Section~\ref{sec:stresslet-level} for further details).

Furthermore, the two icosahedral groups, $I$ and $I_h$, remain truly invisible at this level. Therefore, one can recover isotropic helicoids by adapting Kelvin or Larmor constructions at the icosahedral level. The icosahedral and dodecahedral proposals on figure~\ref{fig:helicoids} all have $I$ symmetry without $I_h$ symmetry; since $\hat{I}^{\mathcal{R}_2} = \SO(3)$, they are hydrodynamically chiral-isotropic.

\subsection{Quadratic level}
\label{sec:quadratic-resistance}

The last level we will consider in detail is the quadratic coupling for $\mathcal{J}_3$ jet space level. This level adds sixteen elements to the hydrodynamic symmetry group set, such that 
\begin{equation}
    \mathfrak{H}_{3} = \mathfrak{H}_{2} \cup \{ C_{(5,6)}, C_{(5,6)h}, C_{(5,6)v}, D_{(5,6)},D_{(5,6)h}, D_{(4,5)d}, S_{(8,10)}, I, I_h \},
\end{equation} 
where we use the compact notation $\mathbb{J}_{(m,n)}=\mathbb{J}_n \cup \mathbb{J}_m$, and we now have $|\mathfrak{H}_3|=50$.

The most interesting phenomenon at this level is the appearance of the last two polyhedral groups, $I$ and $I_h$. As for $O$ and $O_h$ at the previous level, the invariant subspaces $\mathcal{R}^{I}_{3}$ and $\mathcal{R}^{I_h}_{3}$ have one additional dimension compared with $\mathcal{R}^{\SO(3)}_{3}$ and $\mathcal{R}^{\OO(3)}_{3}$, respectively. 

In particular, a Kelvin-style icosahedral design now generally fails to realise a true chiral-isotropic object at this level. From there, we do not have any polyhedral groups left to hide behind: the vaned-sphere blueprints of Kelvin, and the bevelled polyhedra of Larmor, are now ineffective at realising a hydrodynamically chiral isotropic solid. 

This does not rule out the existence of isotropic helicoids. 
It only indicates that belonging to $\mathcal{R}^{\SO (3)}_k$ can no longer be a direct consequence of a shape $B\in\mathcal{B}$ possessing a symmetry that falls in the hydrodynamic shadow of $\SO (3)$.
From level $k=3$ onwards, chiral-isotropic solids are very likely to still exist, but the isotropy happens instead because of accidental symmetry, \textit{i.e.} fortuitous cancellation of anisotropic parameters inside the resistance operator space $\mathcal{R}_k$. 
Similarly to the dumbbell discussed in Section~\ref{sec:pure-translation}, one may start from a higher-dimensional fixed space and seek such a cancellation. Consider a Kelvin-style vaned icosahedral object $K \in \mathcal{B}$, with symmetry $I$. Since $\dim\mathcal{R}^{I}_{3} = 11$ and $\dim\mathcal{R}^{\SO (3)}_3 = 10$, $\mathcal{R}_3 (K)$ generally has one anisotropic parameter, say $\alpha$, keeping it outside of $\mathcal{R}^{\SO (3)}_3$. Applying slight, symmetry-preserving deformations to the vanes such as uniform twists, bumps or holes, is likely to affect $\alpha$ enough to cancel it while retaining the chiral character of $K$. Similarly, one could consider a bevelled prism $K'$ with $D_8$ symmetry, which collapses to $D_\infty$ at this level, and apply uniform deformations of each lateral edge of the prism to cancel the axial effects, \textit{i.e.} the 31 dimensions between $\mathcal{R}^{\SO (3)}_3$ and $\mathcal{R}^{D_\infty}_3$. We do not know whether such cancellation is possible, and explicit realisation of a quadratic-level isotropic helicoid remains an open question.

On the other side of $\mathcal{R}^{\SO (3)}_3$, note that $\dim\mathcal{R}^{\OO (3)}_3 = 7$: for the first time, the dimension gap between achiral and chiral spherical spaces is larger than one. Three independent parameters now distinguish isotropic helicoids from hydrodynamic spheres at this level; this supports the existence of various shapes belonging to $\mathcal{R}^{\SO (3)}_3$, and especially suggests a wider variety of dynamical behaviours within the isotropic helicoid class.

The rest of our journey through $\mathfrak{H}_3$ becomes more repetitive: we see the cyclic groups appear at level $n=5$ and $n=6$, or at levels $4$ and $5$ for the improper rotation groups $D_{nd}$ and $S_{2n}$. 

\subsection{General polynomial background flow}
\label{sec:general-flow}

The infinite and exceptional families are entirely visible from $k=3$. From higher jet space levels at $k=4$ onwards, the growth of $\mathfrak{H}_k$ takes a repeated pattern:
\begin{multline}
    \mathfrak{H}_{k} = \mathfrak{H}_{k-1} \cup \{ C_{(2k, 2k-1)}, C_{(2k, 2k-1)h}, C_{(2k, 2k-1)v}, D_{(2k, 2k-1)},D_{(2k, 2k-1)h}\} \\
    \cup \{ D_{(2k-1, 2k-2)d}, S_{(4k-2,4k-4)} \},
\end{multline} 
adding 14 groups at each layer, which yields the general expression for the cardinality of $\mathfrak{H}_k$:
\begin{equation}
        \forall k \geqslant 3, | \mathfrak{H}_k | = 14k+8.
\end{equation}
The dimension of invariant spaces grows dramatically at each level ($\sim k^4$ for finite groups and $\sim k^3$ for infinite cyclic groups), presenting little practical interest. One remarkable property of the invariant spaces, already seen in table~\ref{tab:master-observability} for $k=2,3$, is that, for pairs
$P_{2k}$ in $\{ C_{2k}, C_{2k,v}, C_{2k, h}, D_{2k}, D_{2k,h}\}$ and the corresponding infinite cyclic group,
\begin{equation}
    \dim \mathcal{R}^{P_{2k}}_k = \dim \mathcal{R}^{P_{\infty}}_k + 1.
\end{equation}
This can be checked by analytical calculation and means that objects with $2k$-fold symmetry admit exactly one additional parameter in their resistance operator at $k$ level, compared to axially invariant objects. 

Finally, a blueprint to find a $k$-level isotropic helicoid would probably start from a $D_{2k+1}$-symmetric object, collapsing into $D_{\infty}$, followed by cancellation-inducing deformation. Another direction would be to start from a sphere and apply combinations of harmonic deformations that yield a strictly $\SO(3)$ response. 

\section{Particle dynamics}
\label{sec:particle-dynamics}

So far, we have studied the resistance operators $\mathcal{R}_k$ as canonical hydrodynamic observables of the shape $B$. In many practical cases, however, one does not consider the resistance operators themselves but rather the induced dynamics of the particle in a fixed background flow. For a particle with some $\mathbb{J}$-symmetry, the number of relevant parameters for such dynamics may be smaller than the dimension of the $\mathbb{J}$-invariant subspace of the full resistance operator. 
For example, the invariant subspace of $D_{\infty h}$ at level $k=1.5$ has dimension 5. Yet, for particles with $D_{\infty h}$ symmetry, which correspond to fore--aft symmetric bodies of revolution, rotational dynamics in simple shear flow, called Jeffery orbits, are characterised by a single parameter called the Bretherton parameter \citep{jeffery1922motion,bretherton1962motion}.
Similarly, chiral $\mathcal{R}_{1.5}$-helicoidal particles follow chiral Jeffery orbits characterised by two parameters \citep{ishimoto2020helicoidal}.
In this section, we apply again representation-theoretic tools to decipher this parameter count as the dimension of $\mathbb{J}$-invariant spaces induced by the operators $\mathcal{R}_k$. The methodology is essentially the same as the one described in Section~\ref{sec:fixed-space-characterisation}; we concisely highlight the main differences below.

\subsection{Representation of dynamical invariant space}

\subsubsection{Character count}

We illustrate the methodology with the dynamics of a passive particle in a background linear flow. To do so, we must distinguish the particle velocity from the undisturbed background flow. Let $\vect{V}_p$ and $\vect{\Omega}_p$ denote the translational and angular velocities of the particle in the laboratory frame, and let
$\vect{V}^{\infty}=\vect{U}^{\infty}$,
$
\vect{\Omega}^{\infty}
=\frac{1}{2}\nabla\times\vect{U}^{\infty}$
denote the corresponding local velocities of the undisturbed flow at the hydrodynamic centre. We introduce the particle velocities relative to the local background flow,
\[
\delta\vect{V}=\vect{V}_p-\vect{V}^{\infty},
\qquad
\delta\vect{\Omega}=\vect{\Omega}_p-\vect{\Omega}^{\infty}.
\]
In the particle-frame convention of Section~\ref{sec:stokes problem}, the uniform and rotational components of the relative incident-flow jet are therefore $-\delta\vect{V}$ and $-\delta\vect{\Omega}$, whereas the rate-of-strain tensor remains $\vect{E}^{\infty}$.

In the absence of inertia and external effects, the net force and torque on the particle vanish, which gives
\begin{equation}
    \mathcal{R}_{1.5} \begin{pmatrix} -\delta \vect{V} \\ - \delta \vect{\Omega} \\ \vect{E}^{\infty} \end{pmatrix} =0 \quad \Rightarrow  \quad \begin{pmatrix} \delta \vect{V} \\ \delta \vect{\Omega} \end{pmatrix} =  \mathcal{R}_1^{-1}\begin{pmatrix} \vect{\Gamma} \\ \vect{\Lambda} \end{pmatrix} \vect{E}^{\infty} \quad \Rightarrow \begin{array}{l} \delta \vect{V} = \mathcal{P}_V \vect{E}^{\infty} \\ \delta \vect{\Omega} = \mathcal{P}_\Omega \vect{E}^{\infty} \end{array}, 
    \label{eq:shear-motion-projection}
\end{equation}
and the actual particle velocities are given by 
$
\vect{V}_p
=
\vect{V}^{\infty}
+
\mathcal{P}_V\vect{E}^{\infty}$ and $
\vect{\Omega}_p
=
\vect{\Omega}^{\infty}
+
\mathcal{P}_{\Omega}\vect{E}^{\infty}.
$
The operators $\mathcal{P}_V$ and $\mathcal{P}_\Omega$ can be seen as linear maps, or homomorphisms, respectively from $V_2^+$ (space of strain tensors) to $V_1^-$ (polar vectors), and from $V_2^+$  to $V_1^+$ (axial vectors); this is denoted $\mathcal{P}_V \in \operatorname{Hom}(V_2^+,V_1^-)$, $\mathcal{P}_\Omega \in \operatorname{Hom}(V_2^+,V_1^+)$. 

Now, considering a group $\mathbb{J}$ in $\mathfrak{G}$, the same objects and procedures developed in Section~\ref{sec:fixed-space-characterisation} for $\mathcal{R}_k$ may be carried over to the operator $\mathcal{P}_V$ (and identically to $\mathcal{P}_\Omega$): deriving the action $\rho_{\mathcal{P}_V}(A)$ of $A \in \mathbb{J}$, which is straightforward from the irreducible blocks; defining the stabiliser of an element $P_V \in \mathcal{P}_V$; and finally defining the invariant subspace $\mathcal{P}^{\mathbb{J}}_V$. Since $\mathcal{P}_V$ is inherited from $\mathcal{R}_{1.5}$, we can restrict $\mathbb{J}$ to one of the 27 elements of $\mathfrak{H}_{1.5}$. Then, we define, as in Equation~\eqref{eq:fixed-space-projector}, the representation-theoretic averaging operator
\begin{equation}
\begin{array}{l c c c}
      \Pi_{\mathbb J}  : & \mathcal P_V  & \to & \mathcal P_V \\
      & P
  &  \mapsto &
  \frac{1}{|\mathbb J|}
  \sum_{A\in\mathbb J}\rho_{\mathcal{P}_V} (A)P .
  \end{array}
  \label{eq:dynamical-projector}
\end{equation}
The rank of $\Pi_{\mathbb J}$ corresponds to the dimension of the invariant subspace $\mathcal{P}^{\mathbb{J}}_V$, which precisely gives the number of independent parameters that will appear in the translational dynamics equations of a particle in shear. We therefore note $p_V (\mathbb{J}) = \dim \mathcal{P}^{\mathbb{J}}_V$, and compute it using the character formula:
\begin{equation}
    p_V (\mathbb{J}) = \frac{1}{|\mathbb J|} \sum_{A \in \mathbb{J}} \chi_{V_1^-} (A) \chi_{V_2^+} (A^{-1}).
\end{equation}
Since $A$ is here a real, orthogonal transformation, one has $\chi_{\mathcal{V}} (A^{-1}) = \chi_{\mathcal{V}} (A)$ for any $\mathcal{V}$, and the rest of the calculation follows from Section~\ref{sec:fixed-space-characterisation}.

By this methodology, we can compute $p_V(\mathbb{J})$ and $p_{\Omega}(\mathbb{J})$, the numbers of dynamical parameters in the translational and rotational dynamics of a passive particle in shear flow, for any $\mathbb{J} \in \mathfrak{H}_{1.5}$. Similarly, we define $p_S(\mathbb{J})$ for the strain--stresslet parameter count with $\mathbb{J} \in \mathfrak{H}_2$, and $q_V(\mathbb{J})$ and $q_\Omega(\mathbb{J})$ for translational and rotational dynamics in quadratic flows with $\mathbb{J} \in \mathfrak{H}_3$.

\subsubsection{Dynamics nullspace}

For a translation--strain coupling operator $P$ in the linear space $\mathcal{P}_V$, invariance under the action of $A \in \mathbb{J}$ means $\rho_{V_1^-}(A) P = P \rho_{V_2^+}(A)$. With column-wise vectorisation, this is
\begin{equation}
    \left[I_5\otimes\rho_{V_1^-}(A)
    -\rho_{V_2^+}(A)^T\otimes I_3\right]\operatorname{vec}P=0.
    \label{eq:dynamics-equivariance-nullspace}
\end{equation}
Stacking \eqref{eq:dynamics-equivariance-nullspace} for generators or all elements of $\mathbb{J}$ defines a linear operator $\mathbf{L}$. A basis of $\ker\mathbf{L}$ obtained by singular value decomposition identifies the positions of the $p_V(\mathbb{J})$ dynamical parameters in $\mathcal{P}_V$, which we call the normal form of $\mathcal{P}_V^{\mathbb{J}}$. Detailed calculations are provided in Appendix \ref{app:normal-forms}.

\subsubsection{Director dynamics}

When the symmetry group $\mathbb{J}$ is cyclic, the parameter count for rotational dynamics can be reduced further. Indeed, the cyclic symmetry axis induces a distinguished director vector $\vect{d}$, and we can study the director dynamics $\dot{\vect{d}} = \vect{\Omega}_p \times \vect{d}$, which discards the spin angle $\psi$ around $\vect{d}$. In our formalism, this amounts to decomposing $V_1^+$ into $\mathbb{R} \vect{d} \oplus V_{\bot}$, where $V_{\bot}$ is the irreducible representation of $\OO (2)$, and studying the projection onto $V_{\bot}$. This defines director-specific invariant spaces whose dimensions are denoted by $p_d$ for shear-induced dynamics and $q_d$ for dynamics in quadratic flow.

Of note, this does not necessarily mean that the director dynamics $\dot{\vect{d}}$ are decoupled from the spin angle $\psi$. If this additionally occurs, then the orientational dynamics $\dot{\vect{d}}$ form a closed two-dimensional system on $V_{\bot}$, which we call \textit{spin-invariant} and typically write in pitch and yaw Euler-angle coordinates. Reduction to an autonomous system on the two-dimensional sphere $S^2$ precludes chaotic dynamics by the Poincar\'e--Bendixson theorem.
Jeffery equations for spheroids are the most prominent example of such a reduction \citep{jeffery1922motion,bretherton1962motion}, with generalisations to chiral objects by \citet{ishimoto2020helicoidal}.

When the dynamics of $\vect{d}$ are not spin-invariant, the full attitude $Q\in\SO(3)$ of the body-fixed frame must be considered. It satisfies $\dot{Q}=\widehat{\vect{\Omega}}_pQ$, where $\widehat{\vect{\Omega}}\vect{a}=\vect{\Omega}\times\vect{a}$. The director-reduced counts $p_d$ and $q_d$ still indicate the number of independent parameters in the spin-projected subsystem. Attitude dynamics are genuinely three-dimensional and can display irregular or chaotic behaviour, as is well known for triaxial particles in shear \citep{hinch1979rotation,yarin1997chaotic}.

By contrast, the parameter reduction $p_d$, $q_d$ is irrelevant for the polyhedral and spherical groups, for they do not possess a distinguished director. Finally, the particular case of the triaxial groups $D_2$ and $D_{2h}$ requires special notice, because their symmetry classes distinguish three possible directors along three perpendicular axes, requiring an arbitrary choice for $\vect{d}$, although this choice does not affect the invariant spaces and parameter counts $p_d$, $q_d$.

In the remainder of this section, we determine these parameter counts and examine the resulting
normal forms. Section~\ref{sec:shear-dynamics} focuses on dynamics in shear flow, for which we complete the classification undertaken by Jeffery and Ishimoto; in particular, we study a new class of dynamics associated with tetrahedral symmetry.
The stresslet response is addressed in Section~\ref{sec:stresslet-level}.
Section~\ref{sec:quadratic-flow-dynamics} is devoted to quadratic-flow dynamics, with a numerical assessment of chaotic trajectories and a particular focus on octahedral dynamics.

\subsection{Dynamics in shear flow: Jeffery-Ishimoto classification}
\label{sec:shear-dynamics}

\begin{table}
\centering
\footnotesize
\setlength{\tabcolsep}{2pt}
\renewcommand{\arraystretch}{1}
\begin{tabular}{ c c | c c c | c | c }
\toprule
Category & $\mathbb{J} \in \mathfrak{H}_{1.5}$
& $\quad p_V \quad$
& $\quad p_\Omega \quad$
& $\quad p_d \quad$
& $\;$ s.i.$^\dag$ $\;$ 
& Rotational dynamics \\
\hline
\multirow{2}{*}{Spherical}
 & $\OO(3)$   & 0 & 0 & - &  & Spherical \\
 & $\SO(3)$   & 0 & 0 & - &  & Spherical \\
\hline
\multirow{5}{*}{Helicoidal}
 & $C_{\infty}$   & 3 & 3 & 2 & $\checkmark$ & Ishimoto \\
 & $C_{\infty h}$ & 0 & 3 & 2 & $\checkmark$ & Ishimoto \\
 & $C_{\infty v}$ & 2 & 1 & 1 & $\checkmark$ & Jeffery \\
 & $D_{\infty}$   & 1 & 1 & 1 & $\checkmark$ & Jeffery \\
 & $D_{\infty h}$ & 0 & 1 & 1 & $\checkmark$ & Jeffery \\
\hline
\multirow{7}{*}{3-fold}
 & $C_3$    & 5 & 5 & 4 & & Triangular \\
 & $S_6$    & 0 & 5 & 4 & & Triangular \\
 & $C_{3v}$ & 3 & 2 & 2 & & Reduced triangular \\
 & $D_3$    & 2 & 2 & 2 & & Reduced triangular \\
 & $D_{3d}$ & 0 & 2 & 2 & & Reduced triangular$^{\ddagger}$ \\
 & $C_{3h}$ & 2 & 3 & 2 & $\checkmark$ & Ishimoto \\
 & $D_{3h}$ & 1 & 1 & 1 & $\checkmark$ & Jeffery \\
\hline
\multicolumn{7}{c}{New particle types} \\
\hline
\multirow{3}{*}{Tetrahedral}
 & $T$   & 1 & 1 & -- & & Tetrahedral \\
 & $T_h$ & 0 & 1 & -- & & Tetrahedral \\
 & $T_d$ & 1 & 0 & -- & & Spherical \\
\hline
\multirow{6}{*}{2-fold}
 & $C_{2h}$ & 0 & 7 & 4 & & Digonal \\
 & $C_2$    & 7 & 7 & 4 & & Digonal \\
 & $C_s$    & 8 & 7 & 4 & & Digonal \\
 & $C_{2v}$ & 4 & 3 & 2 & & Reduced digonal \\
 & $S_4$    & 4 & 3 & 2 & $\checkmark$ & Ishimoto \\
 & $D_{2d}$ & 2 & 1 & 1 & $\checkmark$ & Jeffery \\
\hline
\multirow{4}{*}{No director}
 & $D_2$    & 3  & 3  & 2$^*$ & & Ellipsoidal \\
 & $D_{2h}$ & 0  & 3  & 2$^*$ & & Ellipsoidal \\
 & $C_1$    & 15 & 15 & --    & & General \\
 & $C_i$    & 0  & 15 & --    & & General \\
\bottomrule
\end{tabular}
\caption{Parameter count $p_V$, $p_\Omega$ and, when applicable, $p_d$ for particle dynamics at $k=1.5$ level. A check mark in the spin-invariance column indicates fully decoupled director dynamics. The first half of the table recalls the Jeffery--Ishimoto orbit classification, comprising both helicoidal symmetry and 3-fold symmetry classes. The new parameter counts for the 13 remaining groups of $\mathfrak{H}_{1.5}$ are given in the bottom half. $^*$ Three possible choices of director. $^\dagger$ s.i.: spin-invariance.  $^\ddagger$ Correcting the inexact definition of $D_{3d}$ used in \citet{ishimoto2020jeffery}.}
\label{table:shear-dynamics}
\end{table}

Table~\ref{table:shear-dynamics} applies the fixed-space calculation to every
group in $\mathfrak{H}_{1.5}$. The table is divided into six sets of groups. The first two sets contains the infinite groups: spherical symmetries unaffected by shear flow, and axial symmetries equivalently representing $\mathcal{R}_{1.5}$-helicoidal symmetry. All helicoidal groups possess spin-invariance of their director dynamics. For $C_{\infty v}$, $D_{\infty}$, and $D_{\infty h}$, the director count $p_d$ is down to a single parameter, which corresponds to the Bretherton parameter $\beta$ in Jeffery equations, whilst the translational count $p_V$ differs for each group. The groups $C_{\infty}$ and $C_{\infty h}$ satisfy $p_d = 2$, corresponding to the helicoidal objects studied by \citet{ishimoto2020helicoidal}, whose rotational dynamics are characterised by the Bretherton parameter $\beta$ and an additional chirality parameter $\gamma$, sometimes called Ishimoto parameter.

The second block gathers 3-fold symmetry groups, studied in detail by \citet{ishimoto2020jeffery}. These groups do not generally possess spin-invariance, with the exception of $C_{3h}$ and $D_{3h}$, whose rotational dynamics reduce to Ishimoto and Jeffery dynamics, respectively. The other groups feature parameters associated with spin-dependent terms in the director dynamics, termed triangular terms in Ishimoto's classification. The dynamics of general triangular objects are characterised by two ``triangularity'' parameters in addition to $\beta$ and $\gamma$. Additional symmetries in $C_{3v}$, $D_{3}$ and $D_{3d}$ cancel the Ishimoto parameter $\gamma$ and one of the two triangularity parameters; we call the associated dynamics \textit{reduced triangular} in table~\ref{table:shear-dynamics}. 

The remainder of table~\ref{table:shear-dynamics} completes the classification with symmetry groups which, except for the triaxial groups $D_{2}$ and $D_{2h}$ \citep{jeffery1922motion,bretherton1962motion,hinch1979rotation,yarin1997chaotic}, have not been examined before, to our knowledge. The twofold set generally has two additional parameters in $p_d$ beyond $\beta$ and $\gamma$. The corresponding terms in the director dynamics are not, however, the triangular terms of the threefold groups. By analogy, we call this class of dynamics \textit{digonal}. Further reduction occurs for $C_{2v}$ ($\gamma = 0$ and a single digonal parameter), $S_4$ (Ishimoto dynamics with nonzero $\beta$ and $\gamma$) and $D_{2d}$, which reduces to Jeffery equations.

As seen in table \ref{table:shear-dynamics}, $T_d$ symmetry induces a spherical behaviour for rotation. Hence, particles with sphere-like rotational dynamics in shear belong to a subset $\mathfrak{S}$ of three classes in $\mathfrak{H}_{1.5}$:
\begin{equation}
    \mathfrak{S} = \{ \OO(3), \SO(3), T_d \}.
\end{equation}
Further, the term ``Jeffery body'' was coined in \citet{dalwadi2024generalisedII} to describe a particle whose rotational dynamics follow Jeffery equations in shear, noting that geometric axisymmetry is not a necessary condition. A Jeffery body is hence defined as a particle whose hydrodynamic symmetry class induces Jeffery dynamics. Table~\ref{table:shear-dynamics} provides a complete characterisation of Jeffery bodies as a subset $\mathfrak{J}$ of $\mathfrak{H}_{1.5}$:
\begin{equation}
    \mathfrak{J} = \mathfrak{S} \cup \{ C_{\infty v} , D_{\infty}, D_{\infty h}, D_{3h}, D_{2d} \},
\end{equation}
while helicoidal bodies characterised by generalised Jeffery-Ishimoto spin-invariant dynamics may similarly be fully described as the set $\mathfrak{T}$ defined by
\begin{equation}
    \mathfrak{T} = \mathfrak{J} \cup \{ C_{\infty}, C_{\infty h}, C_{3h}, S_4 \}.
\end{equation}
Other hydrodynamic symmetry classes induce spin-dependent dynamics in shear flow.

For the three tetrahedral groups, both translational and rotational dynamics are reduced to at most one parameter, which we call $\tau_V$ for translation and $\tau_{\Omega}$ for rotation. Achiral symmetry classes $T_h$ and $T_d$ respectively yield $\tau_V = 0$ and $\tau_\Omega = 0$. Despite being characterised by a single parameter, the rotational dynamics of $T$-symmetric particles cannot be reduced to director dynamics and retain the full $\dot{Q}$ attitude law. The $T$-symmetric law is contained in the triaxial $D_2$ law studied for ellipsoidal particles \citep{jeffery1922motion,bretherton1962motion,hinch1979rotation,yarin1997chaotic}. The additional threefold rotation permutes the three $D_2$ axes and forces the three characteristic parameters to coincide with $\tau_{\Omega}$. This reduces the dynamics to
\begin{equation}
    \dot{Q} = \left [  \vect{\Omega}^{\infty}+ \tau_{\Omega} Q \mathcal{T}(Q^T \vect{E}^{\infty} Q) \right ] ^{\wedge} \, Q,
    \label{eq:tetrahedral-dynamics}
\end{equation}
where $\mathcal{T}(\vect{E}) = \begin{pmatrix} E_{23} & E_{31} & E_{12} \end{pmatrix}^T$ and the wedge denotes the cross-product matrix defined above.

Note that ellipsoids whose three characteristic parameters coincide are necessarily spheres, implying $\tau_{\Omega} = 0$; hence, the true tetrahedral case $\tau_{\Omega} \neq 0$ lies outside the reference studies on ellipsoidal particles in shear \citep{hinch1979rotation,yarin1997chaotic}. In \citet{ishimoto2020jeffery},
a $C_{3v}$ particle made of four rods assembled in a tetrahedral manner is described, but the angle definition does not yield $T$ symmetry. Hence, to our knowledge, the tetrahedral dynamics in \eqref{eq:tetrahedral-dynamics} have not been considered previously.

\begin{figure}
    \centering
    \includegraphics[width=\linewidth]{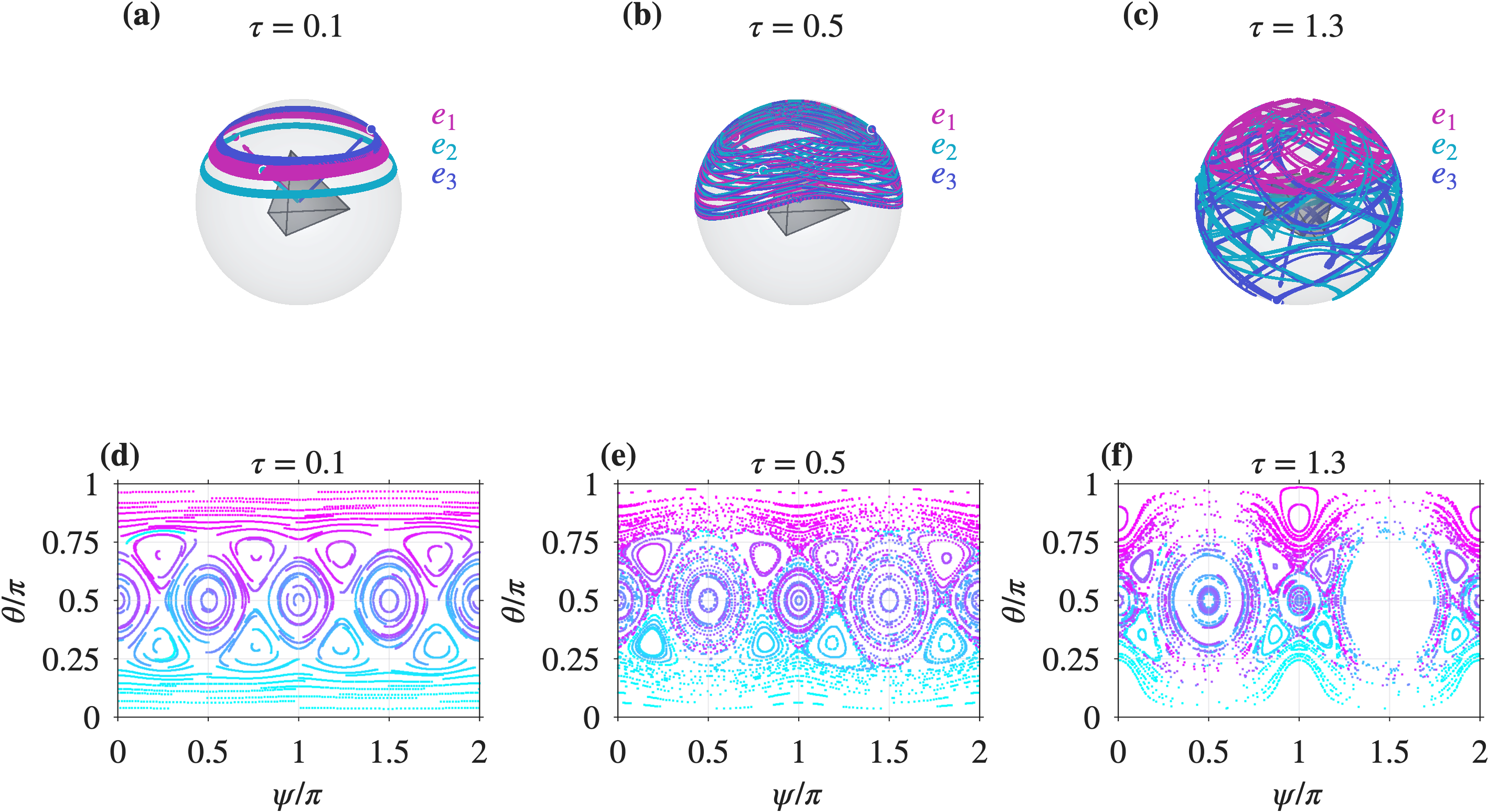}
    \caption{Rotational dynamics of particles with $T$ symmetry in shear flow. The top-row panels show the trajectories of the three body axes $\vect e_1$, $\vect e_2$ and $\vect e_3$ on the unit sphere. The bottom-row panels show finite-time Poincar\'e sections of the same attitude dynamics in the reported angular coordinates.}
    \label{fig:tetrahedral}
\end{figure}

Figure~\ref{fig:tetrahedral} illustrates numerical simulation of the $T$ law for three values of
$\tau_{\Omega}$. 
The body-axis trajectories and Poincar\'e sections pass from recurrent
curves to increasingly dispersed finite-time sections as $\tau_{\Omega}$ and the
initial attitude vary, with further analysis of the underlying dynamical system left to future investigation. 

We conclude this section with two remarks.
First, simple inversion, or centrosymmetry, imposes a simple but useful selection rule on the particle
dynamics, which generalises known results on the resistance matrix \citep{happel2012low}. It reverses polar velocity ($V_1^-$) while
leaving strain ($V_2^+$) and axial angular velocity ($V_1^+$) unchanged. Equation~\eqref{eq:shear-motion-projection} then implies
that $p_V(\mathbb{J})=0$ if $\mathbb{J}$ contains inversion. 
Hence, a force- and torque-free particle with centrosymmetry cannot acquire any strain-induced translation at this order. This is particularly striking for the $C_i$ row in table~\ref{table:shear-dynamics}, for which all translational parameters vanish, but 15 independent parameters for rotational dynamics are retained. 

Second, recall that octahedral symmetry produces no correction from spherical dynamics in simple shear: $O$ and $O_h$ remain indistinguishable from the corresponding isotropic classes at level
$k=1.5$. Their first additional coefficient lies in the stresslet--strain
block and does not alter the force- and torque-free motion in
\eqref{eq:shear-motion-projection}.

\subsection{Stresslet level}
\label{sec:stresslet-level}

The remaining linear-flow observable is the stresslet.
Similarly to Eq. \eqref{eq:shear-motion-projection}, imposing zero force and torque yields the following expression for the stresslet $\vect{S}$:
\begin{equation}
    \mathcal{R}_2
    \begin{pmatrix} - \delta \vect{V} \\ - \delta \vect{\Omega} \\ \vect{E}^{\infty} \end{pmatrix}
    = \begin{pmatrix} 0 \\ 0 \\ \vect{S} \end{pmatrix}
    \quad \Rightarrow \quad
    \vect{S} = \left[
    \vect{\Sigma}
    -\begin{pmatrix}\vect{\Gamma}^T&\vect{\Lambda}^T\end{pmatrix}
    \mathcal{R}_1^{-1}
    \begin{pmatrix}\vect{\Gamma}\\\vect{\Lambda}\end{pmatrix}
    \right]\vect{E}^{\infty}
    = \mathcal{P}_S \vect{E}^{\infty},
    \label{eq:effective-stresslet}
\end{equation}
where $\vect{\Sigma}$ is the stresslet-strain coupling block of $\mathcal{R}_2$ defined in Eq. \eqref{eq:resistance-tensors}.

The operator $\mathcal{P}_S$ retains the same representation structure as the coupling block $\vect{\Sigma}$: a self-adjoint operator on
$V_2^+$. For a given group $\mathbb{J} \in \mathfrak{H}_2$, the dimension of the invariant subspace $\mathcal{P}_S^{\mathbb{J}}$ therefore gives the number of independent
parameters $p_S(\mathbb{J})$ in the constitutive stresslet-strain response law.
Of particular note, the symmetry-invariance classification of the $\operatorname{Sym}^2(V_2^+)$ tensor representation also occurs in solid mechanics, for an elasticity tensor restricted to its deviatoric sector. The parameter counts $p_S$ in table \ref{tab:stresslet-counts} can therefore be found in the corresponding literature \citep{forte1996symmetry,clayton2025symmetries}, although applied to a different problem.

In practice, the stresslet primarily appears in the study of flow properties of dilute suspensions \citep{batchelor1970stress,hinch1972note}. A fluid of viscosity $\mu$ and rate-of-strain tensor $\vect{E}$ has bulk deviatoric stress $\boldsymbol{\sigma}=2\mu\vect{E}$. For a suspension of identical particles at number density $n$, the effective bulk stress at first order is $\boldsymbol{\sigma}'=\boldsymbol{\sigma}+n\langle\vect{S}\rangle$. In a simple shear flow of rate $\dot{\gamma}$, this yields the effective viscosity formula $\mu'=\mu+n\langle S_{12}\rangle/\dot{\gamma}$.

The parameter count $p_S$ tells how many independent parameters characterise the suspension stress or effective viscosity, depending on the hydrodynamic symmetry class of the particles in $\mathfrak{H}_2$, provided they are all aligned in the flow. The results are given in table~\ref{tab:stresslet-counts}. \citet{brenner1964stokesiii} showed that the effective bulk stress of a suspension of spheroids is characterised by three parameters, as in the $p_S=3$ row for the infinite axial groups. For spherical particles, $p_S$ collapses to the single parameter in Einstein's dilute-suspension formula \citep{einstein1906molecular}; for general ellipsoids belonging to the triaxial group $D_{2h}$, six independent parameters are needed.

As expected from the $\mathfrak{H}_2$ level, fourfold symmetry is distinguished from axisymmetry in table~\ref{tab:stresslet-counts}: fourfold objects generally yield a distinct response from helicoidal objects. A suspension of those particles would require one or two additional parameters to characterise their effective stress, provided they all share the same exact orientation, up to spin angle $\psi$. Another distinction occurs for tetrahedral and octahedral groups, for which $p_S=2$, whereas $p_S=1$ for isotropic symmetry. This suggests that a suspension of aligned tetrahedral or octahedral particles may have a distinctive rheological signature compared with a suspension of spheres, including a possible deviation from the Newtonian Trouton ratio \citep{trouton1906viscous}.

\begin{table}
\centering
\footnotesize
\setlength{\tabcolsep}{2pt}
\renewcommand{\arraystretch}{1.1}
\begin{tabular}{ l  c }
\toprule
 $\mathbb{J} \in \mathfrak{H}_{2}$ & $\quad p_S \quad $\\
 \hline
$C_1, C_i$ & 15 \\
$(C_2, C_{2h}); C_s$ & 9\\
$C_{2v},D_2, D_{2h}$ & 6 \\
$(C_3, S_6); (C_4, C_{4h}, S_4)$ & 5 \\
$(D_{2d}, C_{4v}, D_4, D_{4h}); (D_3, D_{3d}); C_{3v}$ & 4  \\
$C_{3h}$, $D_{3h}$, all infinite cyclic & 3  \\
$T, T_d, T_h, O, O_h$ & 2  \\
$\SO (3), \OO(3)$ & 1 
\\
\bottomrule
\end{tabular}
\caption{Stresslet-strain coupling parameter count for the 34 groups in $\mathfrak{H}_2$. Semicolon separators indicate different subspace structure within the same row.}
\label{tab:stresslet-counts}
\end{table}

\subsection{Quadratic flow}
\label{sec:quadratic-flow-dynamics}

Lastly, we consider the effect of the quadratic part of the Stokes jet $\vect{J}$ on the dynamics of a particle in flow. As in Equation~\eqref{eq:shear-motion-projection}, we define $\delta \vect{V}^{(2)}$ and $\delta \vect{\Omega}^{(2)}$ as the corrections to particle velocity induced by the quadratic flow. Then, we can derive the operator coupling quadratic flow to rotational and translational dynamics, assuming zero net force and torque and zero linear background flow:
\begin{equation}
    \begin{pmatrix} \vect{K} & \vect{C} & \vect{\Gamma} & \vect{\Pi} \\ \vect{C}^T & \vect{Q} & \vect{\Lambda} & \vect{\Delta} \end{pmatrix} \begin{pmatrix} -\delta \vect{V} \\ -\delta \vect{\Omega} \\ 0 \\ \vect{J} \end{pmatrix} =0 \quad \Rightarrow  \quad  \begin{pmatrix} \delta \vect{V} \\ \delta \vect{\Omega} \end{pmatrix} = \mathcal{R}_1^{-1}\begin{pmatrix} \vect{\Pi} \\ \vect{\Delta} \end{pmatrix} \vect{J} \quad \Rightarrow \begin{array}{l} \delta \vect{V} = \mathcal{Q}_V \vect{J}, \\ \delta \vect{\Omega} = \mathcal{Q}_\Omega \vect{J}. \end{array}
    \label{eq:quadratic-dynamics-operator}
\end{equation}
The first equation in \eqref{eq:quadratic-dynamics-operator} introduces the explicit notations $\vect{\Pi}$ and $\vect{\Delta}$ for the force-quadratic and torque-quadratic coupling blocks, which are expressed in coordinates as fourth-rank tensors. 

In general, the background flow is not strictly quadratic and contains a linear part $(\vect{V}^{\infty},\vect{\Omega}^{\infty}, \vect{E}^{\infty})$, so that 
\begin{equation}
\vect{V}_p = \vect{V}^{\infty} + \mathcal{P}_V \vect{E}^{\infty} + \mathcal{Q}_V \vect{J}, \qquad \vect{\Omega}_p = \vect{\Omega}^{\infty} + \mathcal{P}_\Omega \vect{E}^{\infty} + \mathcal{Q}_\Omega \vect{J}.
\end{equation}
Moreover, in a quadratic flow, the linear part of the local expansion varies as the particle translates, inducing additional coupling between translational and rotational dynamics. Equation~\eqref{eq:quadratic-dynamics-operator} should therefore be viewed as a local, instantaneous quadratic correction to the linear dynamics, including Jeffery rotation and ambient vorticity; the complete particle dynamics is generally more complex because of translation and these coupling effects.

For $\mathbb{J} \in \mathfrak{H}_3$, the parameter counts $q_{V}(\mathbb{J})$, $q_{\Omega}(\mathbb{J})$, given by the dimension of the $\mathbb{J}$-invariant spaces of $\mathcal{Q}_V$ and $\mathcal{Q}_\Omega$, are then determined by character computation. 

The results are presented in table~\ref{tab:quadratic}. When director reduction is possible, the parameter count $q_d$ is computed as well. To avoid overcrowding, we have omitted low-symmetry groups with twofold symmetry, whose high parameter counts lead to relatively intractable dynamical equations.

A first important remark is that all groups containing inversion, as well as the fivefold groups $D_{5h}$, $D_{5d}$ and $C_{5h}$, satisfy $q_{\Omega}=q_d=0$: for these 19 groups among the 50 in $\mathfrak{H}_3$, the quadratic flow does not affect the rotational dynamics.

We next consider the nine groups in the first block of table~\ref{tab:quadratic}. Although fivefold and sixfold symmetry is visible at this level for the full $\mathcal{R}_3$ operator (see \S\ref{sec:quadratic-resistance}), it is not resolved by the dynamical operators $\mathcal{Q}_V$ and $\mathcal{Q}_{\Omega}$.
Hence, $C_5, C_6$ and their $D_n$ and $C_{nv}$ counterparts follow axisymmetric-like dynamics; we call them $\mathcal{R}_3$-helicoidal. For the same reason, the dynamics of the icosahedral groups $I$ and $I_h$, built on $5$-fold symmetry, are identical to those of the isotropic groups $\SO (3)$ and $\OO (3)$. The distinction is only seen in the higher-coupling $(\vect{G}, \vect{J})$ block.

Among the non-inversion axial groups, the $\mathcal{R}_3$-helicoidal groups are exactly those for which the director dynamics is decoupled from spin, inducing a two-dimensional regular system of Jeffery--Ishimoto type for the $\vect{d}$ dynamics. 

\begin{figure}
    \centering
    \includegraphics[width=\linewidth]{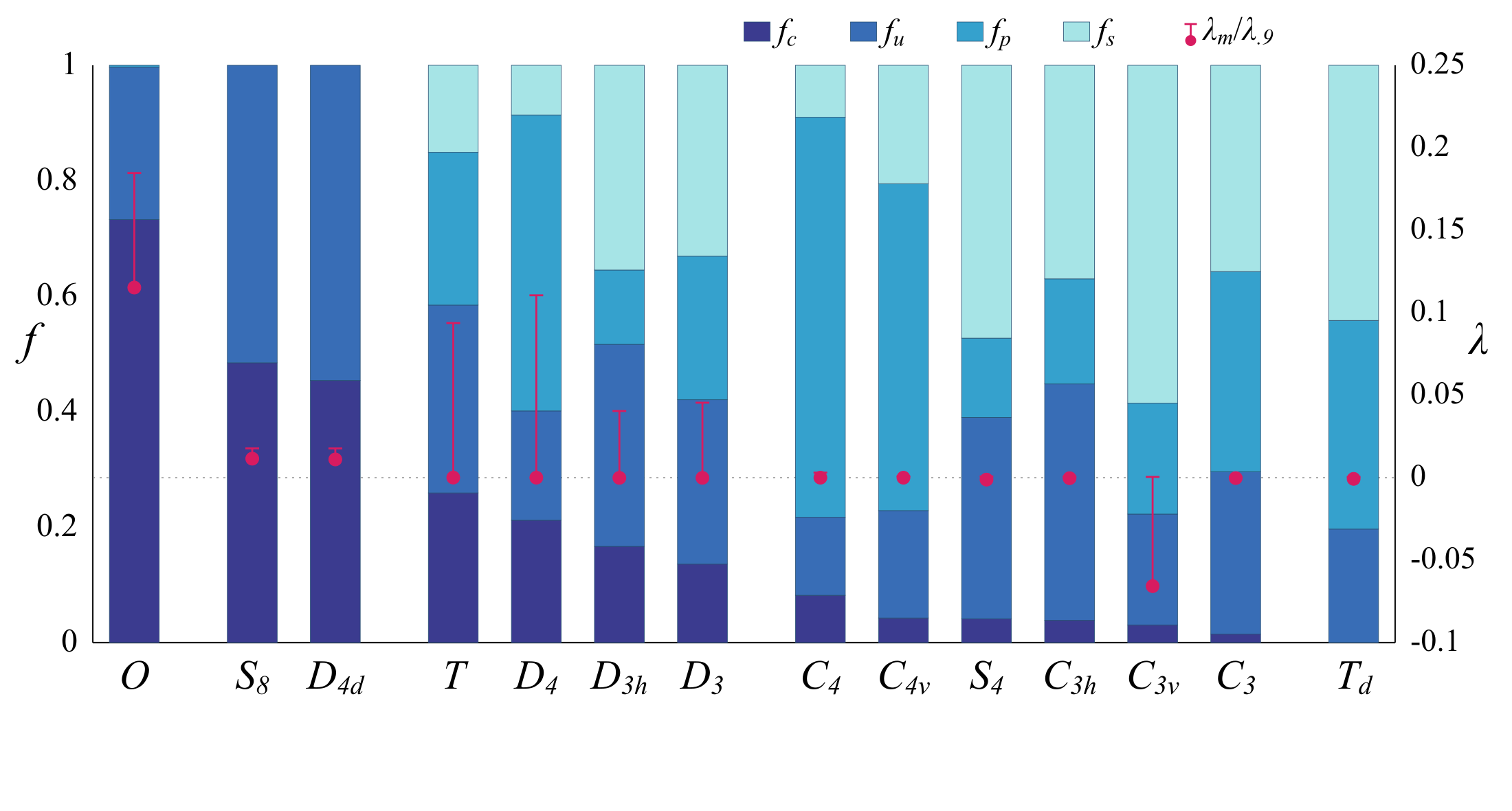}
    
    \vspace{-2em}
    \caption{Numerical exploration of the rotational dynamics of the 14 full-attitude symmetry classes selected from table~\ref{tab:quadratic}. For each group, 1500 simulations are classified into four fractions, shown as a bar plot: chaotic $f_c$, periodic $f_p$, steady $f_s$ and unclassified $f_u$. The median finite-time Lyapunov exponent $\lambda_m$ and its ninth decile $\lambda_{0.9}$ are shown by red points and horizontal bars, respectively, with values read on the right $y$-axis.}
    \label{fig:lyapunov}
\end{figure}

On the other hand, threefold and fourfold symmetry is not helicoidal at this level. The threefold and fourfold block of table~\ref{tab:quadratic}, as well as the polyhedral groups, retain full attitude dynamics. Whilst a full programme investigating every group-wise attitude equation is beyond the scope of the present paper, we provide a qualitative numerical analysis of emergent behaviour in figure~\ref{fig:lyapunov}. For each of the 14 groups featured in the figure, we used 1500 deterministic coefficient--initial-attitude pairs. The coefficient vectors comprise all signed coordinate directions followed by reproducible Gaussian mixed directions, each normalised to Euclidean norm $1.5$; the initial attitudes form the same quasi-uniform Haar design for every group. The rotational dynamics were simulated over $0\leq t\leq1600$ with \textsc{Matlab}'s adaptive \texttt{ode113} solver, using relative and absolute tolerances $10^{-10}$ and $10^{-12}$, respectively, and maximum step $0.25$. The attitude is represented by a unit quaternion $q$, satisfying
$\dot q=\tfrac12(0,\vect{\Omega})\otimes q$. For each run, we compute a finite-time largest Lyapunov exponent from a shadow attitude initially placed at geodesic distance $10^{-7}$, with renormalisation every 10 time units; the first $35\,\%$ of the trajectory is discarded, and the retained interval is split into two windows \citep{benettin1980lyapunov,pikovsky2016lyapunov}. A trajectory is classified as chaotic if the exponent exceeds $0.008$ on both windows and the two estimates satisfy the stated convergence test. Otherwise, the trajectory is tested for periodicity or convergence to a fixed attitude. The remaining trajectories are marked as ``unclassified'' and may correspond to quasiperiodic behaviour, slow transients or unresolved numerical behaviour. Further details are given in Appendix~\ref{app:numerical-dynamics}.

The results in figure~\ref{fig:lyapunov} show a range of dynamical behaviours. The most regular groups display periodic or steady trajectories in a broad majority of cases; in particular, no trajectory in the $T_d$ sample meets the finite-time chaos criterion. Irregular or chaotic trajectories, as well as occasional large values of $\lambda$, occur for $D_3$, $D_{3h}$, $D_4$ and $T$. Periodic and steady trajectories are absent from the $S_8$ and $D_{4d}$ samples, which also have positive median exponents. Their dynamics therefore appear weakly chaotic in this finite-time screen, despite involving only two and one independent attitude-dynamics parameters.

The octahedral case deserves special notice. Its two-parameter rotational normal form may be written explicitly by choosing the body-fixed axes along the three fourfold axes of the particle. Let
$
J^b_{ijk}=Q_{ai}Q_{bj}Q_{ck}J_{abc}
$
denote the quadratic jet expressed in this frame, where $Q$ maps body-fixed coordinates to laboratory coordinates. Under the zero-linear-background assumption of Equation~\eqref{eq:quadratic-dynamics-operator}, the angular velocity and attitude dynamics then take the form
\begin{equation}
\vect{\Omega}_p
=
Q\left[
o_1\mathcal{L}(\vect{J}^b)
+
o_2\mathcal{O}(\vect{J}^b)
\right],
\qquad
\dot Q=\widehat{\vect{\Omega}_p}Q.
\label{eq:octahedral-quadratic-dynamics}
\end{equation} Here, we have introduced the two vector-valued contractions
\begin{equation}
\mathcal{L}(\vect{J}^b)
=
\begin{pmatrix}
\sum_a J^b_{1aa} \\
\sum_a J^b_{2aa} \\
\sum_a J^b_{3aa}
\end{pmatrix},
\mathcal{O}(\vect{J}^b)
=
\begin{pmatrix}
J^b_{111}-\frac{1}{5}\sum_a J^b_{1aa}\\
J^b_{222}-\frac{1}{5}\sum_a J^b_{2aa}\\
J^b_{333}-\frac{1}{5}\sum_a J^b_{3aa}
\end{pmatrix}.
\label{eq:octahedral-contractions}
\end{equation}
The first contraction is rotationally invariant and corresponds to
$\Delta\vect{U}^{\infty}$, whereas the second extracts the component of the harmonic third-order jet that transforms as a vector under the octahedral group. 
In \eqref{eq:octahedral-quadratic-dynamics}, $o_1$ therefore describes the isotropic part of the quadratic rotational response, which is also permitted by $\SO(3)$ symmetry, while $o_2$ is the genuinely octahedral contribution. In particular, setting $o_2=0$ reduces Equation~\eqref{eq:octahedral-quadratic-dynamics} to the isotropic normal form.

\begin{figure}
    \centering
    \includegraphics[width=\linewidth]{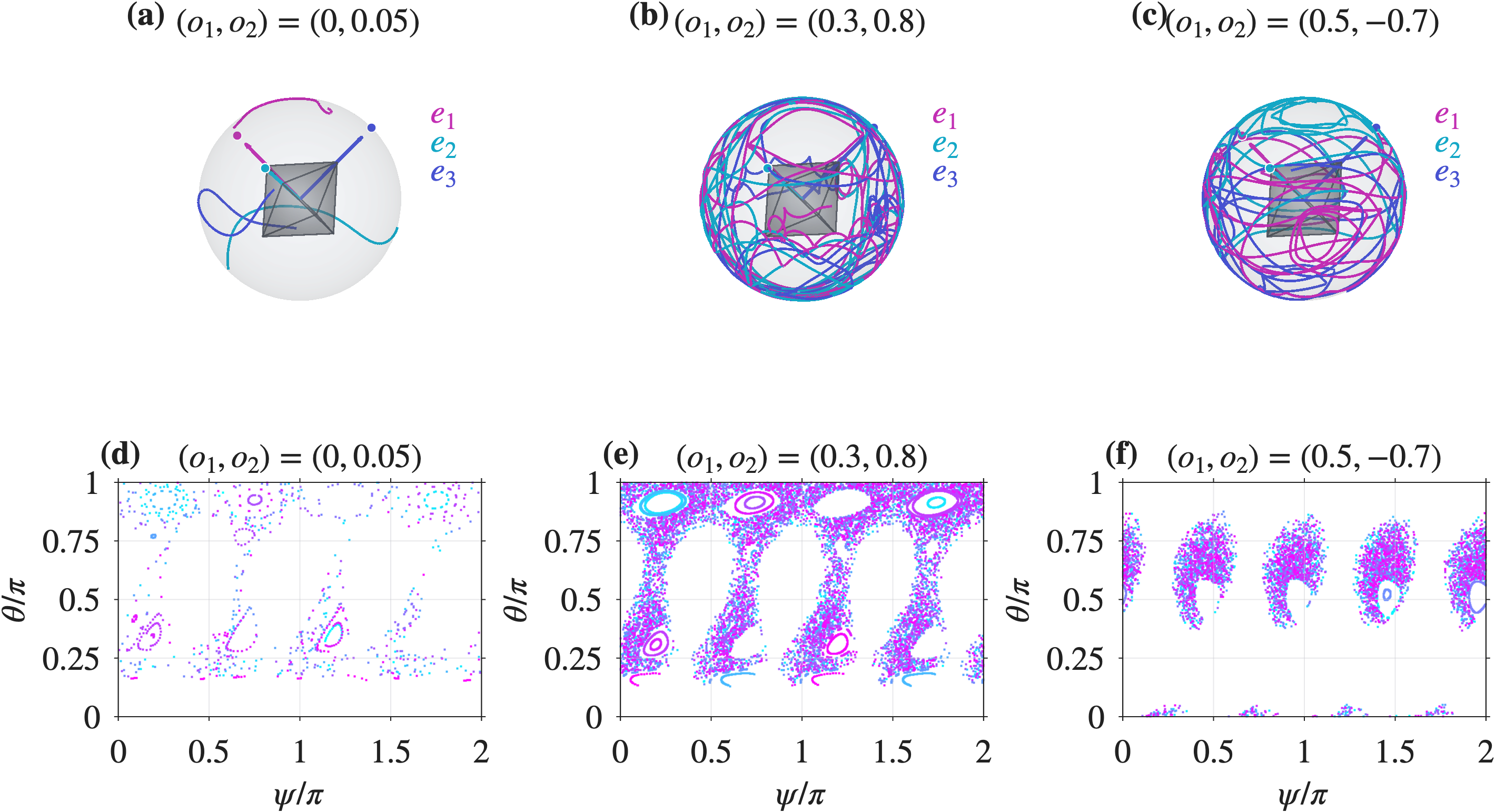}
    \caption{Representative trajectories and Poincar\'e sections for octahedral attitude dynamics in quadratic flow.}
    \label{fig:octahedral}
\end{figure}

In spite of this simple closed form for its quadratic dynamics, the octahedral group O displays the clearest finite-time instability in the present numerical screen. Approximately 73\% of the sampled coefficient–initial-attitude pairs satisfy our chaos criterion, and the median finite-time Lyapunov exponent is approximately 0.115, substantially larger than for the other groups considered. These values are specific to the prescribed sampling and normalization and are not intended as intrinsic statistical measures of octahedral dynamics. Nevertheless, together with the representative trajectories and Poincaré sections shown in Figure~\ref{fig:octahedral}, they provide strong evidence that the octahedral normal form admits chaotic regimes over the sampled parameter range. Indeed, the trajectories spread over a broad region of attitude space when the second octahedral parameter $o_2$ is introduced.

\begin{table}
\centering
\renewcommand{\arraystretch}{1}
\begin{tabular}{ c c | c | c c c }
\toprule
Category
& $\mathbb{J}\in\mathfrak{H}_{3}$
& $\quad q_V\quad$
& $\quad q_\Omega\quad$
& $\quad q_d\quad$
& Qualitative assessment \\
\hline

\multirow{3}{*}{Helicoidal}
& $C_5,C_6,C_\infty$
& 9 & 9 & 6 & Spin-invariant \\
& $C_{5v},C_{6v},C_{\infty v}$
& 5 & 4 & 3 & Spin-invariant \\
& $D_5,D_6,D_\infty$
& 5 & 5 & 3 & Spin-invariant \\
\hline

\multirow{11}{*}{3- and 4-fold}
& $C_3$    & 15 & 15 & 10 & Quasi-regular \\
& $C_4$    & 11 & 11 & 8  & Quasi-regular \\
& $S_4$    & 11 & 12 & 8  & Quasi-regular \\
& $C_{3v}$ & 8  & 7  & 5  & Quasi-regular \\
& $C_{3h}$ & 9  & 6  & 4  & Quasi-regular \\
& $C_{4v}$ & 6  & 5  & 4  & Quasi-regular \\
& $D_3$    & 8  & 8  & 5  & Irregular \\
& $D_4$    & 6  & 6  & 4  & Irregular \\
& $D_{3h}$ & 5  & 3  & 2  & Irregular \\
& $S_8$    & 9  & 2  & 2  & Chaotic \\
& $D_{4d}$ & 5  & 1  & 1  & Chaotic \\
\hline

\multirow{4}{*}{Polyhedral}
& $T$        & 4 & 4 & -- & Irregular \\
& $T_d$      & 2 & 2 & -- & Regular \\
& $O$        & 2 & 2 & -- & Highly chaotic \\
& $I,\SO(3)$ & 1 & 1 & -- & Regular \\
\hline

\multirow{9}{*}{\shortstack{Inversion \\ (or assimilated)}}
& $S_6$
& 15 & \multicolumn{3}{c}{\multirow{9}{*}{No quadratic rotation}} \\
& $C_{4h}$
& 11 & & & \\
& $C_{5h},C_{6h},C_{\infty h},S_{10}$
& 9 & & & \\
& $D_{3d}$
& 8 & & & \\
& $D_{4h}$
& 6 & & & \\
& $D_{5h},D_{5d},D_{6h},D_{\infty h}$
& 5 & & & \\
& $T_h$
& 4 & & & \\
& $O_h$
& 2 & & & \\
& $I_h,\OO(3)$
& 1 & & & \\
\bottomrule
\end{tabular}
\caption{Selected parameter counts $q_V$, $q_\Omega$ and, when
applicable, $q_d$, for the new translational and rotational responses
to a quadratic imposed-flow jet. 
The qualitative assessments concern the autonomous
rotational subsystem driven by a prescribed fixed laboratory-frame
jet at its expansion point.}
\label{tab:quadratic}
\end{table}

\section{Discussion}
\label{sec:discussion}

\subsection{Infinite limit}

The hierarchy of resistance operators $\mathcal{R}_k$ corresponds to progressively refined finite-dimensional observations of particle geometry through its response to the ambient flow. By considering only levels up to $k=3$, the present study is therefore restricted, in a sense, to the coarsest levels of observation, although linear and quadratic moments are undoubtedly sufficient for the vast majority of practical applications in fluid mechanics. The asymptotic behaviour of the hierarchy is nevertheless mathematically interesting. For a fixed particle $B$, let $\Lambda_B$ denote the exterior-Stokes Dirichlet-to-Neumann map taking an admissible boundary velocity to the corresponding surface traction. It is natural to ask whether the finite-order operators $\mathcal{R}_k(B)$, suitably realised as restrictions or projections of a common infinite-dimensional operator, recover $\Lambda_B$ as $k\rightarrow\infty$. The associated inverse problem asks: ``can one feel the shape of a particle with the Stokes fluid flow?'' Related uniqueness results are known for bounded domains containing bounded obstacles with Lipschitz boundaries \citep{alvarez2005obstacles}. A convergence result based on the density of polynomial jet spaces appears plausible, but would require a precise choice of trace spaces, projections and topology; its rigorous treatment is left to future studies.

A different limiting question concerns the non-geometric groups appearing in the hydrodynamic classification. As discussed in \S\ref{sec:classification}, every group in the catalogue $\mathfrak{G}$ occurs in $\mathfrak{H}_k$ from some finite level onward, suggesting the formal set limit $\mathfrak{H}_k\to\mathfrak{G}$. From here, the four-element gap between $\mathfrak{G}$ and $\mathfrak{G}_{\mathrm{geom}}$ raises an intriguing compactness question.

Consider, for example, a sequence of chiral particles $H_k$ such that
$\mathcal{R}_k(H_k)\in\mathcal{R}_k^{\SO(3)}$ for every $k$. By the nested structure of the hierarchy, $\mathcal{R}_j(H_k)$ is then isotropic for every fixed $j\leqslant k$. Suppose that a subsequence of $H_k$ converges to a regular particle $H_\infty$ in a topology for which every finite-order resistance operator depends continuously on the shape. It would follow that $\mathcal{R}_j(H_\infty)$ is isotropic for every finite $j$. If the complete hierarchy recovers a shape-determining boundary operator $\Lambda_{H_\infty}$, that operator would consequently be rotationally invariant, forcing $H_\infty$ itself to be rotationally invariant. The sequence must therefore lose its chirality in the limit. In particular, the chiral coupling $\alpha_k$ in the $\vect{C}$ block would have to vanish. If $\alpha_k$ remained bounded away from zero, no such regular convergent subsequence could exist. Thus, a sequence of increasingly high-order isotropic helicoids must either lose its chiral response, fail to converge in the required topology, or degenerate towards an object outside the regularity class for which the Stokes problem and the inverse result are well-posed. Determining which of these alternatives occurs would clarify how non-geometric hydrodynamic symmetries disappear in the infinite-order limit.

\subsection{Model extensions}

The present study focuses on passive particles in Newtonian Stokes flow. Naturally, the resistance operator structure and the representation-theoretic methodology developed here can be extended to various model refinements, as long as the governing equations remain equivariant under some spatial group such as $\SO (3)$ and the selected response, such as a resistance operator, admits a compatible representation.

Active and prescribed-slip particles provide the closest extension. Their
force moments contain an active contribution in addition to the passive
resistance law, and the relevant symmetry acts jointly on the body and on its
slip or actuation pattern \citep{elfring2017force,nasouri2018higher}.
A sphere
with an anisotropic stick--slip pattern, for instance, may have a stresslet and
rheological signature controlled by the pattern rather than by the geometric
shape alone \citep{premlata2022anisotropic}. Hydrodynamic symmetry must then be understood for the particle's active pattern, in addition to its geometry. 

Other fluid models alter the algebra more deeply. 
In a parity-breaking or odd
fluid \citep{hosaka2023lorentz}, Lorentz reciprocity need not impose the same self-adjointness relations as in a regular Stokes fluid,
so antisymmetric response blocks may appear. 
Weak fluid or particle inertia
introduces orientation-dependent corrections absent from the Stokes operator
and can modify the rotational law
\citep{candelier2015inertia,sundberg2026fluid}. 
Finally, viscoelasticity
adds memory: the response becomes history- or frequency-dependent, and the
group action must then include the internal constitutive variables \citep{grimm2011maxwell}. 
Establishing analogues of the hydrodynamic symmetry-group sets $\mathfrak{H}_k$ for these extended models and highlighting their specificities constitute an interesting research avenue. 

\section{Conclusion}

The central contribution of this study is the sequence of hydrodynamic symmetry-group sets $\mathfrak{H}_k$. The associated invariant spaces at each resistance level $\mathcal{R}_k$, although based on classical representation-theoretic tools, place several well-known results in Stokes fluid mechanics within a common framework: from hydrokinetic symmetry and isotropic helicoids introduced in the nineteenth century, through Brenner's helicoidal symmetry in the 1960s, to recent generalisations of Jeffery's equations.

Besides this conceptual framework, the description of $\mathfrak{H}_k$ at each level provides a classification of particle responses into $|\mathfrak{H}_k|$ possible hydrodynamic symmetry classes, represented by normal forms of the resistance operators.

The classification induces an exhaustive list of possible dynamical equations in terms of independent parameter counts. From there, we were able to complete the dynamical classification of Jeffery--Bretherton dynamics of particles in shear flow, partially realised by Ishimoto, and extend it to stresslet coupling and quadratic corrections. Amongst several new normal forms identified by the classification, we have examined the rotational dynamics of chiral-tetrahedral particles in shear flow and chiral-octahedral particles, which, to our knowledge, had not been considered before. Despite having a high level of symmetry, these particle types seem to follow remarkably irregular dynamics, warranting further investigation of bifurcations and chaos signatures in their dynamical systems, as well as experimental and numerical realisations. 

Indeed, although this study provides a structure for the space of hydrodynamic responses, it does not populate the structure with actual particles. The inverse problem of exhibiting a shape that realises prescribed independent parameters within a symmetry class remains open. Within the general programme of generating a varied family of shapes for a given level $k$, symmetry class $\mathbb{J}$ and possible parameter values in $\mathcal{R}_k^\mathbb{J}$, several problems are, in my opinion, of particular interest: (a) constructing the isotropically helicoidal icosahedral particle described in \S\ref{sec:quadratic-resistance}; (b) determining typical parameter values for tetrahedral particles in shear and octahedral particles in quadratic flow (\S\S\ref{sec:shear-dynamics} and \ref{sec:quadratic-flow-dynamics}); and (c) illustrating the distinction between $\mathfrak{G}_{\mathrm{geom}}$ and $\mathfrak{H}_k$ by exhibiting particles with an $\OO(3)$-invariant response at level $k$ but an arbitrarily chosen geometric symmetry $\mathbb{J}\in\mathfrak{G}_{\mathrm{geom}}$. These questions could be addressed using Stokes solvers designed specifically for complex particles and microswimmers \citep{gissinger2026resistance,Cass2026} combined with symmetry-preserving shape optimisation \citep{moreau2025shapes} and level-set methods.

\begin{bmhead}[Acknowledgements]
I would like to thank Prof. Kenta Ishimoto for fruitful discussions.
\end{bmhead}

\begin{bmhead}[Funding information]
I acknowledge funding by the Pulsar programme from R\'egion Pays de la Loire.
\end{bmhead}

\begin{bmhead}[Data availability]
The code used for this paper is available at \url{https://github.com/Clementmoreau/stokes-resistance-symmetry}.
\end{bmhead}

\begin{bmhead}[Competing interests]
I declare no conflict of interest.
\end{bmhead}

\begin{bmhead}[Declaration of AI usage]
Any use of generative AI in this manuscript adheres to ethical guidelines for use and acknowledgement of generative AI in academic research. I assume responsibility for the integrity of my contributions. AI (OpenAI ChatGPT 5.5) was used in drafting the Matlab codes supporting this manuscript. 
\end{bmhead}

\enlargethispage{20pt}

\begin{appen}
\renewcommand{\theHsection}{appendix.\Alph{section}}

\section{Dimension of the general reduced resistance space}
\label{app:general-dimension}

This appendix gives the calculation behind \eqref{eq:reduced-space-dimension}.  For an
integer \(k\geqslant3\), let
\(\mathcal H_q\) denote the homogeneous degree-\(q\) Stokes sector and set
\begin{equation}
  W_k:=\mathcal J_{k-1}
  =\mathcal H_0\oplus\mathcal H_1\oplus\cdots\oplus\mathcal H_{k-2}
\end{equation}
be the space of imposed Stokes jets through homogeneous polynomial degree
\(k-2\).  Thus \(W_3=V_1^-\oplus V_1^+\oplus V_2^+\) is the full linear-flow
space and \(W_4=W_3\oplus\mathcal H_2\) includes the quadratic jet.  Here
\(\mathcal J_{k-1}\) is the truncation used in the main text, whereas
\(\mathcal H_q\) denotes one homogeneous sector.  The unreduced reciprocal
response space is
\begin{equation}
  \mathcal R_{k-1}^{\rm raw}=\operatorname{Sym}^2(W_k^*).
  \label{eq:app-raw-resistance-space}
\end{equation}

The dimensions of the homogeneous jet sectors follow from their harmonic
decomposition.  The constant and linear sectors are
\begin{equation}
  \mathcal H_0\simeq V_1^-,\qquad
  \mathcal H_1\simeq V_1^+\oplus V_2^+,
\end{equation}
and therefore have dimensions \(3\) and \(3+5=8\).  For every \(q\geqslant2\),
the homogeneous degree-\(q\) Stokes sector decomposes as
\begin{equation}
  \mathcal H_q\simeq
  V_{q-1}^{(-1)^{q+1}}\oplus
  V_q^{(-1)^{q+1}}\oplus
  V_{q+1}^{(-1)^{q+1}}.
  \label{eq:app-stokes-jet-decomposition}
\end{equation}
The common parity records the sign under inversion.  Since
\(\dim V_\ell=2\ell+1\),
\begin{equation}
  \dim\mathcal H_q=(2q-1)+(2q+1)+(2q+3)=6q+3.
\end{equation}
Consequently,
\begin{align}
  D_k:=\dim W_k
  &=3+8+\sum_{q=2}^{k-2}(6q+3) \\
  &=3k^2-6k+2.
  \label{eq:app-jet-dimension}
\end{align}
The check \(D_4=26\) recovers the \(3+3+5+15\) components of the
quadratic-flow jet used in \S\ref{sec:quadratic-flow-dynamics}.

Reciprocity makes the resistance operator a symmetric bilinear form, so
\begin{align}
  \dim\mathcal R_{k-1}^{\rm raw}
  &=\frac{D_k(D_k+1)}2 \\
  &=\frac92k^4-18k^3+\frac{51}{2}k^2-15k+3.
  \label{eq:app-raw-dimension}
\end{align}
The hydrodynamic-centre condition removes the three coordinates associated
with translating the reference point.  For a generic body, diagonalising the
translational resistance then fixes the three rotational coordinates of the
adapted frame.  Hence
\begin{align}
  \dim\tilde{\mathcal R}_{k-1}
  &=\dim\mathcal R_{k-1}^{\rm raw}-3-3 \\
  &=\frac92k^4-18k^3+\frac{51}{2}k^2-15k-3,
  \qquad k\geqslant3,
\end{align}
which is \eqref{eq:reduced-space-dimension}. Note that the two subtractions have
different meanings: the first is an origin gauge and the second is an
adapted-frame quotient.  This distinction becomes important once a symmetry
group itself fixes some directions, as detailed next.

\section{Closed dimension formulae for the subgroup families}
\label{app:family-dimensions}

We derive here the family formulae announced in
\S\ref{sec:fixed-space-characterisation}. 
Standard accounts of character projection for finite groups and Haar
projection for compact groups are given by \citet{serre1977linear} and
\citet{brocker2003representations,folland2016course}.

\subsection{Characters of the Stokes jet and its symmetric square}

For a proper rotation through angle \(\theta\), define the usual
\(\SO(3)\) character
\begin{equation}
  \psi_\ell(\theta)
  =\frac{\sin((\ell+\tfrac12)\theta)}{\sin(\theta/2)}
  =1+2\sum_{m=1}^{\ell}\cos(m\theta).
\end{equation}
Equation \eqref{eq:app-stokes-jet-decomposition} gives the character of \(W_k\):
\begin{equation}
  \chi_k(\theta)
  =3\sum_{\ell=1}^{k-3}\psi_\ell(\theta)
   +2\psi_{k-2}(\theta)+\psi_{k-1}(\theta),
  \qquad k\geqslant3.
  \label{eq:app-proper-jet-character}
\end{equation}
Equivalently,
\begin{equation}
  \chi_k(\theta)=b_0+2\sum_{m=1}^{k-1}b_m\cos(m\theta),
  \label{eq:app-axial-weights}
\end{equation}
where
\begin{equation}
  b_0=3k-6,\qquad
  b_m=3(k-m-1)\ (1\leqslant m\leqslant k-2),
  \qquad b_{k-1}=1.
  \label{eq:app-weight-multiplicities}
\end{equation}
These \(b_m\) are the multiplicities of the axial Fourier weights.  In
particular, the largest weight in \(W_k\) is \(k-1\), and the largest weight in
its symmetric square is \(2k-2\).  This gives the collapse threshold used in
\S\ref{sec:general-flow}: a finite axial sample of order greater
than \(2k-2\) cannot alias a non-zero weight to zero.

Improper elements also require the inversion parity in
\eqref{eq:app-stokes-jet-decomposition}.  If \(Q=(-I)R_\theta\), define
\begin{equation}
  \eta_k(\theta)=
  -\psi_1(\theta)+\psi_1(\theta)+\psi_2(\theta)
  +\sum_{q=2}^{k-2}(-1)^{q+1}
   \{\psi_{q-1}(\theta)+\psi_q(\theta)+\psi_{q+1}(\theta)\}.
  \label{eq:app-improper-jet-character}
\end{equation}
The first two terms display the cancellation between the polar constant-flow
sector and the axial part of the linear-flow sector.

For \(Q\in\OO(3)\), let
\begin{equation}
  \Xi_k(Q)=
  \begin{cases}
  \chi_k(\theta(Q)),&\det Q=1,\\
  \eta_k(\theta(-Q)),&\det Q=-1.
  \end{cases}
\end{equation}
The symmetric-square identity in \eqref{eq:symmetric-square-character} then gives,
for every finite point group \(H\),
\begin{equation}
  d_k^{\rm raw}(H)
  :=\dim\bigl(\mathcal R_{k-1}^{\rm raw}\bigr)^H
  =\frac1{2|H|}\sum_{Q\in H}
  \left\{\Xi_k(Q)^2+\chi_k(\theta(Q^2))\right\}.
  \label{eq:app-universal-fixed-dimension}
\end{equation}
This is already a closed formula for every finite row in the Schoenflies
catalogue.  It also makes the leading growth transparent.  Since only the
identity has character \(D_k=3k^2-6k+2\) of order \(k^2\),
\begin{equation}
  d_k^{\rm raw}(H)
  =\frac{9}{2|H|}k^4-\frac{18}{|H|}k^3+O(k^2)
  \label{eq:app-finite-leading-polynomial}
\end{equation}
for a fixed finite group \(H\).  The lower coefficients are quasi-polynomial:
they depend on \(k\) modulo the orders of the rotations in \(H\).

\subsection{Cyclic and improper cyclic families}

For \(C_r=\langle R_{2\pi/r}\rangle\),
\begin{equation}
  d_k^{\rm raw}(C_r)=\frac1{2r}\sum_{j=0}^{r-1}
  \left\{\chi_k(2\pi j/r)^2+\chi_k(4\pi j/r)\right\}.
  \label{eq:app-Cr-class-sum}
\end{equation}
The following residue form turns this trigonometric sum into integer
arithmetic.  Extend \(b_{-m}=b_m\), put \(L=k-1\), and define
\begin{equation}
  \mathcal B_s^{(r)}=
  \sum_{\substack{-L\leqslant m\leqslant L\\m\equiv s\ ({\rm mod}\ r)}}
  b_{|m|},
  \qquad
  \mathcal T_r=\sum_{\substack{s\in\mathbb Z/r\mathbb Z\\2s=0}}
  \mathcal B_s^{(r)}.
\end{equation}
Then
\begin{equation}
  d_k^{\rm raw}(C_r)=\frac12\left(
  \sum_{s\in\mathbb Z/r\mathbb Z}\mathcal B_s^{(r)}
  \mathcal B_{-s}^{(r)}+\mathcal T_r\right).
  \label{eq:app-Cr-residue}
\end{equation}
For a completely explicit floor-function evaluation, set
\begin{equation}
  m_a=\begin{cases}a,&a>0,\\r,&a=0,\end{cases}
  \qquad
  K_a=\max\left(0,1+\left\lfloor\frac{k-2-m_a}{r}\right\rfloor\right),
\end{equation}
and
\begin{equation}
  F_a^{(r)}(k)=3K_a(k-1)
  -\frac{3K_a}{2}\{2m_a+(K_a-1)r\}
  +\boldsymbol{1}_{k-1\equiv a\ ({\rm mod}\ r)}.
\end{equation}
One has
\begin{equation}
  \mathcal B_s^{(r)}=
  \boldsymbol{1}_{s=0}(3k-6)+F_s^{(r)}(k)+F_{-s}^{(r)}(k).
  \label{eq:app-residue-coefficient}
\end{equation}
Equations \eqref{eq:app-Cr-residue}--\eqref{eq:app-residue-coefficient} give the polynomial on
each residue class of \(k\) modulo \(r\), without listing the subcases.

The continuous axial limit keeps only equal, rather than congruent, weights:
\begin{align}
  d_k^{\rm raw}(C_\infty)
  &=\frac{b_0(b_0+1)}2+\sum_{m=1}^{k-1}b_m^2 \\
  &=3k^3-9k^2+3k+7.
  \label{eq:app-Cinf-polynomial}
\end{align}
For vertical mirrors, every improper element has \(-Q\) a half-turn and
\(Q^2=I\), which yields
\begin{equation}
  d_k^{\rm raw}(C_{rv})
  =\frac12d_k^{\rm raw}(C_r)+\frac12(2k^2-5k+3),
  \label{eq:app-Crv}
\end{equation}
and hence
\begin{equation}
  d_k^{\rm raw}(C_{\infty v})
  =\frac{3k^3-7k^2-2k+10}{2}.
  \label{eq:app-Cinfv-polynomial}
\end{equation}

For horizontal mirrors, write
\(\eta_k(\theta)=e_0+2\sum_{m=1}^{k-1}e_m\cos(m\theta)\), set
\(g_m=(-1)^me_m\), and form the residue sums
\begin{equation}
  \mathcal G_s^{(r)}=
  \sum_{\substack{-L\leqslant m\leqslant L\\m\equiv s\ ({\rm mod}\ r)}}
  g_{|m|}.
\end{equation}
The coefficients \(e_m\) are obtained directly by collecting the
\(\psi_\ell\) terms in \eqref{eq:app-improper-jet-character}.  The closed residue formula is
\begin{equation}
  d_k^{\rm raw}(C_{rh})=\frac14\left[
  \sum_s\mathcal B_s^{(r)}\mathcal B_{-s}^{(r)}+
  \sum_s\mathcal G_s^{(r)}\mathcal G_{-s}^{(r)}+2\mathcal T_r\right].
  \label{eq:app-Crh}
\end{equation}
Its continuous limit is the parity polynomial
\begin{equation}
  d_k^{\rm raw}(C_{\infty h})
  =\frac32k^3-\frac92k^2+\frac72k+
  \begin{cases}1,&k\ \text{even},\\-\tfrac12,&k\ \text{odd}.
  \end{cases}
  \label{eq:app-Cinfh-polynomial}
\end{equation}

Finally, let \(S_{2r}=\langle R_{\pi/r}\sigma_h\rangle\).  With
\(\alpha=\pi/r\), the even powers are proper and the odd powers improper, so
\begin{equation}
  d_k^{\rm raw}(S_{2r})=\frac1{4r}\sum_{j=0}^{2r-1}
  \left\{X_j^2+\chi_k(2j\alpha)\right\},
  \quad
  X_j=\begin{cases}
  \chi_k(j\alpha),&j\ \text{even},\\
  \eta_k(\pi+j\alpha),&j\ \text{odd}.
  \end{cases}
  \label{eq:app-S2r}
\end{equation}

\subsection{Dihedral, polyhedral and spherical families}

It is convenient to define the proper-element contribution
\begin{equation}
  A_k(\theta)=\frac12\{\chi_k(\theta)^2+\chi_k(2\theta)\}
\end{equation}
and the improper-element contribution
\begin{equation}
  B_k(\theta;\varphi)=\frac12\{\eta_k(\theta)^2+\chi_k(\varphi)\}.
\end{equation}
Adding the \(r\) perpendicular half-turns to \(C_r\) gives
\begin{align}
  d_k^{\rm raw}(D_r)
  &=\frac12d_k^{\rm raw}(C_r)+\frac12A_k(\pi),
  \label{eq:app-Dr}\\
  d_k^{\rm raw}(D_{rh})
  &=\frac12d_k^{\rm raw}(C_{rh})
    +\frac14\{A_k(\pi)+B_k(\pi;0)\}.
  \label{eq:app-Drh}
\end{align}
The antiprismatic family is generated, in the convention used throughout the
paper, by \(R=R_z(2\pi/r)\), \(C=R_x(\pi)\), and a diagonal reflection
\(\sigma_d\) whose plane makes angle \(\pi/(2r)\) with a perpendicular
twofold axis.  Its \(4r\) elements are
\begin{equation}
  \{R^j,\,CR^j,\,\sigma_dR^j,\,C\sigma_dR^j:
  0\leqslant j<r\}.
  \label{eq:app-Drd-elements}
\end{equation}
Substitution of this explicit list into
\eqref{eq:app-universal-fixed-dimension} is the closed formula for
\(D_{rd}\); it expands into polynomials on residue classes modulo \(2r\).
The continuous dihedral limits simplify to
\begin{equation}
  d_k^{\rm raw}(D_\infty)=
  \begin{cases}
  \frac32k^3-\frac{15}{4}k^2+5,&k\ \text{even},\\
  \frac32k^3-\frac{15}{4}k^2+\frac{17}{4},&k\ \text{odd},
  \end{cases}
  \label{eq:app-Dinf-polynomial}
\end{equation}
and
\begin{equation}
  d_k^{\rm raw}(D_{\infty h})=
  \begin{cases}
  \frac34k^3-\frac{11}{8}k^2-\frac14k+2,&k\ \text{even},\\
  \frac34k^3-\frac{11}{8}k^2-\frac14k+\frac78,&k\ \text{odd}.
  \end{cases}
  \label{eq:app-Dinfh-polynomial}
\end{equation}

The exceptional groups are most compactly reproduced from their conjugacy
classes.  Put \(A_0=A_k(0)\), \(A_2=A_k(\pi)\),
\(A_4=A_k(\pi/2)\), and
\(A_5^\Sigma=A_k(2\pi/5)+A_k(4\pi/5)\).  Then
\begin{align}
  d_k^{\rm raw}(T)&=\frac{A_0+3A_2}{12},\\
  d_k^{\rm raw}(O)&=\frac{A_0+6A_4+9A_2}{24},\\
  d_k^{\rm raw}(I)&=\frac{A_0+15A_2+12A_5^\Sigma}{60}.
  \label{eq:app-proper-polyhedral}
\end{align}
For the full and improper polyhedral groups, let
\(B_{00}=B_k(0;0)\), \(B_{20}=B_k(\pi;0)\),
\(B_4=B_k(\pi/2;\pi)\), and
\begin{equation}
  B_5^\Sigma=B_k(2\pi/5;4\pi/5)+B_k(4\pi/5;2\pi/5).
\end{equation}
The remaining class sums are
\begin{align}
  d_k^{\rm raw}(T_d)
  &=\frac{A_0+3A_2+6B_{20}+6B_4}{24},\\
  d_k^{\rm raw}(T_h)
  &=\frac{A_0+3A_2+B_{00}+3B_{20}}{24},\\
  d_k^{\rm raw}(O_h)
  &=\frac{A_0+6A_4+9A_2+B_{00}+6B_4+9B_{20}}{48},\\
  d_k^{\rm raw}(I_h)
  &=\frac{A_0+15A_2+12A_5^\Sigma+B_{00}+15B_{20}+12B_5^\Sigma}{120}.
  \label{eq:app-full-polyhedral}
\end{align}
These formulae contain only evaluations of
\eqref{eq:app-proper-jet-character} and
\eqref{eq:app-improper-jet-character}.  They therefore reproduce all
polyhedral congruence subcases without a separate lookup table.  Note that
the threefold rotation and rotoreflection classes are absent from all these
class sums: from \eqref{eq:app-axial-weights} one finds
\(\chi_k(2\pi/3)=-1\) and \(\eta_k(2\pi/3)^2=1\) for every
\(k\geqslant3\), so their contributions \(A_k(2\pi/3)\) and
\(B_k(2\pi/3;2\pi/3)\) vanish identically and are omitted.

For the spherical rows, Schur's lemma \citep{serre1977linear} counts one scalar coefficient for every
pair of equivalent irreducible copies.  The multiplicities in
\eqref{eq:app-stokes-jet-decomposition} give directly
\begin{equation}
  d_k^{\rm raw}(\SO(3))=6k-14,
  \qquad
  d_k^{\rm raw}(\OO(3))=4k-9.
  \label{eq:app-spherical-polynomials}
\end{equation}
At \(k=4\), these are \(10\) and \(7\), as in
Table~\ref{tab:master-observability}.

\subsection{From raw formulae to the numbers in the master table}

The origin and frame reductions must be applied after the fixed-space
calculation.  First,
\begin{equation}
  d_k^{\rm ctr}(H)=d_k^{\rm raw}(H)-\dim(V_1^-)^H.
  \label{eq:app-centre-correction}
\end{equation}
Thus the origin correction is \(3\) for \(C_1\), \(2\) for \(C_s\), \(1\)
for \(C_r,C_{rv}\) and their continuous limits, and zero for the remaining
standard rows.  Second, let
\begin{equation}
  \mathfrak r(H)=
  \operatorname{Lie}N_{\OO(3)}(H)/\operatorname{Lie}H
\end{equation}
be the continuous residual freedom of an adapted frame.  For a generic
\(R\in(\mathcal R_{k-1}^{\rm ctr})^H\), set
\begin{equation}
  s_k(H)=\operatorname{rank}
  \{X\mapsto\rho_k(X)R:X\in\mathfrak r(H)\}.
  \label{eq:app-residual-orbit-rank}
\end{equation}
The reduced dimension printed for a visible group is
\begin{equation}
  \mu_k(H)=d_k^{\rm ctr}(H)-s_k(H).
  \label{eq:app-reduced-family-dimension}
\end{equation}
Generically, \(s_k=3\) for \(C_1,C_i\), \(s_k=1\) for \(C_s\), and
\(s_k=1\) for a visible finite \(C_r,C_{rh}\) or \(S_{2r}\) row.  It is zero
for \(C_{rv}\), all dihedral and polyhedral groups, and all continuous rows;
in the latter case the axial phase or spherical rotation already belongs to
the stabilising group.  If a finite row has collapsed to a continuous shadow,
the shadow's value is used instead.

\section{Numerical audits of dimensions, reductions and collapses}
\label{app:numerical-audits}

The systematic counts in the paper were checked by two independent
computations: a direct nullspace audit and a character audit.  

\subsection{Concrete representation bases}

Every elementary space is represented in an orthonormal Euclidean basis.
Polar vectors use \(Q\), axial vectors use \(\det(Q)Q\), and strain tensors use
an orthonormal basis of symmetric traceless \(3\times3\) matrices.  The
quadratic jet is first embedded in \(\mathbb R^{27}\).  We impose
\begin{equation}
  J_{ijk}=J_{ikj},\qquad J_{iik}=0,
  \label{eq:app-quadratic-constraints}
\end{equation}
and take an orthonormal basis of the nullspace of these constraints, producing
the expected \(15\)-dimensional space.  The three-index action
\begin{equation}
  (Q\cdot J)_{ijk}=Q_{ia}Q_{jb}Q_{kc}J_{abc}
\end{equation}
is then restricted to that basis.

For a symmetric block on a representation \(U\), we construct the induced
action on an orthonormal basis of \(\operatorname{Sym}^2U\).  For a rectangular
map \(M:U\to V\), the coordinate convention of the resistance matrices gives
\begin{equation}
  M\longmapsto \rho_V(Q)^T M\rho_U(Q).
  \label{eq:app-cross-action}
\end{equation}
The centred translation--rotation block is treated as a symmetric
pseudotensor and therefore carries an additional factor \(\det Q\).  A response
level is the direct sum of its block actions.

\subsection{Groups and direct nullity}

Finite axial groups are generated from the standard rotation, perpendicular
half-turn and horizontal, vertical or diagonal reflection specified by their
Schoenflies symbol.  Matrix products are closed until no new element appears,
and the resulting order is checked against the expected \(r,2r\), or \(4r\).
Polyhedral groups are recovered as all orthogonal maps preserving a vertex
set, with the determinant selecting the proper subgroup. 

For a representation space \(V\) and generators or full group elements
\(Q_1,\ldots,Q_p\), we stack the invariance equations
\begin{equation}
  L_H=
  \begin{pmatrix}
  \rho(Q_1)-I\\ \vdots\\ \rho(Q_p)-I
  \end{pmatrix},
  \qquad
  V^H=\ker L_H.
  \label{eq:app-stacked-nullspace}
\end{equation}
The direct audit evaluates
\begin{equation}
  \dim V^H=\dim V-\operatorname{rank}L_H
\end{equation}
by SVD, with a relative numerical tolerance of \(10^{-8}\) in the present
orthonormal scaling.  The singular spectrum is inspected whenever the answer
changes under a modest tolerance perturbation.  Continuous axial groups are
audited either with their Lie generator plus the required reflections, or
with a cyclic sample whose order exceeds the largest Fourier weight.  Generic
rotations about non-parallel axes, supplemented by inversion for \(\OO(3)\),
audit the spherical fixed spaces.

\subsection{Independent character audit and reduction audit}

The character audit never forms \(L_H\).  For each group element it computes
the traces of the elementary polar, axial, strain and jet representations,
combines them with
\begin{equation}
  \chi_{U\otimes V}=\chi_U\chi_V,
  \qquad
  \chi_{\operatorname{Sym}^2U}(Q)
  =\frac{\chi_U(Q)^2+\chi_U(Q^2)}2,
\end{equation}
and averages the resulting level character.  The numerical average must lie
within \(10^{-7}\) of an integer.  At every tabulated group and level, this
integer is compared with the nullity in
\eqref{eq:app-stacked-nullspace}.  The spherical entries are supplied by the
exact irreducible calculation \eqref{eq:app-spherical-polynomials}, rather
than by a finite sample masquerading as a continuous integral.

The origin reduction is audited twice: by replacing the raw polar--axial
cross-block with the symmetric pseudotensor representation, and by checking
the difference \(\dim(V_1^-)^H\) in
\eqref{eq:app-centre-correction}.  For the residual adapted-frame quotient, a
deterministic generic vector is formed in the centred fixed-space basis.  The
infinitesimal orbit columns
\begin{equation}
  \frac{\rho(\exp(\varepsilon X_j))R-R}{\varepsilon},
  \qquad \varepsilon=10^{-6},
\end{equation}
are assembled for a basis of \(\mathfrak r(H)\), and their rank gives
\(s_k(H)\) in \eqref{eq:app-residual-orbit-rank}.  Several deterministic
generic coefficient vectors are used to guard against an accidental
lower-stabiliser sample.

Finally, two groups are declared to have the same hydrodynamic shadow at a
given level only when their fixed-space projectors have the same range.  Numerically, this is tested through the mutual residuals
\begin{equation}
  \|(I-\Pi_H)\Pi_K\|,
  \qquad
  \|(I-\Pi_K)\Pi_H\|.
\end{equation}
Subgroup monotonicity and the expected finite-to-continuous collapse above the
highest axial weight provide further structural checks.

\section{Construction of detailed normal forms}
\label{app:normal-forms}

Dimensions say how many coefficients are allowed; a dynamical equation needs
to know where they sit.  This appendix gives a constructive procedure for
obtaining a detailed normal form for any group and any response block.

\subsection{Equivariant map spaces}

Let \(U\) be an input representation, \(V\) an output representation, and
\(M\in\operatorname{Hom}(U,V)\).  The \(H\)-equivariance condition is
\begin{equation}
  \rho_V(h)M=M\rho_U(h),\qquad h\in H.
  \label{eq:app-equivariance}
\end{equation}
After vectorisation,
\begin{equation}
  \left[I_U\otimes\rho_V(h)-\rho_U(h)^T\otimes I_V\right]
  \operatorname{vec}M=0.
  \label{eq:app-vectorised-equivariance}
\end{equation}
It is enough to stack these equations for a generating set of \(H\).  For a
continuous component, one differentiates at the identity.  If \(X\) is a Lie
algebra generator, the corresponding equation is
\begin{equation}
  \mathrm d\rho_V(X)M-M\mathrm d\rho_U(X)=0.
  \label{eq:app-lie-equivariance}
\end{equation}
Discrete reflections or half-turns are then added to distinguish, for example,
\(C_\infty\), \(C_{\infty v}\), \(D_\infty\), and \(D_{\infty h}\).

For a reciprocal diagonal block, \(U=V\) and \(M=M^T\); one solves
\eqref{eq:app-equivariance} directly in an orthonormal basis of
\(\operatorname{Sym}^2U\).  For the complete resistance operator, the
solutions for all diagonal and reciprocal cross-blocks are placed into the
block matrix \eqref{eq:resistance-tensors}.  

Signed-permutation groups such as \(D_2,T,O\) permit exact rational
row-reduction.  Groups involving fifth-order axes or diagonal planes can be
treated in the corresponding algebraic number field, or numerically by SVD
followed by recognition of simple radicals.  In either case, a candidate basis
\(M_1,\ldots,M_p\) is accepted only after checking
\begin{equation}
  \max_{h\in\mathcal G}\|\rho_V(h)M_j-M_j\rho_U(h)\|=0
  \label{eq:app-normal-form-residual}
\end{equation}
exactly or to the stated numerical tolerance for the chosen generators
\(\mathcal G\), and after confirming that \(p\) equals the character count.
Sparse bases are preferred for exposition, but any invertible change of basis
in coefficient space describes the same normal form.

\subsection{From a normal form to the particle equations}

For linear flow, let \(Q\) map body coordinates to laboratory coordinates.
The strain seen in the body is
\begin{equation}
  \vect{E}_b=Q^T\vect{E}^\infty Q.
\end{equation}
If \(M_1^\Omega,\ldots,M_{p_\Omega}^\Omega\) is a basis of
\(\operatorname{Hom}_H(V_2^+,V_1^+)\), then the most general
symmetry-allowed angular correction is
\begin{equation}
  \vect{\gamma}_b(\vect{E}_b)
  =\sum_{j=1}^{p_\Omega}\beta_jM_j^\Omega\vect{E}_b,
  \qquad
  \vect{\Omega}=\vect{\Omega}^\infty+Q\vect{\gamma}_b.
  \label{eq:app-linear-angular-normal-form}
\end{equation}
This is the direct route to the Jeffery--Ishimoto and tetrahedral laws in
\S\ref{sec:shear-dynamics}.  For an axial director
\(\vect{d}=Q\vect{e}_3\), one projects away the body spin parallel to
\(\vect{e}_3\) before writing
\(\dot{\vect{d}}=\vect{\Omega}\times\vect{d}\).

For a quadratic incident jet, the body components are
\begin{equation}
  (\vect{J}_b)_{ijk}=Q_{ai}Q_{bj}Q_{ck}(\vect{J}^\infty)_{abc}.
\end{equation}
A basis \(M_j^J\in\operatorname{Hom}_H(\mathcal H_2,V_1^+)\) gives
\begin{equation}
  \vect{\gamma}_b^{(J)}(\vect{J}_b)
  =\sum_{j=1}^{q_\Omega}\beta_j^J M_j^J\vect{J}_b.
  \label{eq:app-quadratic-angular-normal-form}
\end{equation}
Equations \eqref{eq:app-linear-angular-normal-form} and
\eqref{eq:app-quadratic-angular-normal-form}, together with the ambient
vorticity, determine the attitude equation
\begin{equation}
  \dot Q=\widehat{\vect{\Omega}}Q.
  \label{eq:app-attitude-matrix-equation}
\end{equation}

If one begins with a full resistance normal form rather than a projected map,
partition its force--torque block as \(\mathcal R_1\) and its incident-flow
coupling as \(B\).  Force- and torque-free motion gives
\begin{equation}
  \begin{pmatrix}\vect{V}\\\vect{\Omega}\end{pmatrix}
  =-\mathcal R_1^{-1}B\,\mathcal U_b^\infty.
  \label{eq:app-eliminate-force-torque}
\end{equation}

\section{Attitude integration, Poincar\'e sections and finite-time Lyapunov exponents}
\label{app:numerical-dynamics}

This appendix expands the numerical procedure summarised in
\S\ref{sec:particle-dynamics}.

\subsection{Attitude representation and integrators}

We represent the attitude by a unit quaternion \(q=(q_0,\vect{q})\) and
use the convention
\begin{equation}
  \dot q=\frac12(0,\vect{\Omega})\otimes q.
  \label{eq:app-quaternion-ode}
\end{equation}
The angular velocity is reconstructed in the laboratory frame from the
body-frame normal form as described in Appendix~\ref{app:normal-forms}. One
fixed-step update of size \(\Delta t\), used for the tetrahedral and
octahedral trajectory and Poincar\'e calculations, uses classical RK4. Each provisional stage is
normalised before evaluating the next right-hand side, and the final
quaternion is normalised again.  Since \(q\) and \(-q\) represent the same
rotation, the sign is selected so that its scalar product with the preceding
quaternion is positive.  

The large finite-time Lyapunov screen in Figure~\ref{fig:lyapunov} instead
integrates the reference and shadow quaternions together with the adaptive
\texttt{ode113} solver. Note that Euler angles are never integrated -- they are used only
to construct the initial attitudes and report section coordinates.

\subsection{Shadow-trajectory Lyapunov estimate}

Let \(q\) be a reference solution and \(q'\) a shadow solution. Initially,
\begin{equation}
  q_0'=\left(\cos\frac{\delta_0}{2},
  \sin\frac{\delta_0}{2}\vect{a}_0\right)\otimes q_0,
  \qquad \delta_0=10^{-7},
\end{equation}
where \(\vect{a}_0=(1,\sqrt2,\sqrt3)/\sqrt6\). Both attitudes are advanced by the
same adaptive solve, so that their local errors remain correlated. At
intervals of 10 time units, form the relative quaternion
\begin{equation}
  q_{\rm rel}=q'\otimes q^{-1}
\end{equation}
and choose its sign with non-negative scalar part.  If
\(q_{\rm rel}=(w,\vect{v})\), the geodesic separation and direction are
\begin{equation}
  \delta=2\operatorname{atan2}(\|\vect{v}\|,w),
  \qquad \vect{a}=\vect{v}/\|\vect{v}\|.
\end{equation}
The logarithmic stretch \(\log(\delta/\delta_0)\) is accumulated after the
transient, and the shadow is reset to
\begin{equation}
  q'=\left(\cos\frac{\delta_0}{2},
  \sin\frac{\delta_0}{2}\vect{a}\right)\otimes q.
\end{equation}
For a retained duration \(T_r\), the finite-time largest exponent is
\begin{equation}
  \lambda_T=\frac1{T_r}\sum_j\log\frac{\delta_j}{\delta_0}.
  \label{eq:app-ftle}
\end{equation}
The same sum is recorded separately over the first and second halves,
producing \(\lambda^{(1)}\) and \(\lambda^{(2)}\). This is a finite-amplitude
Benettin estimate on \(\SO(3)\). The computation uses relative and absolute
tolerances \(10^{-10}\) and \(10^{-12}\), respectively, with maximum solver
step \(0.25\). Tightening these tolerances, changing \(\delta_0\) within the
linear-separation regime, and varying the renormalisation interval provide
the natural numerical-resolution checks.

In the survey of Figure~\ref{fig:lyapunov}, a trajectory is classified as
chaotic when both half-window estimates exceed \(8\times10^{-3}\) and
\begin{equation}
  |\lambda^{(2)}-\lambda^{(1)}|<
  \max(0.02,0.75|\lambda_T|).
  \label{eq:app-lyapunov-convergence}
\end{equation}
Otherwise, steady motion requires both a late mean angular speed below
\(10^{-4}\) and a quotient-attitude spread below \(10^{-2}\). Periodicity is
tested using a neutral late exponent, a low spectral entropy, a close return,
and median and upper-decile cycle errors below \(10^{-2}\) and
\(3\times10^{-2}\), respectively. Trajectories that meet none of these tests
are marked unclassified.

\subsection{Poincar\'e section and initial-attitude design}

The section is the \(ZXZ\) surface \(\phi=0\pmod{2\pi}\), reported in
\((\psi,\theta)\).  Direct event detection in Euler angles would be fragile at
their coordinate singularities.  Instead, for the rotation matrix \(Q\), we
use
\begin{equation}
  Q_{13}=\sin\theta\sin\phi,
  \qquad Q_{23}=-\sin\theta\cos\phi.
\end{equation}
A sign change of \(Q_{13}\) brackets a crossing. Linear interpolation gives
the reported section coordinates; the branch condition \(Q_{23}<0\) selects
\(\phi=0\) rather than \(\phi=\pi\), and crossings with
\(\sqrt{Q_{13}^2+Q_{23}^2}<10^{-7}\) are discarded as Euler-coordinate
singularities.  The plotted coordinates are then
\begin{equation}
  \theta=\arccos Q_{33},\qquad
  \psi=\operatorname{atan2}(Q_{31},Q_{32})\pmod{2\pi}.
\end{equation}
The sign of the bracketed change in \(Q_{13}\) defines the crossing direction.

Initial attitudes are deterministic and quasi-uniform.  A three-dimensional
Halton sequence in bases \(2,3,5\) gives \(u_1,u_2,u_3\in[0,1)\), from which
\begin{equation}
  \alpha=2\pi u_1,\qquad
  \beta=\arccos(1-2u_2),\qquad
  \gamma=2\pi u_3.
  \label{eq:app-haar-euler}
\end{equation}
Uniform \(\alpha,\gamma\) and uniform \(\cos\beta\) are precisely the
normalised Haar volume in $ZXZ$ coordinates.  The same attitude set is used
for every column of a figure, so changes in the section reflect the
coefficients rather than a changed initial ensemble.

\subsection{Parameters used in the three computations}

For Figure~\ref{fig:tetrahedral}, the representative body-axis trajectories
use \(\Delta t=0.02\), \(T=600\), and save every five steps.  The Poincar\'e
maps use \(160\) initial attitudes, \(\Delta t=0.02\), \(T=1000\), save every
five steps, discard the first \(15\%\), and retain both crossing directions.

Figure~\ref{fig:lyapunov} retains 14 genuinely full-attitude laws and omits
the roll-invariant axial rows and the attitude-independent \(I\) and
\(\SO(3)\) controls. Each retained class uses 1500 coefficient vectors: all
signed coordinate axes followed by deterministic Gaussian mixed directions,
with every vector normalised to Euclidean norm \(1.5\). The vectors are paired
with a common 1500-point deterministic Haar attitude design. The adaptive
\texttt{ode113} integration runs to \(T=1600\), uses relative and absolute
tolerances \(10^{-10}\) and \(10^{-12}\), maximum step \(0.25\), and a
Lyapunov-renormalisation interval of 10; the first \(35\%\) is discarded.
Group medians, upper deciles and classification fractions are computed from
these 1500 finite-time estimates.

For Figure~\ref{fig:octahedral}, the representative trajectories use
\(\Delta t=0.025\), \(T=400\), and discard \(25\%\).  The Poincar\'e maps use
\(100\) initial attitudes, \(\Delta t=0.03\), \(T=10^4\), save every four
steps, discard \(20\%\), and retain only positive crossings.  The code reports
the minimum, median and maximum numbers of crossings per initial attitude.
\end{appen}


\bibliographystyle{jfm}

\bibliography{hydroshape}

\end{document}